\documentclass[twocolumn,showpacs,showkeys,preprintnumbers,superscriptaddress,amsmath,floatfix,amssymb,aps,secnumarabic,nofootinbib,floatfix]{revtex4-1}
\usepackage[colorlinks=true]{hyperref}
\usepackage{graphicx}
\usepackage{dblfloatfix}    
\usepackage{placeins}
\usepackage{braket}
\usepackage{amsmath}
\usepackage{epsfig}
\usepackage{float}
\usepackage{nicefrac}
\usepackage{xcolor}
\usepackage{color}
\usepackage{bm}
\usepackage{subfigure}
\usepackage{chngcntr}

\def\({\left(}
\def\){\right)}
\def\[{\left[}
\def\]{\right]}

\newcommand{\ve}[1]{\boldsymbol{#1}}

\newcommand{\diag}[1]{ {\rm diag} \, \left( #1 \right) }
\newcommand{\Tr}{ {\rm Tr} \, }

\newcommand{\beq} {\begin{eqnarray}}
\newcommand{\eeq} {\end{eqnarray}}

\newcommand{\nn}{ \nonumber}

\newcommand{\comment}[1]{}

\begin{document}
\sloppy

\title{Instanton gas approach to the Hubbard model}

\author{Maksim~Ulybyshev}
\email{Maksim.Ulybyshev@physik.uni-wuerzburg.de}
\affiliation{Institute for Theoretical Physics, Julius-Maximilians-Universit\"at W\"urzburg,
   97074 W\"urzburg, Germany}

\author{Christopher~Winterowd}
\email{winterowd@itp.uni-frankfurt.de}
\affiliation{Johann Wolfgang Goethe-Universit\"at Frankfurt am Main,Frankfurt am Main, Germany}

\author{Fakher~Assaad}
\email{Fakher.Assaad@physik.uni-wuerzburg.de}
\affiliation{Institute for Theoretical Physics, Julius-Maximilians-Universit\"at W\"urzburg,
   97074 W\"urzburg, Germany}
\affiliation{W\"urzburg-Dresden Cluster of Excellence ct.qmat, Julius-Maximilians-Universit\"at W\"urzburg,
   97074 W\"urzburg, Germany}

\author{Savvas~Zafeiropoulos}
\email{Savvas.Zafeiropoulos@cpt.univ-mrs.fr}
\affiliation{ Aix Marseille Univ, Universit\'e de Toulon, CNRS, CPT, Marseille, France}

\begin{abstract}
 In this  article    we  consider   a  path  integral  formulation of the  Hubbard model  based on a  Hubbard-Stratonovich  transformation  that couples the auxiliary field to  
the  local  electronic  density.   This  decoupling is   known to have a  saddle-point  structure   that  shows a   remarkable regularity:   the  field  configuration at  each  saddle  point  can  be understood  in terms of  a  set of elementary   field  configurations localized in  space  and  imaginary  time  which  we coin instantons.    The  interaction between instantons is  short  ranged.    Here, we  formulate  a   classical  partition function for  the  instanton  gas that  has  predictive power.  For  a given  set of physical parameters,  we  can predict  the   distribution of   instantons  and   show  that the  instanton   number is  sharply defined in the  thermodynamic limit,    thereby  defining    a  unique   dominant saddle point.     Decoupling  in the  charge  channel,  conserves   SU(2)   spin  symmetry for  each  field  configurations.  Hence,  the   instanton approach  provides  an  SU(2)  spin-symmetric  approximation to  the  Hubbard model.   It  fails, however,  to capture the  magnetic   transition  inherent  to the Hubbard model on the honeycomb lattice despite being able to describe  local moment  formation.  In  fact,  the   instanton itself    corresponds  to local moment  formation and concomitant short-ranged  anti-ferromagnetic correlations.  
This   aspect is  also seen  in the  single particle  spectral  function   that  shows  clear  signs of the  upper and lower Hubbard \textit{bands}.      Our instanton approach  bears   remarkable similarities  to  local  dynamical approaches,   such  as   dynamical  mean  field  theory,   in the sense  that it has  the unique  property of   allowing  for  local moment formation without  breaking  the  SU(2) spin  symmetry. In contrast to local approaches, it captures  short-ranged magnetic fluctuations.  Furthermore, it  also offers  possibilities   for  systematic  improvements  by  taking into account fluctuations around  the  dominant saddle point.   Finally,   we   show  that the  saddle point structure  depends  upon the  choice of lattice geometry.  For  the  square lattice  at half-filling,  the  saddle point structure  reflects the  itinerant  to  localized  nature  of the magnetism  as  a function of  the coupling strength.    The  implications of  our  results  for  Lefschetz thimbles  approaches  to  alleviate the  sign  problem are also discussed.   
\end{abstract}

\maketitle

\section*{\label{sec:Intro}Introduction}

A strong local  Coulomb  repulsion between electrons  leads to  the  localization   of  charge  degrees of  freedom  and  to the  formation of local 
  magnetic  moments. 
As  shown in Anderson's   seminal paper \cite{Anderson61}, local  moment  formation  in metals can  be  captured  at  the  mean-field  level    by   breaking the  spin-rotational  symmetry.     Generically,  however,  local moment  formation is a  dynamical  effect  in   which  the  net  moment   averages to zero  
over  time,  thus  restoring  the spin  rotational  symmetry.     The  success  of  the  so-called   dynamical  mean-field  theory  (DMFT)    \cite{Metzner89,Georges96},   is  that it  captures this  phenomena.   local moment  formation and  the associated  short-range  magnetic  fluctuations in   metals   presents a   key  challenge  in  the  understanding of strongly-correlated  electron systems,  and  has   important  implications  for  the   understanding of    transition metal oxides   such as  
high-temperature  superconductors  \cite{Imada_rev} or     rare-earth   heavy   fermion  materials    \cite{Fisk_rev,Coleman07_rev}.

The  aim  of  this  paper  is  to provide  a   framework  that  captures  local moment  formation  in metallic  environments.   In  contrast  to   DMFT,     both  temporal  and  spatial  fluctuations    will be   taken into account. 
We  will concentrate  on the  Hubbard model  on the square  and  honeycomb  lattices, working  within a path-integral  formulation in order to derive  our   approximation from the 
saddle point structure.  Clearly,  the path-integral  formulation of   the  Hubbard model  on a given lattice  is not unique and    the  saddle point  structure  will depend  on the  specific  treatment of the interaction term. For instance,  one    can   use a  decoupling  where the real  scalar  field  couples to the 
local magnetization.  As  a  consequence,  the  saddle  point  structure  will  correspond to states where the spin symmetry  is broken.    However,  if   the integration over  the scalar  field    is  carried out  exactly, the  final  result   will  be  spin  rotational symmetric and  independent of the decoupling  channel.

As   we  want   to  describe local moment  formation   without  explicitly breaking  the  spin symmetry, we  will   adopt   a path-integral  formulation  where a real  space-  and  time- dependent  scalar    field  couples to the   local   charge  degree of  freedom.     For   this  choice  of  Hubbard-Stratonovich (HS) transformation,   SU(2)   spin symmetry is present  for  all  field  configurations.  Solving  the  saddle  point  equations  under  the assumption of  fields which are constant in space  and  time reduces  to  the  paramagnetic mean-field  approximation to the Hubbard model \cite{Hirsch85}    in  which  the  field  vanishes.

 \begin{figure}[]
   \centering
  \includegraphics[width=0.35\textwidth,clip]{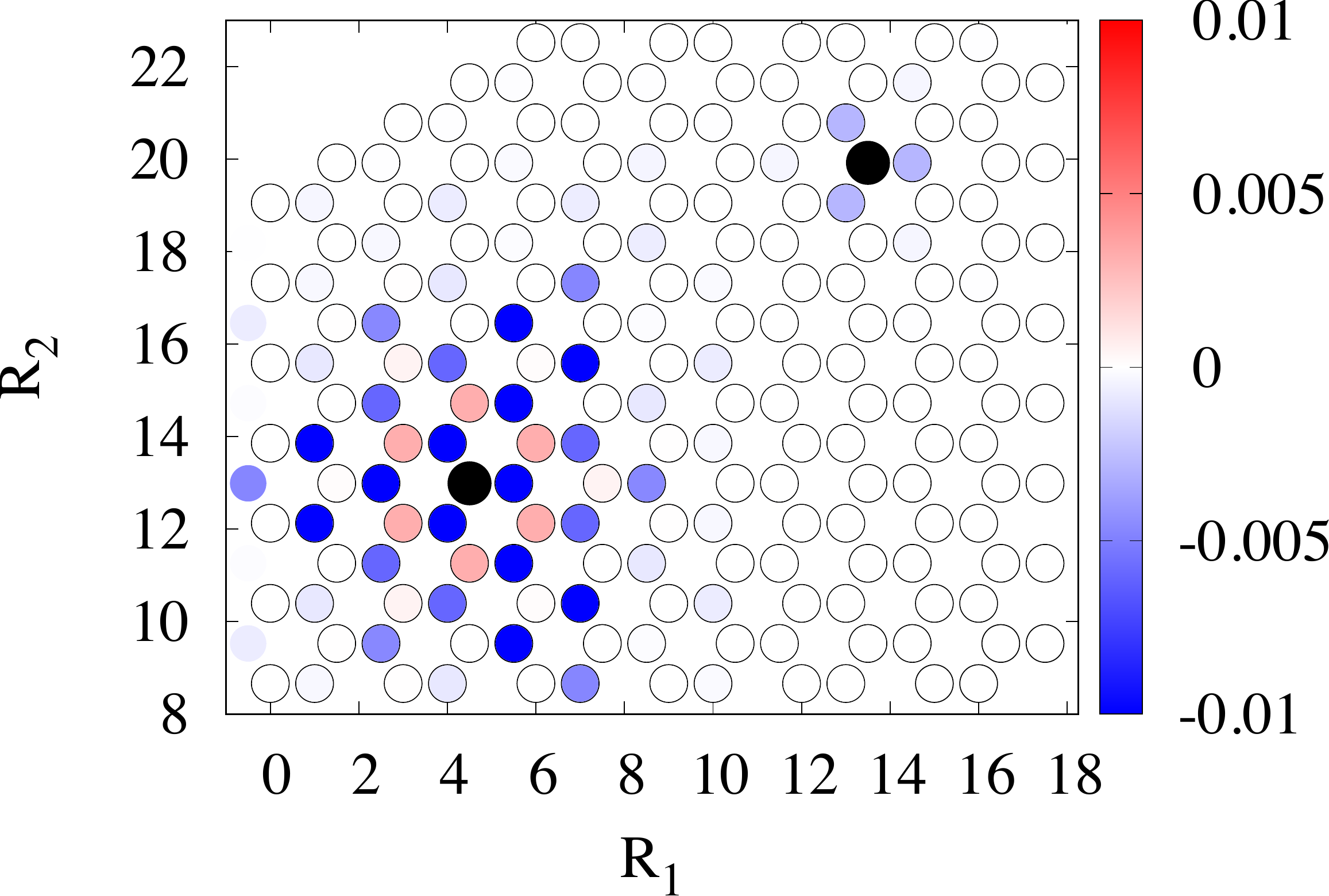}
         \caption{
Spin-spin correlations, $\frac{1}{3}\langle \hat{\ve{S}}_{\ve{x}_0}(T) \cdot \hat{\ve{S}}_{\ve{x}_0  +  \ve{x}} (T) \rangle$,  for a field  configuration  with one instanton at space time   $(X,T)$.   We  consider two  values of   $\ve{x}_0$.   The left  black  circle  corresponds  to  $\ve{x}_0 = X$.    The other value of  $\ve{x}_0$  (right black  circle)  is  \text{far}   from the  instanton.  $R_1$ and $R_2$ are two Cartesian coordinates of the lattice sites, displayed in the units of the distance between nearest neighbours. These calculations were performed on a $12\times12$ lattice at interaction strength $U=2.0\kappa$ (see sections \ref{subsec:Model} and \ref{subsec:path_int} for the notation). } 
          \label{fig:SpinCorrInstantonVacuumMap}
  \end{figure}

  \begin{figure}[]
   \centering
   \subfigure[]{\label{fig:spectralFunctionsU6QMC}\includegraphics[width=0.35\textwidth , angle=0]{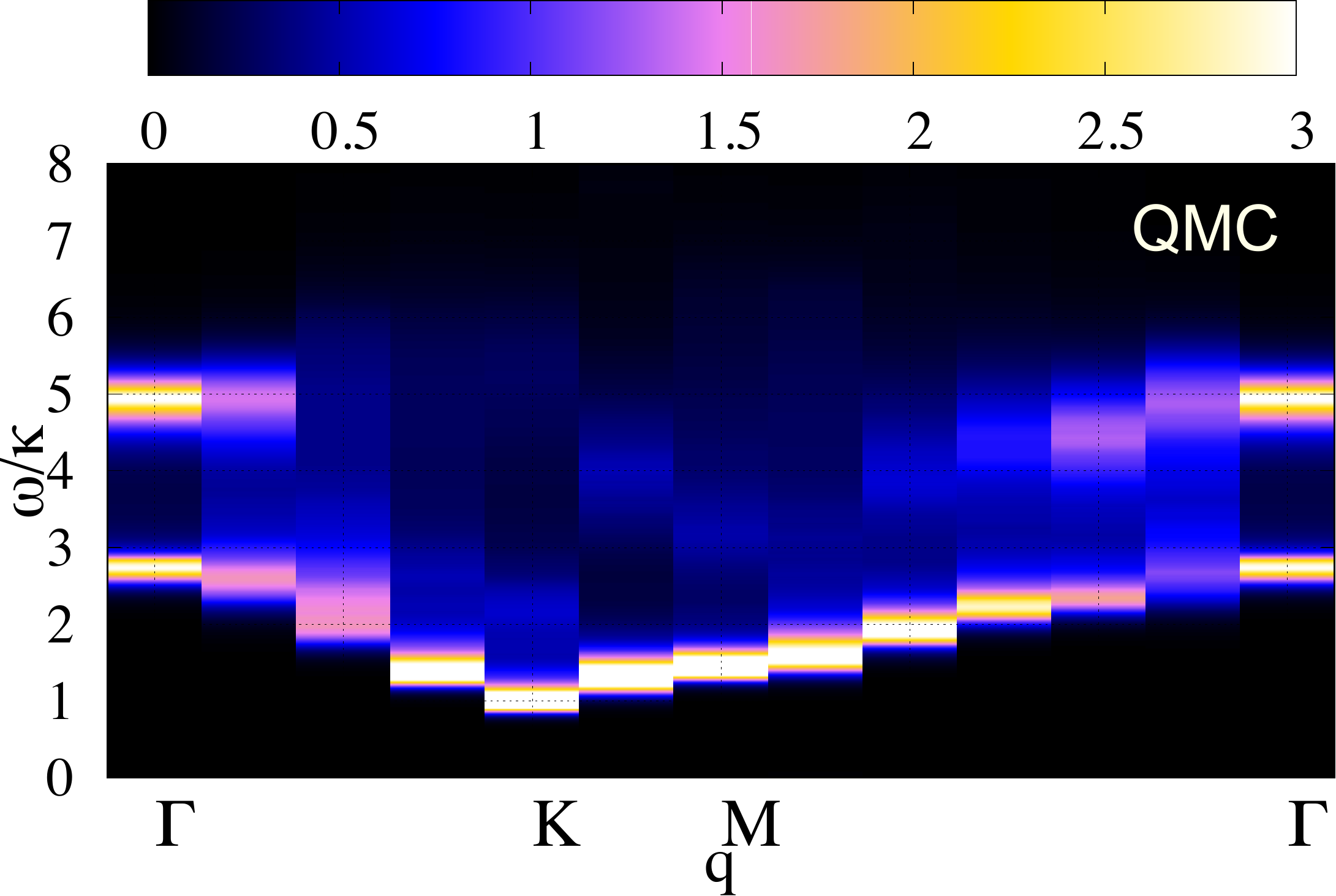}}
   \subfigure[]{\label{fig:spectralFunctionsU6Saddles}\includegraphics[width=0.35\textwidth , angle=0]{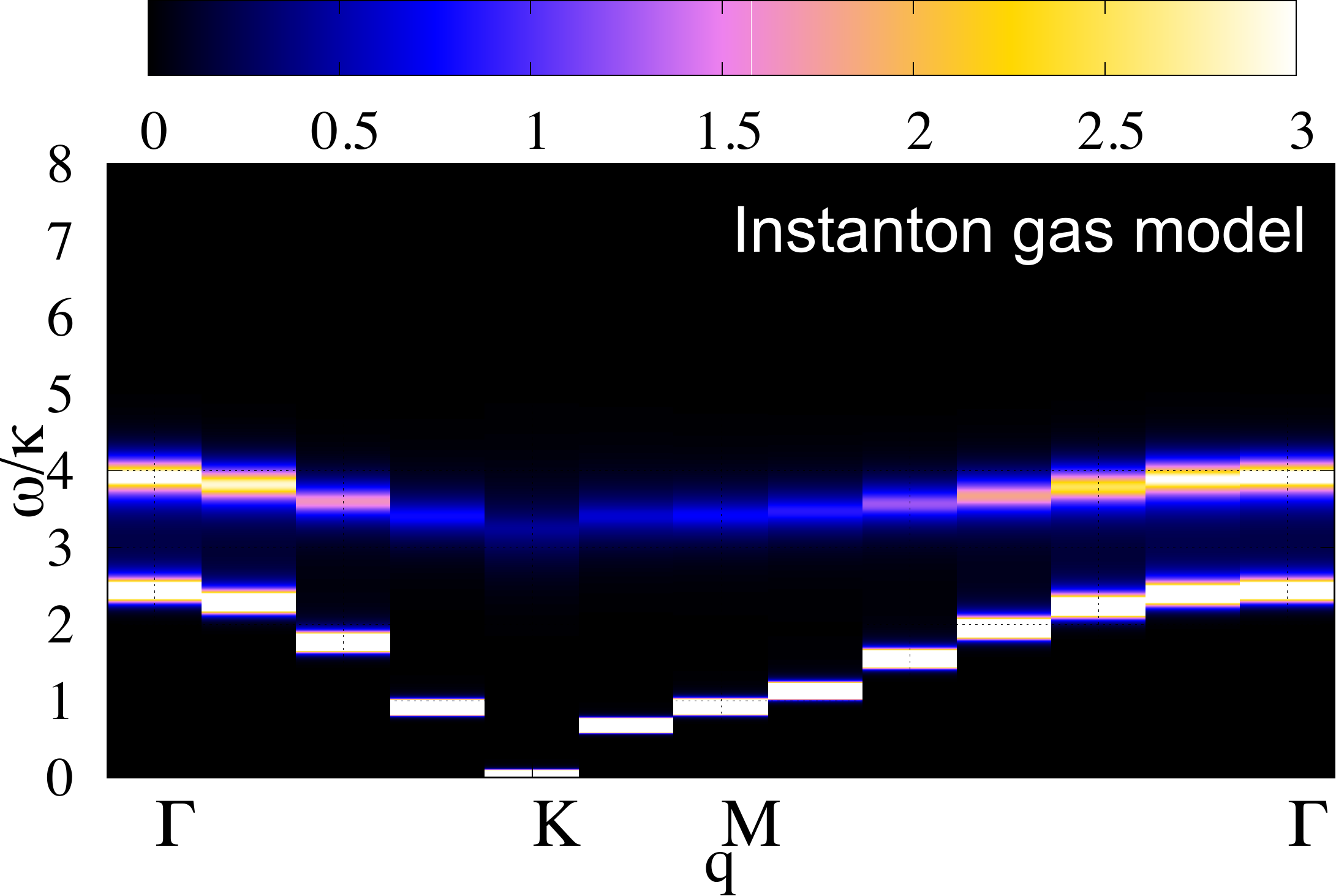}}
   \subfigure[]{\label{fig:spectralFunctionsU6Z}\includegraphics[width=0.35\textwidth , angle=0]{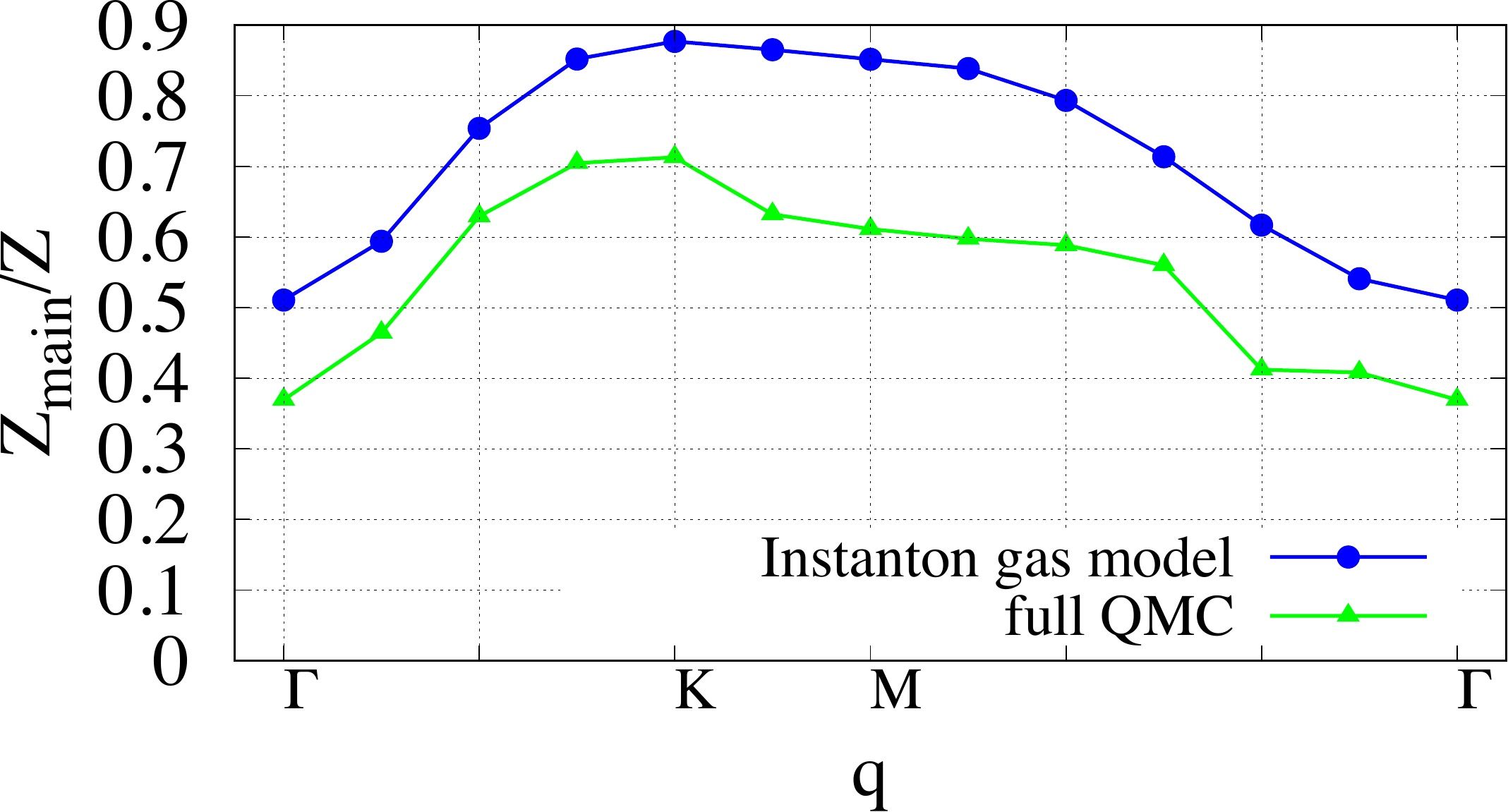}}
         \caption{(a) Spectral functions in momentum space using the  ALF \cite{ALF_v2}   implementation of the  auxiliary  field QMC.  (b) The same spectral functions obtained with instanton gas model. (c) The share of the lower peak in the overall spectral weight along the same profile in momentum space. Calculations were done for $12\times12$ lattice with $N _{\tau}=256$ and $\beta \kappa=20$ (see sections \ref{subsec:Model} and \ref{subsec:path_int} for the notation). The interaction strength is equal to $U=6.0 \kappa$, which is equal to the bandwidth. 
         }
    \label{fig:SpectralFunctionsU6Comparison}
\end{figure}

We would like  to  go  beyond  this  trivial solution, and in particular,  provide a map of all  saddle    points  without  the  restriction to fields which are constant in space  and time.  We note that   since  the action is  not  necessarily real,   one   generically   has to  continue the real scalar field 
to the complex plane to  achieve  this goal. The motivation to do so  is at least  twofold.    On one  hand,    the saddle point   structure  is  
necessary  to formulate  the so-called Lefschetz thimble  decomposition \cite{Witten:2010zr, Witten:2010cx}  that has  the potential of  alleviating  the severity  of the negative sign  problem  \cite{Cristoforetti:2012su, Alexandru:2015xva}. In particular,  each  thimble is  attached to a saddle point,  and  the imaginary  part  of  the action is  constant  within the  thimble.  On the other  hand, the  very  structure of the (complex)  saddle points,  can   yield  valuable   approximation schemes  that can  be   improved at will.   Here we will   consider  the  latter   but   concentrate  on  cases   where  the action is  real,  as  realized  at 
the  particle-hole  symmetric point.  In  this  case,  the complexification of the field  is  not   required.

Finding  saddle points is   a  daunting  task.    Here  we   use     auxiliary field  quantum Monte Carlo simulations  to  sample   the  fields,  and for  each independent  configuration,   stop  the Monte Carlo  sampling      and  integrate   the   steepest  descent   differential    equation   so    as to flow to  the saddle point.   This  provides a complete map.     Remarkably, as was shown in \cite{PhysRevD.101.014508}, for the   honeycomb lattice  at  any  coupling  and for   the square  lattice   at strong  coupling,  the saddle point structure is quite regular. All saddles  can  be understood  in terms of   an   elementary configuration,  an   instanton,  in  which  the   fields    differs  from   zero  only  in  a  small  
space  time  region.     Physically,  it  corresponds  to the  formation of a local moment  at  a  given space-time   point  and   concomitant  short  ranged  anti-ferromagnetic  fluctuations  around  this point  (see Fig.~\ref{fig:SpinCorrInstantonVacuumMap}).   Under  the     assumption of  spatial locality,   and  as shown in   Appendix \ref{sec:AppendixB},  the   instanton is  characterized by  a topological    winding  number    

This instanton approach provides an interesting link between the structure of the path integral for the Hubbard model and long-established methods in quantum chromodynamics (QCD). 
Instantons were introduced almost fifty years ago in the context of Yang-Mills theory ~\cite{Belavin:1975fg} and are defined as topologically non-trivial solutions of the classical field equations in Euclidean space with finite action. They very quickly found many applications even in quantum mechanics, where they describe the tunneling processes from one vacuum to another. This is also the case in Yang-Mills theories where they describe tunnelling processes between different degenerate vacua which are labeled by different values of the winding number, a topological index. In the context of QCD, they play an important role in the explanation of the mechanism of spontaneous breaking of chiral symmetry and applications of instantons can be found in the solution of the U(1) problem and the strong CP problem. Beyond the context of strong interactions, instantons have related counterparts in the electroweak sector, where the so-called sphalerons can lead to processes that violate the baryon and lepton number conservation and could potentially describe rare processes of baryon decay. The instanton calculus has proven to be extremely powerful in supersymmetric gauge theories where it allowed for example the calculation of the exact $\beta$-function. For more details on field theoretical applications, we refer the interested reader to ~\cite{Coleman:1978ae,Vandoren:2008xg}.
Beyond physics, instantons also have fascinating applications in mathematics where, for example, they can be used for the classification of four-manifolds~\cite{DonaldsonKronheimerbook}.

Returning to QCD, the introduction of the concept of an instanton led to the modelling of the QCD partition function as a gas of instantons~\cite{tHooft:1976snw} that could allow for analytical treatment. Later, however, it was understood that correlations between instantons are extremely important for numerous phenomena in QCD and one is forced to go beyond the mean-field approximation and study numerically a liquid of instantons where the ’t Hooft interaction is included to all orders~\cite{Shuryak:1980tp}. Guided by these ideas, we will try to adapt this approach in the framework of the Hubbard model in order to demonstrate that this kind of approximations can lead to interesting and highly non-trivial results in strongly-correlated electron systems.

The  key   result of the paper is  that we  can define  a classical  model    of  the instanton gas  that    reproduces  the  saddle  point  structure  of  the path  integral for the Hubbard model for a HS field coupling to the charge density.   The only inputs  needed in order to completely  define   this  classical  model  are the characteristics of a single instanton and the two-body interaction between these semiclassical objects. This description of the physics of instantons through a pairwise short-ranged interaction appears naturally through   an analysis  of  the one-  and  two-instantons   configurations.     With   \textit{ simple } classical  simulations, we    can  then  generate  saddle  point   field configurations,  which can then  determine  physical properties of the Hubbard model.

 \begin{figure}[]
   \centering
  \includegraphics[width=0.35\textwidth,clip]{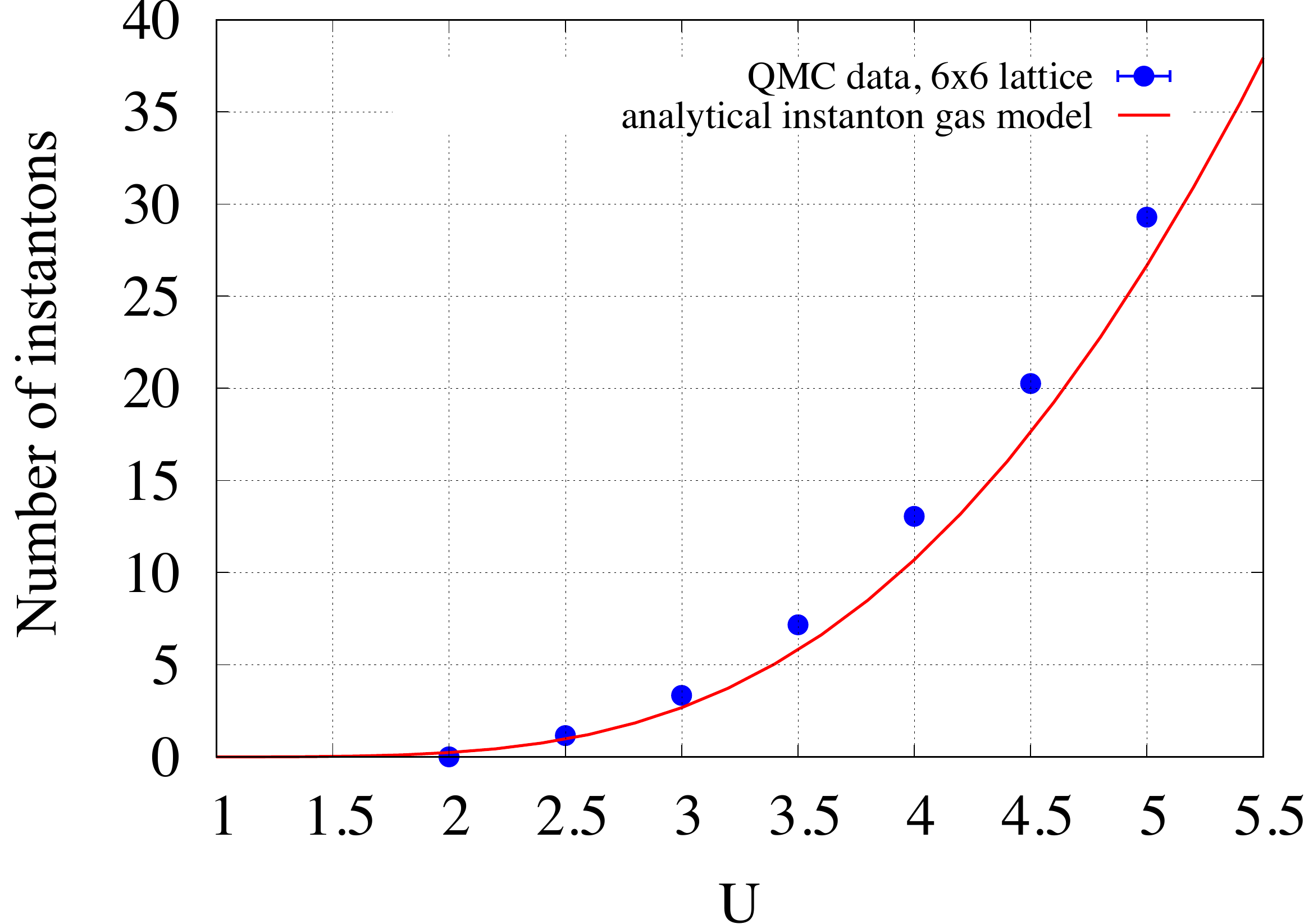}
         \caption{A comparison of the average number of instantons obtained from real QMC data with the analytical instanton gas model. The QMC data corresponds to a $6\times 6$ spatial size, with $N _{\tau}=512$ and $\beta  \kappa=20$.} 
          \label{fig:DistributionComparisonCenterFirst}
  \end{figure}

Before  plunging  into  the  technical details of the  approach  we    summarize our  key  results.   
Fig.~\ref{fig:SpinCorrInstantonVacuumMap} shows  the  connection between  the instanton  and  enhanced  short-ranged  anti-ferromagnetic (AFM) correlations.  Here  we  consider  a  configuration with  a  single  instanton   saddle  point, with the instanton  located  at  the space-time point $(X,T)$.   We  then plot  
the  spin-spin correlations   $\frac{1}{3}\langle \hat{\ve{S}}_{\ve{x}_0}(T) \cdot \hat{\ve{S}}_{\ve{x}_0  +  \ve{x}} (T) \rangle$,  where $\hat{\ve{S}}_{\ve{x}}$ is  the spin operator.   One  will  see  that for  $ \ve{x}_0 = X$   (lower left black  circle)   substantial  short-ranged  correlations are present.  On the other  hand, \textit{far}   from  the instanton,  where  $ \ve{x}_0  $ corresponds  to the upper right black  circle, no  anti-ferromagnetic order  is   observed  beyond  one lattice site.

We  expect local moment  formation to  show  up as  upper  Hubbard \textit{bands}  in the  single-particle spectral function.  Fig.~\ref{fig:SpectralFunctionsU6Comparison}    shows a comparison of   results for  this  quantity between the   instanton gas  approach  and a   full 
auxiliary field  QMC  simulation  \cite{Blankenbecler81,White89} for the  Hubbard  model on the honeycomb lattice   in the  magnetically  ordered  phase   at  $U/U_c  = 1.6$   where  $U_c$   denotes  the  critical value of the interaction where   anti-ferromagnetism sets  it  and  a  mass gap  is  generated \cite{AssaadHerbut2013,Sorella2012}.    Our  instanton  approach  does 
not   capture  the   SU(2)  symmetry  breaking  and  hence  no mass  gap is generated  at  the  Dirac  point  $\ve{k} = K$.   However,   local moment   formation  and concomitant 
short-ranged  anti-ferromagnetic  correlations capture  the  high-energy  properties,  encoded in the so-called  upper  Hubbard band.   

Finally,  Fig.~\ref{fig:DistributionComparisonCenterFirst}   shows  that  the  instanton number obtained in  the QMC   simulation and in the  analytical  instanton gas model show very good agreement.  Thus, the instanton gas model can be used to predict the dominant saddle point without performing costly QMC simulations.

The paper is organized as follows: in Section \ref{Sec:1}, we give a brief description of the specific path integral formulation for the Hubbard model employed in this study. We also give a short introduction to the Lefschetz thimble formalism. The second section (\ref{sec:StructureFromQMC}) is devoted to the description of the structure of the saddle points for the Hubbard model on the hexagonal lattice. This includes a detailed account of the one-instanton as well as many-instanton solutions. The third section  (\ref{sec:GasModel}) covers the construction of a semi-analytical instanton gas model. In the fourth section  (\ref{sec:GasPhysics}),  we describe the physics following from the instanton gas model. The last section  (\ref{sec:Square}) presents preliminary results for the saddle point approximation to the Hubbard model on the square lattice which is relevant for  high-$T_c$ superconductivity.       We  have  included  appendices  that   discuss in full detail ergodicity  issues in  the  HMC  (App.~\ref{sec:AppendixA}),   analytical solutions for  individual  instantons   with emphasis on  the topological winding number  interpretation of the instanton (App.~\ref{sec:AppendixB}),  Hessians for $N$-instanton saddle points (App.~\ref{sec:AppendixC}),   details of the grand-canonical   Monte Carlo simulation for the   classical instanton model (App.~\ref{sec:AppendixD}), and  finally  the  relation of the instanton  to  the  Gutzwiller projection (App.~\ref{sec:AppendixE}). 
 
\section{Background}
\label{Sec:1}
This study builds on previous work \cite{PhysRevD.101.014508} which employed methods from lattice gauge theories to elucidate the physics of the Hubbard model, both at half-filling and at finite density. The aim of this section is to recall the basic definitions and setup in order to motivate the study of the saddle points and understand the physics which they encode. This will motivate the formulation of an instanton gas model which captures much of the physics of the Hubbard model.
\subsection{ \label{subsec:Model} Hubbard Model}
In this work, the Hubbard model on a bipartite (square and hexagonal) lattice is considered. The SU(2)-spin symmetric form of the Hamiltonian is given by  
\begin{eqnarray}
  \label{eq:Hamiltonian}
  \!\!\!\hat{{H}}\! =\! -\kappa \sum_{\langle {\ve{x}},{\ve{y}}\rangle} (  \hat a^\dag_{\ve{x}} \hat a_{\ve{y}} \!+\! \hat b^\dag_{\ve{x}} \hat b_{\ve{y}} \!+\! \mbox{h.c} ) \!+\! \frac{U}{2} \sum_{{\ve{x}}} \hat q_{\ve{x}}^2 \!+\! \mu  \sum_{\ve{x}} \hat q_{\ve{x}},
\end{eqnarray}
where $\hat a^\dag_{{\ve{x}}}$ and  $\hat b^\dag_{{\ve{x}}}$ are creation operators for electrons and holes,  $\hat q_x=\hat n_{{\ve{x}}, \text{el.}} - \hat n_{{\ve{x}}, \text{h.}}={\hat a^\dag_{x}} \hat a_{{\ve{x}}}- {\hat b^\dag_{{\ve{x}}}} \hat b_{{\ve{x}}}$ is the charge operator, $\kappa$ is the hopping parameter, $U>0$ is the Hubbard interaction, and $\mu$ is the chemical potential. From now on, we will express all dimensional parameters like $U$, inverse temperature $\beta$, etc. in the units of hopping $\kappa$. This form of the Hamiltonian will be useful for the functional integral approach where the interaction term will be decomposed by the introduction of an auxiliary bosonic field. Away from half-filling, $\mu=0$, the theory suffers from the notorious sign problem. This is a generic feature of a large class of many-body theories and in order to deal with this problem, a variety of different methods and techniques have been devised  ~\cite{Fujii:2015bua,Tanizaki:2015rda,Kanazawa:2014qma,Alexandru:2015xva,Alexandru:2015sua,DiRenzo:2017igr, Alexandru:2016ejd,Alexandru:2018brw,Alexandru:2018ngw,Alexandru:2018ddf}. The case of finite chemical potential will only briefly be commented on, while the case of half-filling will be the main focus in all subsequent numerical and analytical calculations. 

At half-filling, this model is known to exhibit a semimetal-to-insulator transition (\cite{AssaadHerbut2013,Sorella2012}). At large $U$, the Hubbard model on the hexagonal lattice exhibits AFM order while at small $U$ it is a Dirac semimetal with no long-range order. The critical coupling, $U_c$, at which this transition takes place, defines an appropriate physical scale for the interaction strength. In the functional integral approach, not only can one take into account all quantum fluctuations which accurately describe both phases, but one can also employ semi-classical methods. These methods rely on knowledge of the stationary points of the action and fluctuations around the solutions to these saddle-point equations. One, in principle, could ask how the character and importance of these saddle-point solutions vary as the system passes through the phase transition. This is one of the questions we have addressed in this paper.

\subsection{ \label{subsec:path_int}Path Integral Formulation}
This study involves the path integral formulation of the Hubbard model. Previous studies have detailed this construction \cite{ALF_v2, Conformal_PhysRevB.99.205434}, which we briefly review here. The approach starts   with the standard expression for the partition function as the trace of the quantum Boltzmann weight
\begin{eqnarray} \label{eq:partition_function_trace}
\mathcal{Z} =\Tr (e^{ - \beta \hat H }).
\end{eqnarray}
Denoting the hopping term in (\ref{eq:Hamiltonian}) as $\hat{H}_0$ and the Hubbard term as $\hat{H}_U$, one performs the following Trotter decomposition of the Boltzmann weight (\ref{eq:partition_function_trace}) 
\beq  \label{eq:boltzmann_weight_trotter}
\Tr \left( e^{ - \beta \hat H } \right)  = \Tr \left(   e^{-\Delta \tau \hat H_0 } e^{-\Delta \tau \hat H_U } \right)^{N_\tau} + O(\Delta \tau^2)
\eeq   
where the Euclidean time step $\Delta \tau \equiv \beta/N_{\tau}$ has been introduced and on the right-hand side of  
Eq.~(\ref{eq:boltzmann_weight_trotter}) there are $N_{\tau}$ repetitions of the exponential factors involving the kinetic and the Hubbard term. In turn, $2 N_{\tau}$ Grassmann resolutions of the identity are introduced, one between each exponential factor, and the matrix elements of the exponential factors are then computed. This is straightforward for the kinetic term, since $\hat H_0$ is bilinear in the fermionic operators. To deal with the four-fermion interaction term, continuous auxiliary bosonic fields are introduced at each Euclidean time slice through the usual Gaussian Hubbard-Stratonovich (HS) transformation
\begin{eqnarray}
\label{eq:continuous_HS_imag}
  e^{-\frac{\Delta \tau}{2} U \hat q^2_{\ve{x}}} \cong \int d \phi_{\ve{x}}  e^{- \frac{  \phi^2_{\ve{x}}}{2 U \Delta \tau}  + i\phi_{\ve{x}} \hat q_{\ve{x}}}.
\end{eqnarray}
 After applying this to each factor of $e^{-\Delta \tau \hat{H}_U}$ in the Trotterized Boltzmann factor and integrating out the Grassmann variables, one obtains the following expression for the functional integral
\begin{eqnarray}
  \mathcal{Z} = \int \mathcal{D} \phi e^ {-S_B[\phi]}  \det M_{el.}[\phi] \det M_{h.}[\phi], \nonumber \\
   S_B[\phi] = \sum_{{\ve{x}},\tau}  \frac  {\phi_{{\ve{x}},\tau}^2 } {2 U \Delta \tau},
  \label{eq:Z_continuous}
\end{eqnarray}  
where $M_{el.}$ and $M_{h.}$ are the fermionic operators for the electrons and holes respectively. The determinants of these operators can conveniently be expressed as
\begin{eqnarray}
 \det M_{el.} = \det \left[ I +\prod^{N_\tau}_{\tau=1} D_{2\tau-1} D_{2\tau} \right], \nonumber \\
 \det M_{h.} = \det \left[ I +\prod^{N_\tau}_{\tau=1} D_{2\tau-1} D^*_{2\tau}\right], 
  \label{eq:M_continuous}
\end{eqnarray}
where $D_{2\tau} \equiv \diag{e^{i\phi_{{\ve{x}},\tau}} }$ and $D_{2\tau+1} \equiv e^{-\Delta \tau h}$ have been introduced.
Both of these are $N_S \times N_S$ matrices, where $N_S$ is the total number of spatial lattice sites. We have also introduced $h$, which is the matrix characterizing the tight-binding Hamiltonian $\hat{H}_0$. From the form of the determinants in (\ref{eq:M_continuous}), one can show that the integrand of the functional integral (\ref{eq:Z_continuous}) is real and positive-definite at half-filling since $\det M_{el.} = \det M_{h.}^*$.

\subsection{Lefschetz Thimbles and the Gradient Flow}
In order to construct an effective theory based on a semiclassical approach to the path integral for the repulsive Hubbard model on both the hexagonal and square lattices, one  must first
 quantitatively understand the saddle points of the theory. The Lefschetz thimbles decomposition of the partition function serves as the mathematical basis for a precise study of these saddle-points. The idea of the Lefschetz thimbles approach is to complexify the space of fields over which we integrate in the functional integral. It is especially useful, when the action is complex and its oscillatory phase precludes the use of importance sampling methods. Picard-Lefschetz theory, a generalization of Morse theory to complex manifolds, provides a framework by which this poorly-behaved integral is converted into a sum of strictly convergent integrals. Considering the most general form of the functional integral for a generic lattice theory with $N$ bosonic fields one can write \cite{Witten:2010zr,Witten:2010cx}:
\begin{align}
\label{eq:thimbles_sum_and_integral}
&\;\;\mathcal{Z} = \int_{\mathbb{R}^N} \mathcal{D} \Phi\, e^{-S[\Phi]}=\sum_\sigma k_\sigma \mathcal{Z_\sigma},&\nonumber\\
\text{where}\quad&\mathcal{Z_\sigma} = \int_{\mathcal{I}_\sigma} \mathcal{D} \Phi\, e^{-S[\Phi]},&
\end{align}
and $\sigma$ labels all complex saddle-points $z_\sigma \in \mathbb{C}^N$ of the action. Here $\mathcal{I_\sigma}$ are the thimble manifolds attached to the saddle points. These manifolds, defined below, are the generalization of the contours of steepest descent in the theory of asymptotic expansions. This is what is known as the Lefschetz thimble decomposition of the functional integral. The saddle points are determined by the condition 
\begin{equation}
  {\left.\frac{\partial S}{\partial \Phi}\right| }_{\Phi=z_\sigma} = 0,  
\end{equation}
while the integer-valued coefficients $k_\sigma$ encode the intersection of a manifold which we call the anti-thimble with the original domain of integration. At half-filling, all saddles lie in the original, real space of fields. We stress here that if the saddle points are non-degenerate (${\left.\det \partial^2 S/\partial \Phi' \partial \Phi\right| }_{\Phi=z_\sigma} \neq 0$) and isolated, the relation (\ref{eq:thimbles_sum_and_integral}) holds (for a generalization to the case of gauge theory see \cite{Witten:2010cx}).

The Lefschetz thimble is a manifold associated with a given saddle point. Let us  endow  the  fields  with  an additional, non-physical  temporal   parameter $t$,  and  define the  
gradient  flow  (GF)  equation  as:
\begin{equation}
\label{eq:flow}
\frac{d\Phi}{d t}=\overline{ \frac{\partial S}{\partial \Phi}},
\end{equation}
where the bar denotes complex conjugation.
The  Lefschetz thimble   is  the union  of   all   fields   $\Phi(t=0)$   that  satisfy  the   boundary  condition:  $~\Phi(t=0) \in\mathcal{I}_\sigma $  if  $   \Phi( t \rightarrow -\infty) \rightarrow z_\sigma$.  Just as we have made an analogy between the thimble and the contour of steepest ascent, there is a second manifold associated with each saddle point which is analogous to the contour of steepest descent. This manifold is known as the anti-thimble, $\mathcal{K}_\sigma$, and consists of all possible    $\Phi(t=0)$   which end up at a given saddle point $z_\sigma$:  $\Phi(t=0) \in\mathcal{K}_\sigma  $  if $ \Phi(t \rightarrow + \infty) \rightarrow z_\sigma$.  As previously stated, $k_\sigma$ counts the number of intersections of  $\mathcal{K}_\sigma$ with $\mathbb{R}^N$, $k_\sigma = \langle \mathcal{K}_\sigma, \mathbb{R}^N \rangle$. Along a given thimble, the imaginary part of the action is constant, and thus one can rewrite the Lefschetz decomposition of the functional integral as 
\beq \label{eq:thimbles_sum_and_integral_constant_phase}
\mathcal{Z} = \sum_\sigma k_\sigma e^{-i \operatorname{Im} S} \int_{\mathcal{I}_\sigma} \mathcal{D} \Phi\, e^{-\operatorname{Re}S[\Phi]},
\eeq 
which makes the previously mentioned claim of converting an oscillatory integral to a sum of convergent ones abundantly clear. Early success with this method centered around the study of toy models without fermions. Recently, however, it has been used to address the sign problem in both non-trivial, low-dimensional relativistic field theories \cite{Alexandru:2016ejd, Fujii:2015bua, Alexandru:2018ngw} as well as in two-dimensional many-body systems  \cite{Alexandru:2018ddf,PhysRevD.100.114510}. 

As evident from Eq. (\ref{eq:thimbles_sum_and_integral_constant_phase}), the application of the thimbles decomposition would be much easier if one knew the structure of the saddle points, $z_\sigma$, in advance. In this case, it would be possible to simplify (\ref{eq:thimbles_sum_and_integral_constant_phase}) by considering only the dominant saddles or by using the Gaussian approximation to the integrals. The instanton gas approach performs exactly this task: it predicts the dominant saddle for the Hubbard model for a wide range of parameters without prior QMC simulations.

\begin{center}
  \begin{figure}[]
   \subfigure[]{\label{fig:histogramU1}\includegraphics[width=0.23\textwidth,clip]{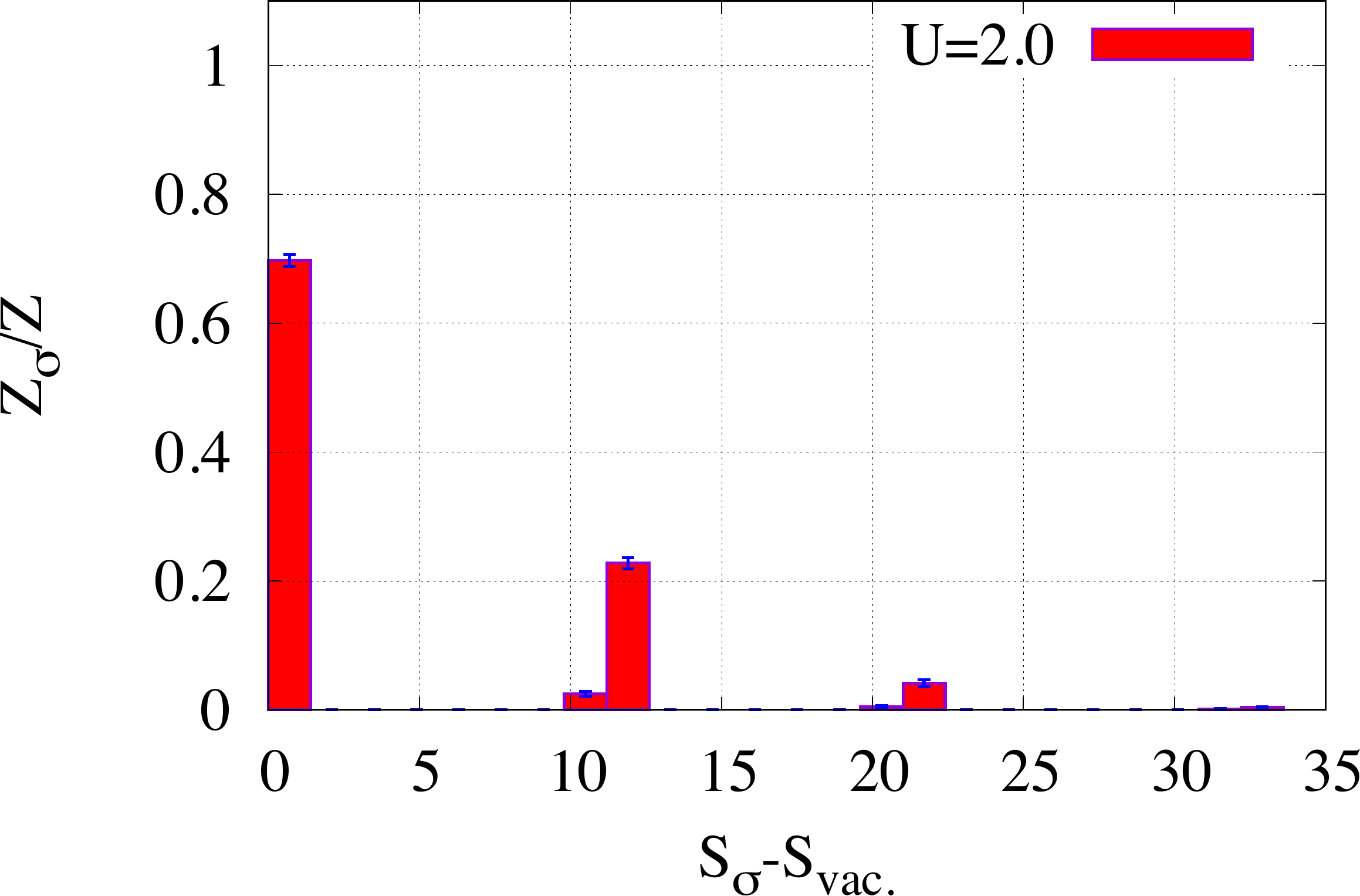}}
   \subfigure[]{\label{fig:histogramU2}\includegraphics[width=0.23\textwidth,clip]{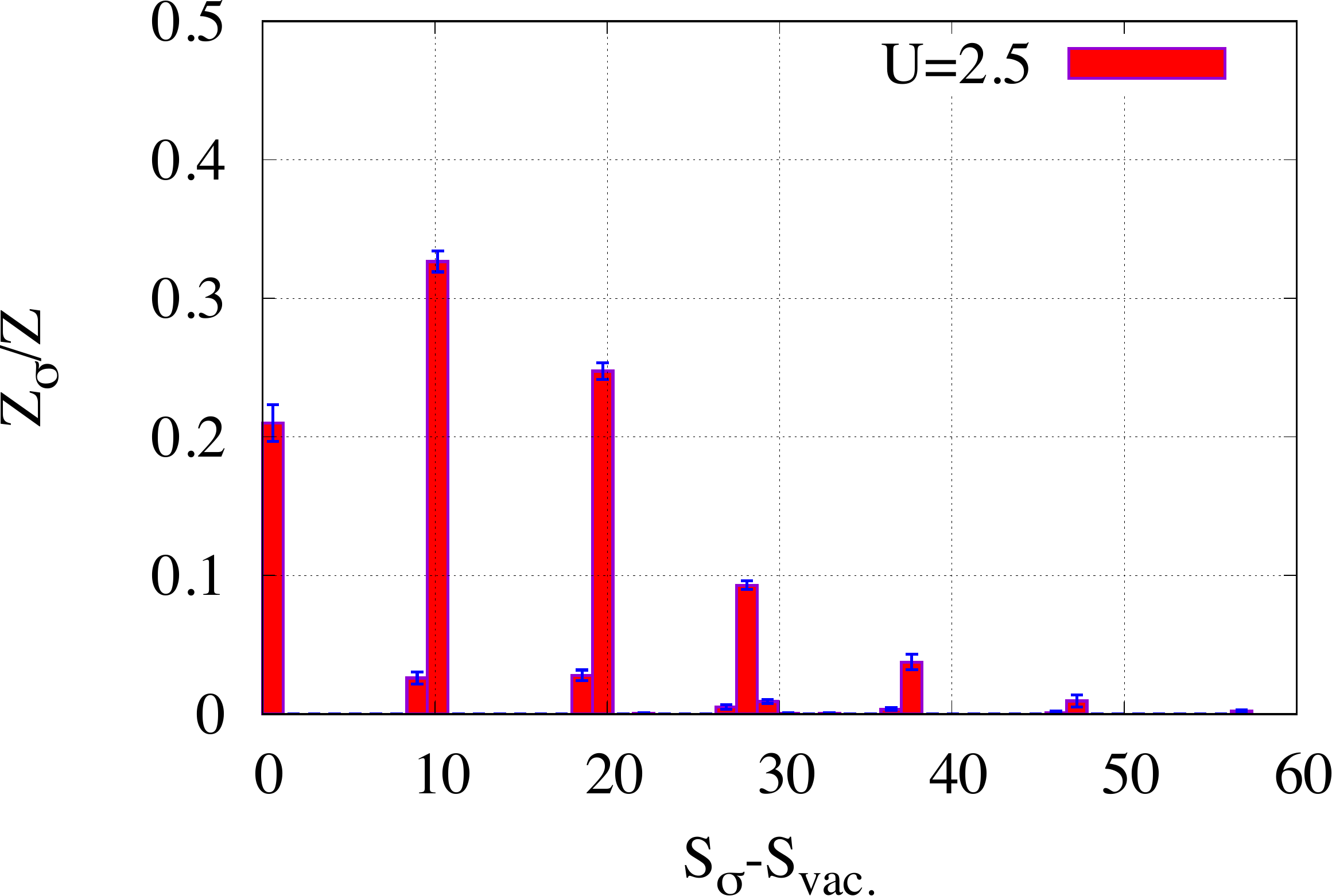}}
   \subfigure[]{\label{fig:histogramU3}\includegraphics[width=0.23\textwidth,clip]{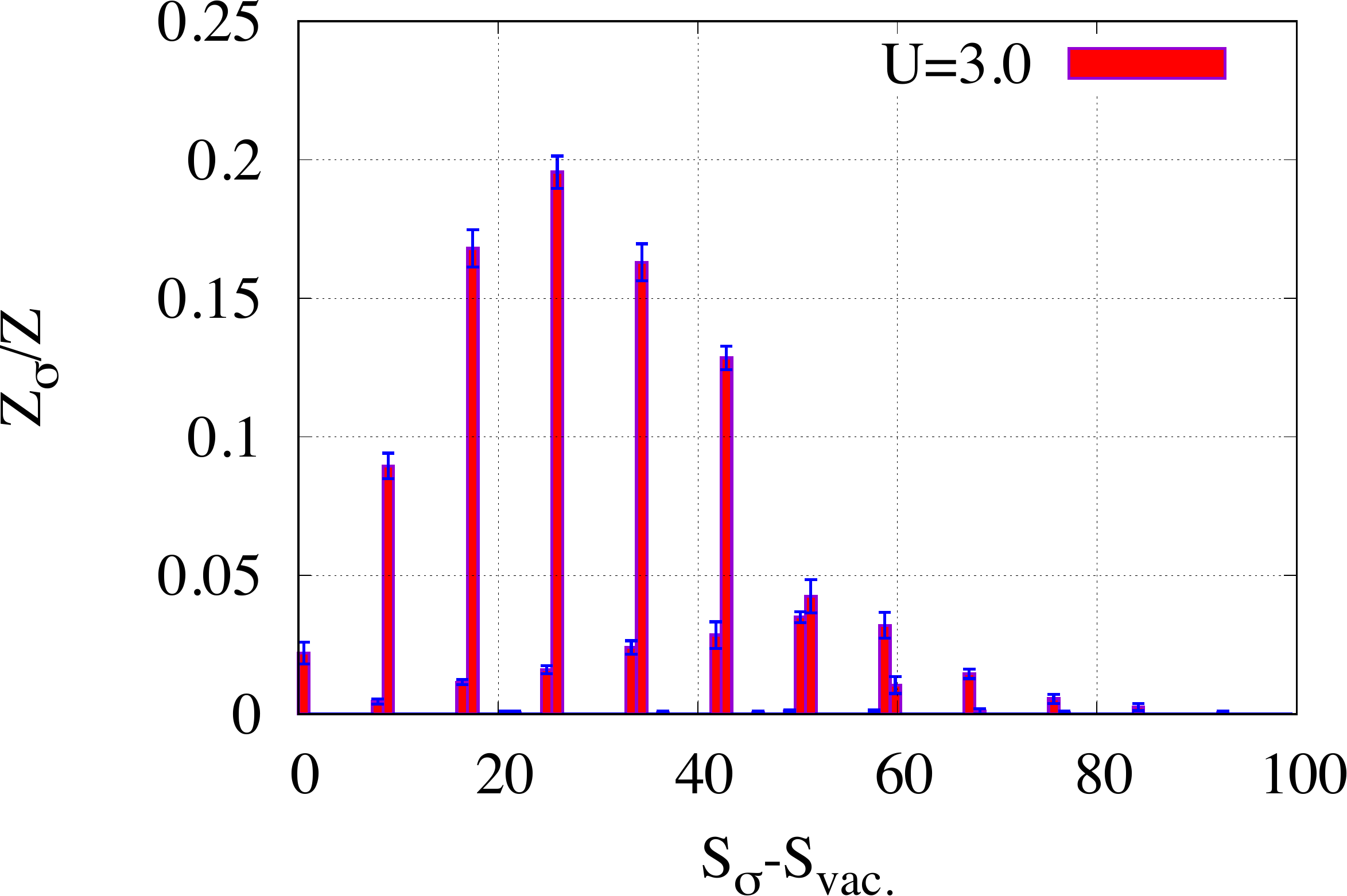}}
    \subfigure[]{\label{fig:histogramU4}\includegraphics[width=0.23\textwidth,clip]{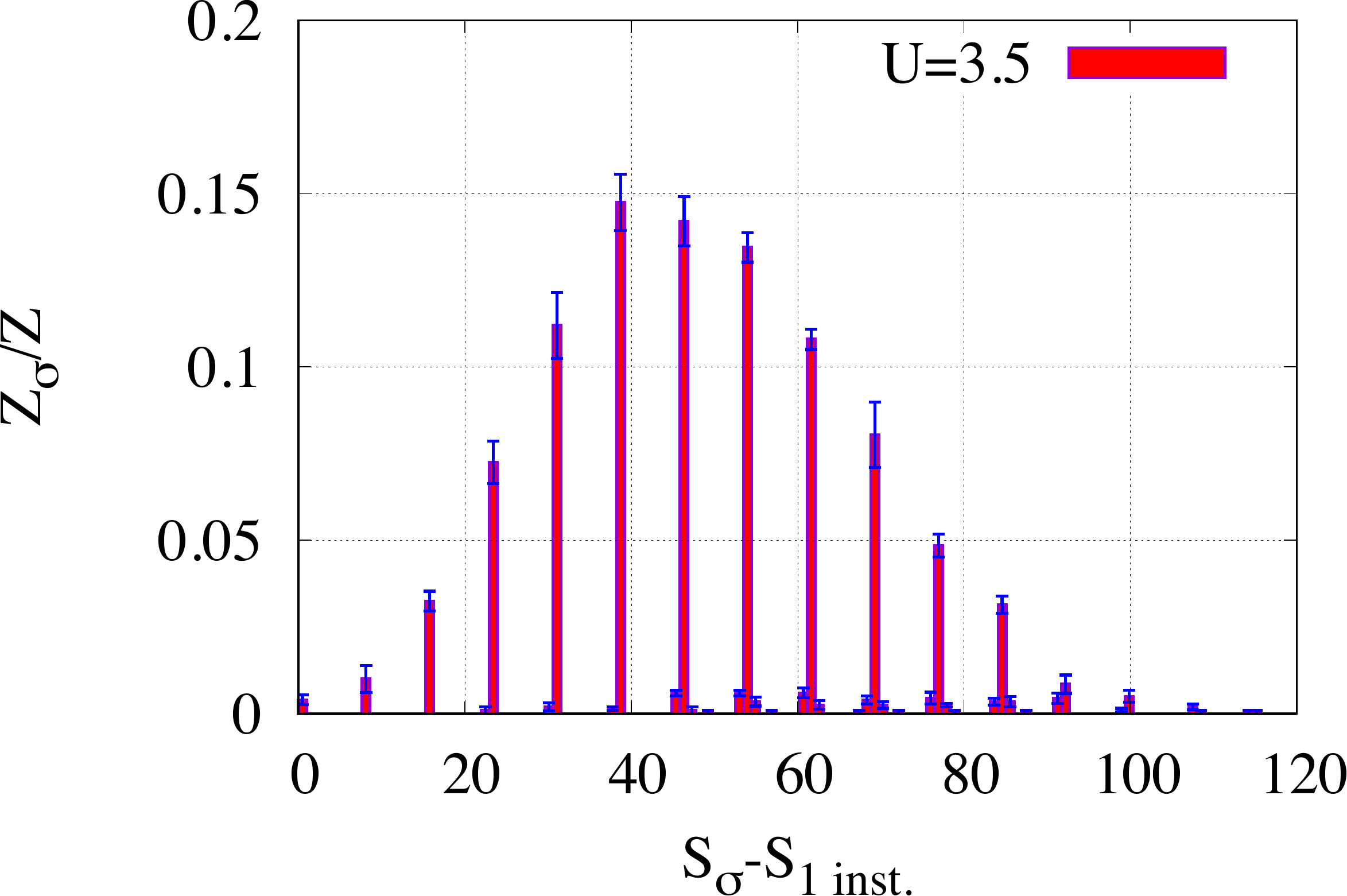}}
   \subfigure[]{\label{fig:histogramU5}\includegraphics[width=0.23\textwidth,clip]{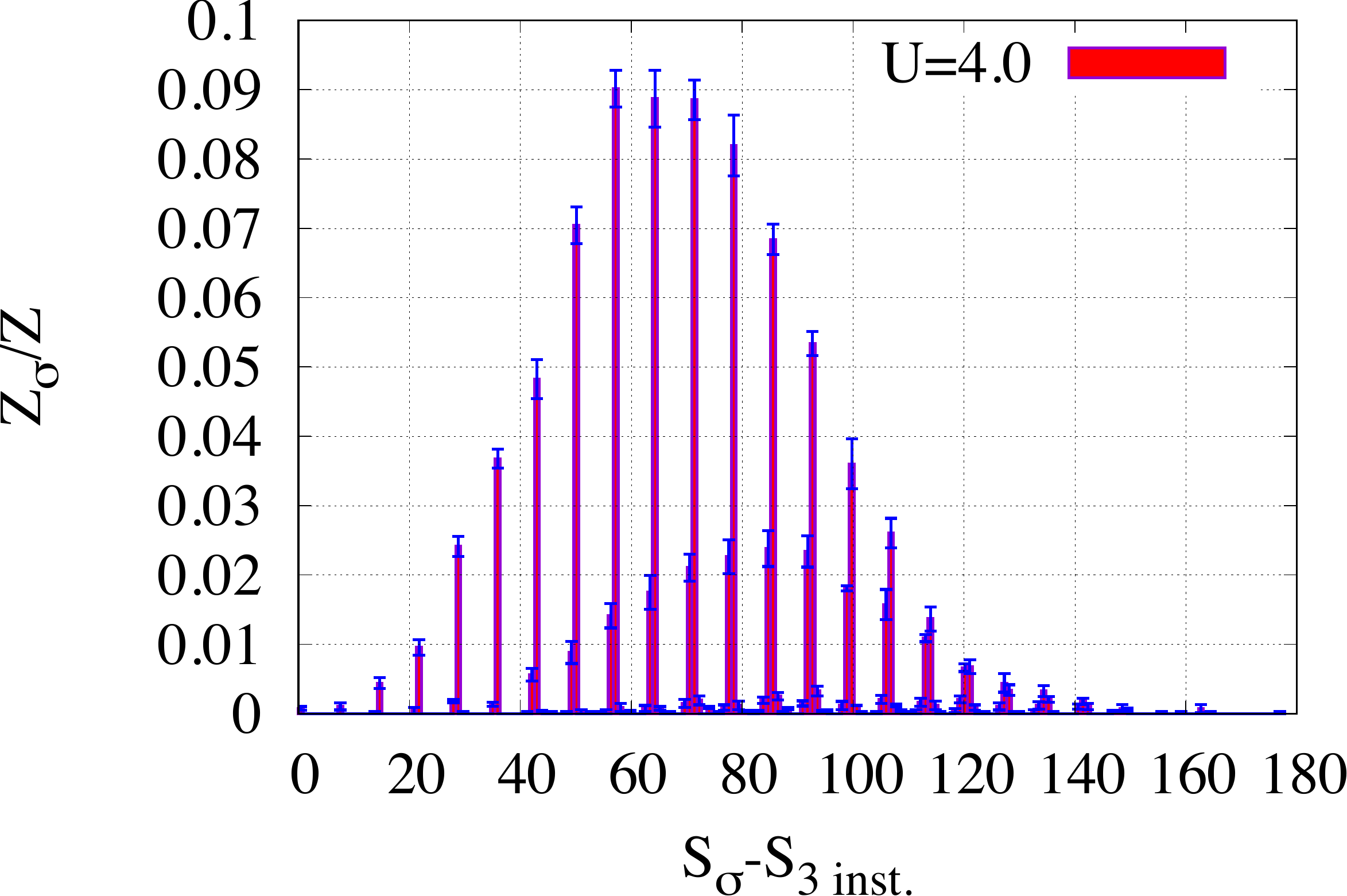}}
   \subfigure[] {\label{fig:histogramU6}\includegraphics[width=0.23\textwidth,clip]{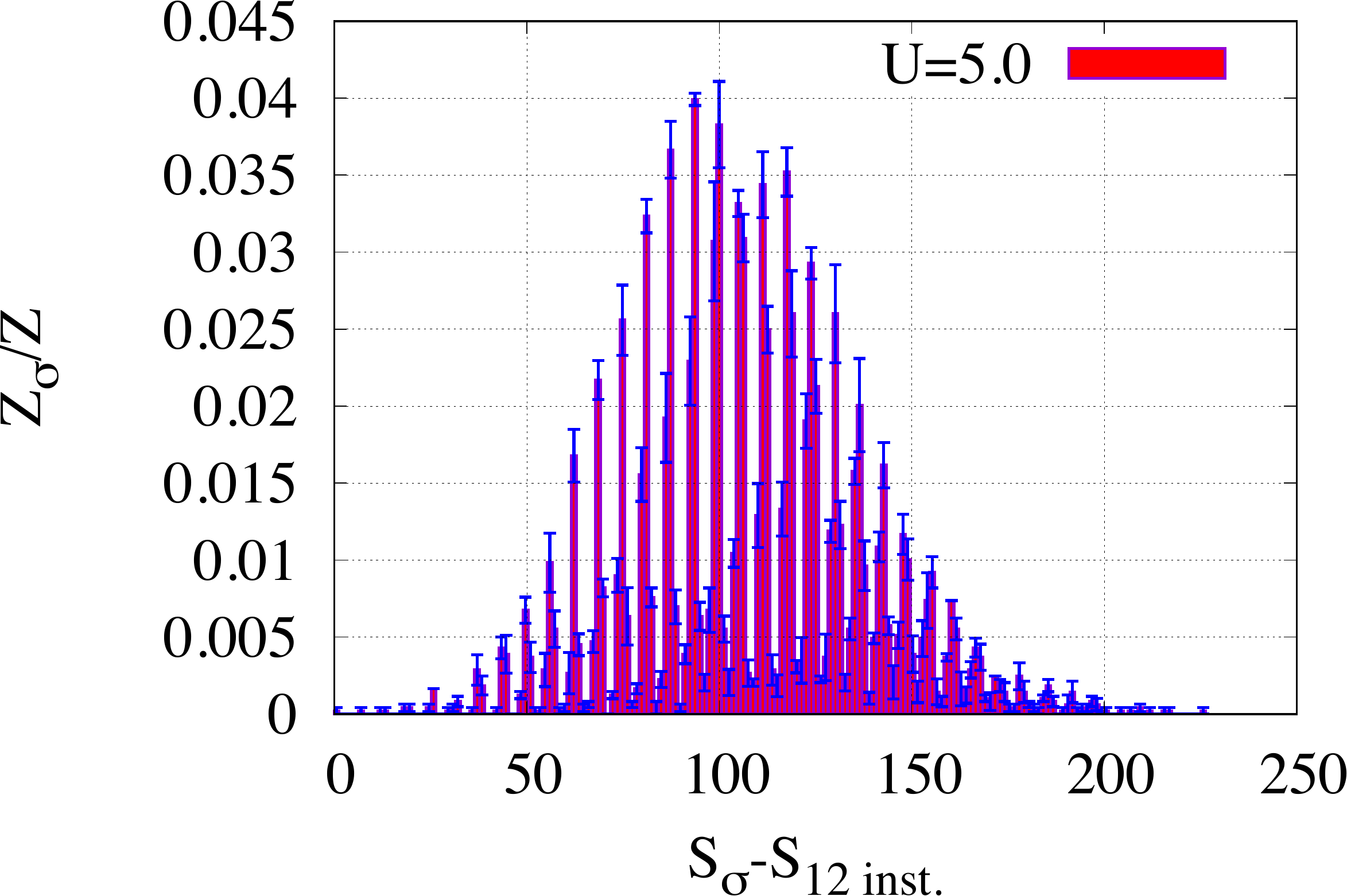}}
         \caption{Histograms depicting the relative contributions of the various $N$-instanton saddles to the full partition function. The horizontal axis corresponds to the action of an $N$-instanton solution, offset by an amount equal to the action of the observed saddle with the least number of instantons. One can clearly see that the minimal number of observed instantons increases with increasing $U$. These calculations were performed on a $6\times6$ lattice with $N_{\tau}=512$, $\beta \kappa=20 $.}
    \label{fig:histogramsU}
\end{figure}
\end{center}

\section{\label{sec:StructureFromQMC}Structure of the saddle points from QMC data}
In previous studies \cite{PhysRevD.101.014508}, it was demonstrated how one can numerically determine the Lefschetz thimbles decomposition (\ref{eq:thimbles_sum_and_integral}) at half-filling, where the sign problem is absent and all thimbles are confined to the real subspace $\mathbb{R}^N$.
We first generate configurations of the continuous bosonic auxiliary fields according to their weight $e^{-S}$, where 
\beq \label{eq:full_action}
S=S_B-\ln (\det M_{el.}\det M_{h.}).
\eeq
In the next stage, we evolve the auxiliary fields according to the gradient flow equations in the inverse direction
\begin{equation}
\label{eq:inverse_flow}
\frac{d\Phi}{d t}=-{ \frac{\partial S}{\partial \Phi}}
\end{equation}
starting from each of these QMC-generated field configurations. 
These flows converge to the local minima of the action within $\mathbb{R}^N$, which are, of course, just the relevant saddle points. At the end of such a procedure, we obtain a set of saddle point field configurations, distributed according to their relative weight in the full partition function: $\mathcal{Z}_\sigma/\mathcal{Z}$. This distribution can be plotted as the  histogram of the actions of these various saddle point field configurations. 
The technical details of this procedure as well as some additional checks (e.g. the question of ergodicity of QMC generator and the continuum limit) can be found in Appendix \ref{sec:AppendixA}.

In general, the number and the form of the saddle point configurations critically depend on the way in which we introduce the auxiliary fields \cite{PhysRevD.101.014508} . In this paper, we are interested in an analytical saddle point approximation. Thus, we employ the specific HS decomposition, 
where the scalar auxiliary field $\phi$ is coupled to the charge density. In this particular case, the saddle points are especially simple, as their histogram can be seen to be a collection of equidistant discrete peaks, as clearly displayed in Fig.~\ref{fig:histogramsU}. This regular saddle structure makes the creation of an analytical saddle-point approximation relatively straightforward.

\begin{figure}[]
    \centering
             {\label{fig:histogramNs}\includegraphics[width=0.35\textwidth,clip]{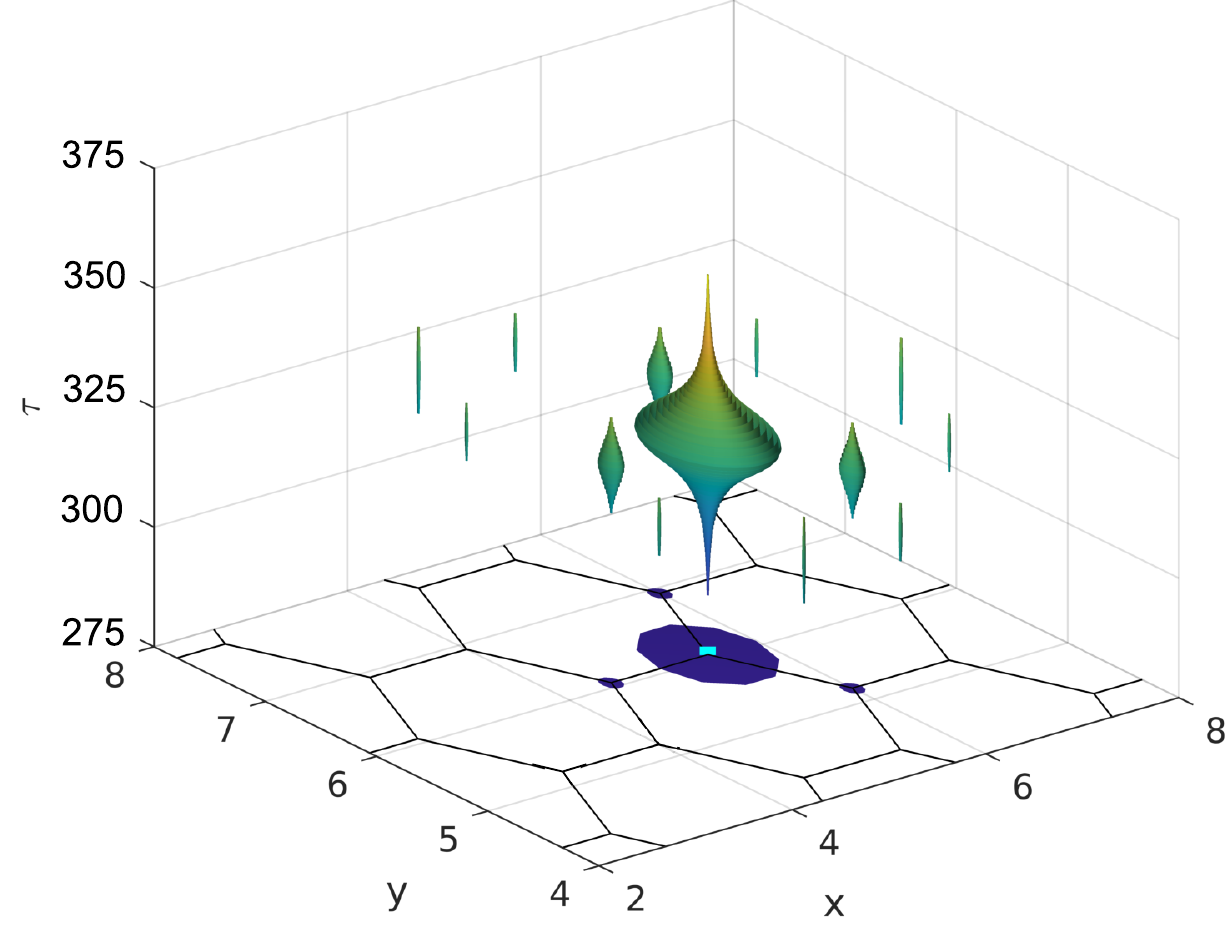}}
      \caption{Visualization of the $\phi_{{\ve{x}},\tau}$ field for  the saddle point configuration with one instanton. The widths of the vertical spindles correspond to the value of $|\phi_{{\ve{x}},\tau}|$  at a given spatial lattice site and time step in Euclidean time. For clarity, we only draw the spindles  if $|\phi_{{\ve{x}},\tau}|>\epsilon$, where  $\epsilon$ is some suitably small threshold. In order to clearly illustrate the spatial positions of the spindles within the lattice, we also draw their projections on the $\tau=275$ plane. Calculations were  carried out  on a  $6\times6$ lattice with interaction strength $U=5.0 \kappa$, $N _{\tau}=512$ and $\beta \kappa=20$.}
   \label{fig:single_instanton}
\end{figure}

 \begin{figure}[]
   \centering
   \subfigure[]{\label{fig:histogramBeta}\includegraphics[width=0.3\textwidth,clip]{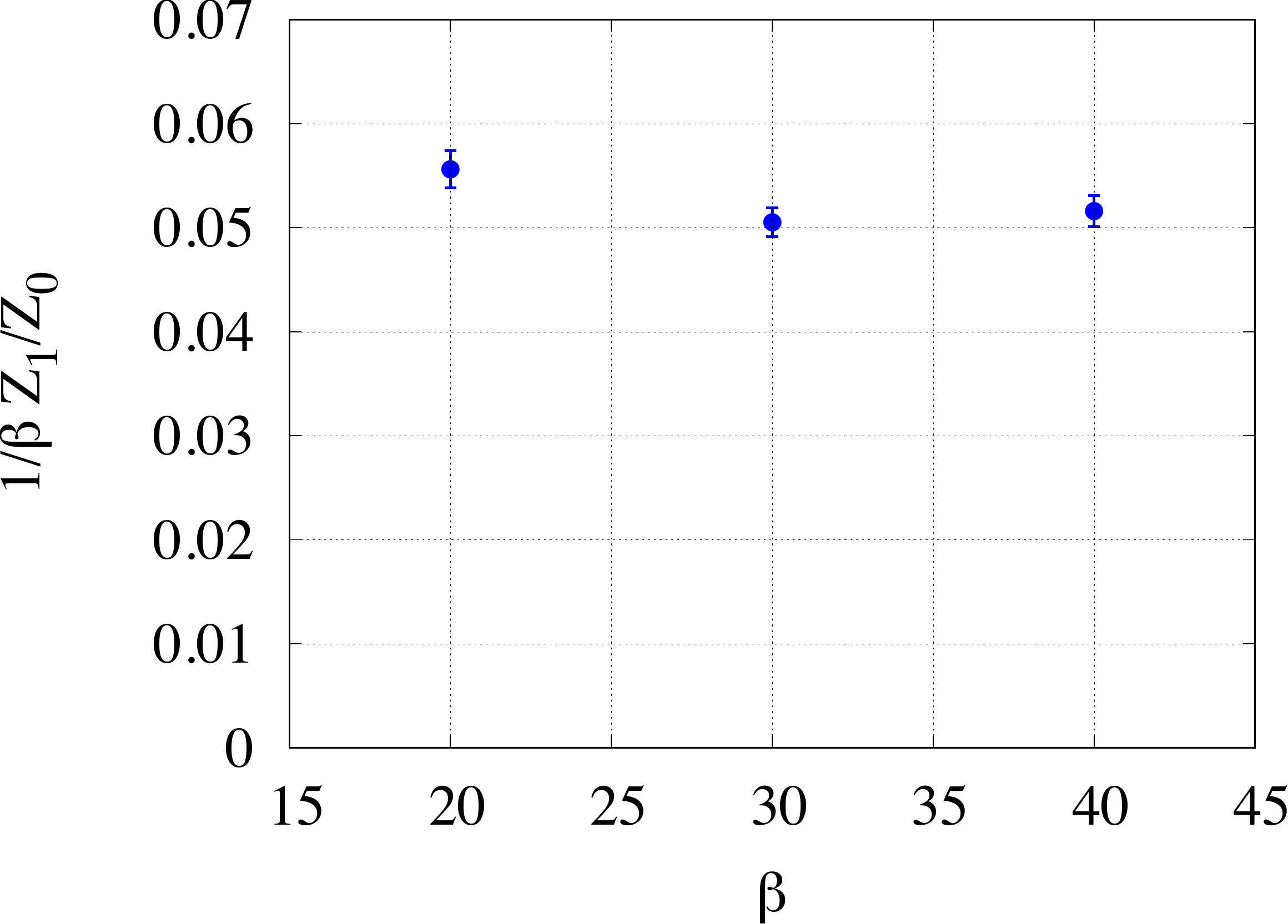}}
   \subfigure[]{\label{fig:histogramNs}\includegraphics[width=0.3\textwidth,clip]{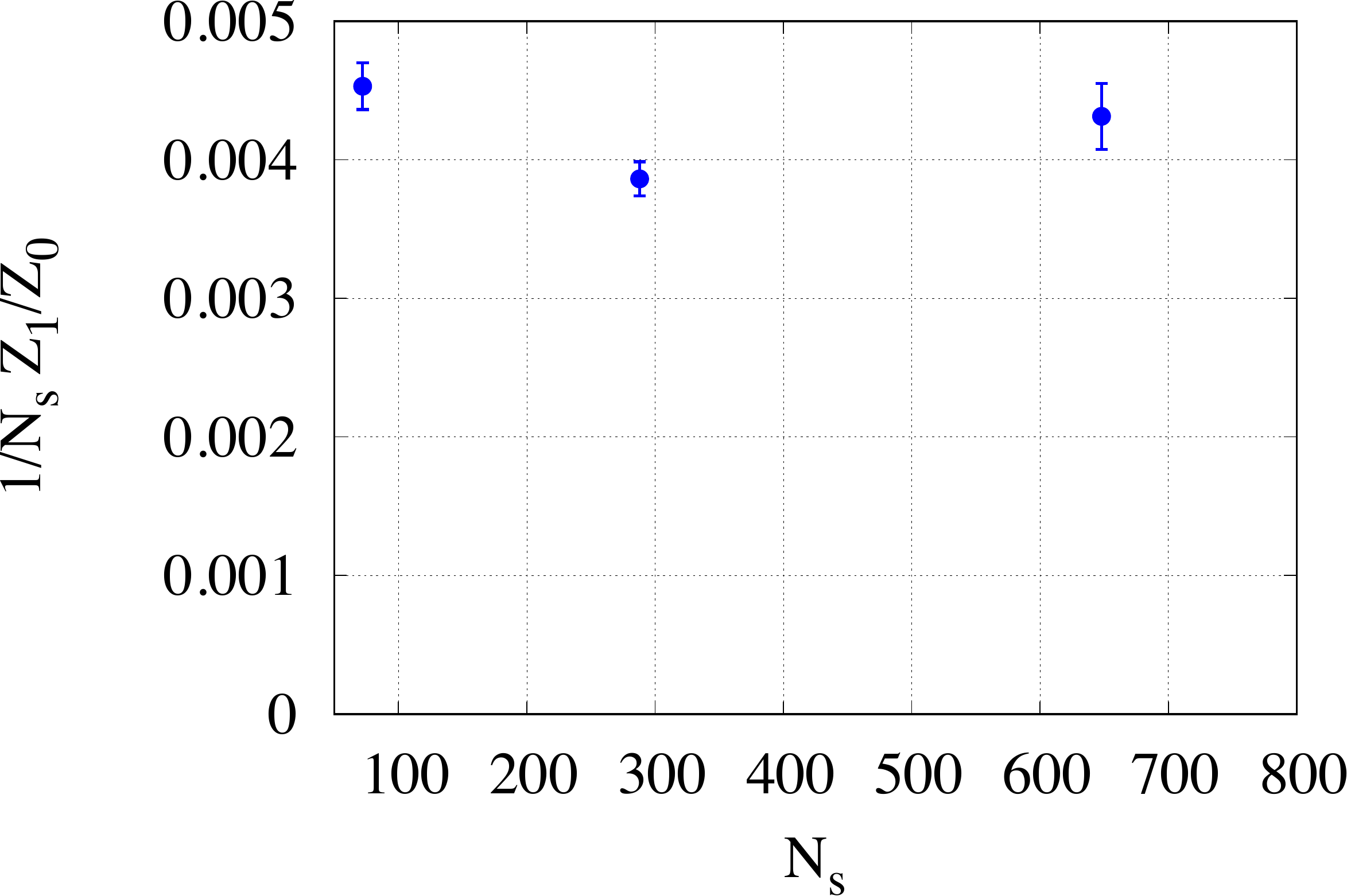}}
         \caption{(a) The relative weight of the first non-vacuum thimble with respect to the full partition function as a function of the inverse temperature. The calculations were
         carried  out  on a $6\times6$ lattice,  with a Euclidean time-step corresponding to $N _{\tau}=512$ at $\beta \kappa=20$. (b) The scaling of the relative weight of the first non-vacuum thimble with the spatial system size at fixed $\beta \kappa =20$ and $N _{\tau}=512$. The interaction strength is fixed at $U=2.0 \kappa$ for all plots.}
   \label{fig:histograms_N_beta}
\end{figure}

\subsection{\label{subsec:Previous}Individual instantons}

The discrete structure of the histograms which characterize the values of the action of the saddle points has a particularly simple explanation. As was already shown in our previous work \cite{PhysRevD.101.014508}, all non-vacuum saddle points for this particular choice of the HS transformation (\ref{eq:continuous_HS_imag}) are formed  by  a collection of individual localized field configurations.  For convenience, we repeat here the plot, showing this type of configuration for the auxiliary bosonic field (Fig. \ref{fig:single_instanton}). One can clearly see that $\phi_{{\ve{x}},\tau}$ is localized both in Euclidean time and in space. This field configuration is the solution for the Euclidean equations of motions for the auxiliary field $\phi_{{\ve{x}},\tau}$ following from the action  (\ref{eq:full_action}). We will henceforth refer to this field configuration as an \textit{instanton}. The detailed reasons for this are outlined in Appendix \ref{sec:AppendixB}.  The one subtlety is that we should take into account the back-reaction from the fermionic determinant from the very beginning, as the bosonic part of the action (\ref{eq:Z_continuous}) is purely Gaussian. Each instanton is defined by its location in space (including sublattice), position of its ``center" (where $|\phi_{{\ve{x}},\tau}|$ is largest) in Euclidean time, and the binary instanton-anti-instanton index. The instanton-anti-instanton index reflects the symmetry of the integrand in (\ref{eq:continuous_HS_imag}) with respect to the sign of the auxiliary bosonic field. Thus, the anti-instanton configuration is related to the instanton by simply inverting the sign of the auxiliary field at each spatial lattice site and on all timeslices, $\phi_{{\ve{x}},\tau}\to -\phi_{{\ve{x}},\tau}$. 

With this information at hand, the histograms in Fig. ~\ref{fig:histogramsU} can be easily understood: the first bar, at $S_\sigma=S_{vac.}$, corresponds to the vacuum field configuration ($\phi_{{\ve{x}},\tau}=0$); the next bar, at $S^{(1)}$, is the saddle with just one instanton located at a random position inside the lattice, which is allowed by translational symmetry; the third bar, at $S^{(2)}$, corresponds to the saddle with two instantons, etc. The width of the bars does not substantially increase as the number of instantons increases, which means that the action of the $N-$instanton field configuration is still approximately equal to $S^{(N)}=S_{vac.}+N (S^{(1)}-S_{vac.})$. Thus, the action of the $N$-instanton configurations is only weakly dependent on the relative position of the instantons, at least if the density of instantons is not too large. Therefore, we can effectively describe the saddle points as a gas of weakly interacting instantons. This conjecture is further supported by the data shown in Fig.~\ref{fig:histograms_N_beta}. This plot clearly illustrates that the weight of the one-instanton saddle  is proportional to both the spatial size of the lattice and the inverse temperature
\begin{equation}
\label{eq:z_proportional}
\mathcal{Z}_1/\mathcal{Z} \sim N_S \beta,
\end{equation}
where $\mathcal{Z}_1$ is the sector of the partition function, corresponding to the integral over the thimble attached to the one-instanton saddle.  
Thus, the localized one-instanton field configuration is not sensitive to the lattice size, provided that its dimensions exceed the size of the instanton.

The next step is the study of $N$-instanton saddles and the interaction of instantons. However, before we turn to the instanton interaction, a few words are in order concerning the continuum limit. Unlike the case of relativistic lattice field theories, the limit of zero lattice spacing is only to be taken for the Euclidean time direction. This is needed in order to be sure that the error introduced in our Trotter decomposition of the Boltzmann weight can be neglected.
As we can see from the analysis in Appendix \ref{sec:AppendixA}, the weights of the $N-$instanton saddles are independent of the lattice spacing in Euclidean time, and thus our numerical results are already effectively at the continuum limit. This property should also be a  requirement of the analytical saddle point approximation. However, a certain complication stems from the fact that the saddles, like the one shown in Fig.~\ref{fig:single_instanton}, are degenerate with respect to the continuum symmetry of translations in Euclidean time. Instead of a single saddle, we in fact have a closed valley in configuration space and it appears that the minimal approximation which has a well-defined continuum limit is the Gaussian integral in all directions except that of the zero mode associated with the translational symmetry in Euclidean time. 

The removal of the zero mode taking into account the collective coordinate factor is well known in the instanton calculus (see e.g. \cite{Coleman:1978ae, Dunne_2008}). For the sake of completeness, we nevertheless consider here explicitly the analytic expression for the partition function in the one-instanton sector in Gaussian approximation. Let $\phi^{(X, T)}_{{\ve{x}},\tau}$ be the one-instanton configuration centered at the space time point $(X,T)$, where the coordinate $X=(\nu, \ve{r})$ includes the spatial position of the center of the instanton $\ve{r}$ (including the sublattice index) and the binary instanton-anti-instanton index $\nu=\pm 1$, while the Euclidean time position is denoted by $T\in[0; \beta)$.  All these configurations belong to one valley $\mathcal{O}^{(1)}=\bigcup_{T\in[0;\beta)}\phi^{(X, T)}$, with the instanton center $T$ being its parameter:
\begin{equation}
	 \left. \frac{\partial  S (\phi) } {\partial \phi_{{\ve{x}},\tau}}   \right|_{\phi  = \phi^{(X, T)}, T\in[0; \beta) }   = 0. 
	 \label{eq:valley_def}
\end{equation}
 We now approximate the action by considering Gaussian fluctuations of the field around the saddle
 \begin{eqnarray}
S  \approx  S(\phi^{(X, T)} ) + \nonumber \\  \frac{1}{2}  \left(\phi_{{\ve{x}},\tau_1}- \phi^{(X, T)}_{{\ve{x}},\tau_1} \right)  \mathcal{H}^{(1)}_{ ({\ve{x}},\tau_1),({\ve{y}},\tau_2 )}   \left( \phi_{{\ve{y}},\tau_2}-  \phi^{(X, T)}_{{\ve{y}},\tau_2} \right)  
\label{eq:valley_gauss_action} 
 \end{eqnarray}
 where 
 \begin{eqnarray}
 \mathcal{H}^{(1)}_{ ({\ve{x}},\tau_1),({\ve{y}},\tau_2 )}  = \left.  \frac{\partial^2 S(\phi)} {\partial \phi_{{\ve{x}},\tau_1} \partial \phi_{{\ve{y}},\tau_2} } \right|_{ \phi=\phi^{(X, T)} } 
\label{eq:hessian_def} 
 \end{eqnarray}
 is the Hessian of the one-instanton saddle point. We denote  the eigenvalues of $\mathcal{H}^{(1)}$ as $\lambda^{(1)}_{i}$, $i=0...N_S-1$. This set contains the zero mode, $\lambda^{(1)}_{0}=0$, due to the above-mentioned translational symmetry.
 
Now, $\mathcal{Z}_1$ can be written as the line integral along the curve $\mathcal{O}^{(1)}$ in configuration space:
\begin{equation}
\label{eq:Z1_zero_mod_sep}
\mathcal{Z}_1= 2 N_S \int_{\mathcal{O}^{(1)}} d \tilde{\phi_0} \mathcal{Z}^P_1 (\{\phi^{(X, T)}\}), 
\end{equation}
where $2 N_S$ factor describes the trivial discrete spatial and instanton - anti-instanton degeneracies and $\mathcal{Z}^P_1$ is what we will refer to as the partial partition function. Here we have introduced
$d \tilde{\phi_0}$, which is the differential arc length of the $\mathcal{O}^{(1)}$ curve: 
 \begin{equation}
 d \tilde{\phi_0}  = ||\phi^{(X, T+dT)} -\phi^{(X, T)} ||    
\label{eq:valley_arc_element} 
 \end{equation}
 such  that  the  length  of  the valley  is 
 \begin{equation}
 	 L^{(1)}= \int_{0}^{\beta}  dT   \left| \left|  \frac{ \phi^{(X, T+dT)} -\phi^{(X, T)} } {dT }  \right|   \right|. 
\label{eq:valley_length} 
 \end{equation}
  In practice, $L^{(1)}$  on the lattice is the collection of $N_\tau$ steps, each corresponding to the shift $T \rightarrow T+\Delta \tau $. Thus, according to  (\ref{eq:valley_length}) , $L^{(1)}$ can be approximated by the following finite difference of field values
\begin{equation}
\label{eq:length_discrete}
L^{(1)}=N_\tau \sqrt{\sum_{{\ve{x}},\tau} \left(  \phi^{(X, 0)}_{{\ve{x}},\tau} -  \phi^{(X, \Delta \tau)}_{{\ve{x}},\tau} \right)^2 }.
\end{equation}
Alternatively, we can take into account that the field configuration $\phi^{(X, T)}_{{\ve{x}},\tilde \tau}$ is in fact a function of the difference $\tilde \tau-T$, where the dimensional Euclidean time index: $\tilde \tau \in [0; \beta)$: $\tau=\tilde \tau/\Delta \tau$. Thus
\begin{equation}
\label{eq:phi_derivative_norm}
\frac{L^{(1)}}{\beta}=||\Delta \phi^{(X, T)}||=\sqrt{\sum_{{\ve{x}},\tau} \left( \frac{ \phi^{(X, T)}_{{\ve{x}},\tau+1} -  \phi^{(X, T)}_{{\ve{x}},\tau}}{\Delta \tau} \right)^2 },
\end{equation}
where $||\Delta \phi^{(X, T)}||$ is the norm of the lattice derivative of the one-instanton field configuration with respect to the physical Euclidean time.

The partial partition function $\mathcal{Z}^P_1 (\{\phi^{(X, T)}\})$ describes the Gaussian fluctuations around the configuration $\phi^{(X, T)}$ in all directions except the one corresponding to  the zero mode:
\begin{equation}
\label{eq:Z1_P}
\mathcal{Z}^P_1 (\{\phi^{(X, T)}\})= \int \prod_{i=1}^{N_S N_\tau-1} d \tilde \phi_i e^{-S^{(1)} - \frac{1}{2} \sum_{i=1}^{N_S N_\tau-1} \lambda^{(1)}_{i} \tilde \phi_i^2}.
\end{equation} 
Here, $\tilde \phi_i$ are the coordinates in configuration space in the directions of the corresponding eigenvectors of the Hessian $\mathcal{H}^{(1)}$, computed for the configuration $\phi^{(X, T)}$.  Now, the eigenvalues of the Hessian $ \lambda^{(1)}_{i}$ and the  value of $\mathcal{Z}^P_1 (\{\phi^{(X, T)}\})$ are in fact independent of the coordinates of the instanton center $(X, T)$. This means that the integral (\ref{eq:Z1_zero_mod_sep}) boils down to just 
\beq
\nn
\mathcal{Z}_1&=& 2 N_S  \mathcal{Z}^P_1 (\{\phi^{(X, T)}\}) \int_{\mathcal{O}^{(1)}} d \tilde \phi_0  \\ \label{eq:Z1_zero_mod_sep} &=& 2 N_S  \mathcal{Z}^P_1 (\{\phi^{(X, T)}\}) L^{(1)}.
\eeq

Performing the Gaussian integral in (\ref{eq:Z1_P}), the final expression for $\mathcal{Z}_1$ reads as
\begin{equation}
\label{eq:Z1_final}
\mathcal{Z}_1=2 N_S L^{(1)}  e^{-S^{(1)}}\sqrt{\frac{(2 \pi)^{N_S N_\tau-1}}{\prod'_{i} \lambda^{(1)}_{i}}}.
\end{equation}
Here, the product of the eigenvalues of the Hessian in the denominator excludes the zero mode, for a total of $N_s N_{\tau}-1$ eigenvalues. In order to reproduce the empirical relation given in (\ref{eq:z_proportional}), we restore physical units in Euclidean time according to (\ref{eq:phi_derivative_norm}) in order to obtain
\begin{equation}
\label{eq:Z1_final1}
\mathcal{Z}_1=2 N_S \beta e^{-S^{(1)}} ||\Delta \phi^{(X, T)}|| \sqrt{\frac{(2 \pi)^{N_S N_\tau-1}}{\prod'_{i} \lambda^{(1)}_{i}}}.
\end{equation}
If the inverse temperature $\beta$ is substantially larger than the width of the instanton, the norm is independent of $\beta$ and we reproduce the desired, empirically-determined scaling in (\ref{eq:z_proportional}). 

The absence of the zero mode in the product over eigenvalues in the denominator in Eq.(\ref{eq:Z1_final1}) can be formally expressed as follows  
\begin{equation}
\label{eq:det_H_perp}
\det \mathcal {H}^{(1)}_\perp = \det \left( \mathcal {H}^{(1)} + \mathcal {P}^{(1)} \right)=\prod^{N_s N_{\tau}-1}_{i=1} \lambda^{(1)}_{i},
\end{equation}
where $\det \mathcal {H}^{(1)}_\perp$ corresponds to the result of the Gaussian integral over all directions around the one-instanton saddle point excluding the zero mode, and $\mathcal {P}^{(1)}$ is the projection operator on to the zero mode 
direction in configuration space. 

Finally, for the instanton structure of the partition function (for which the $N$-instanton saddle is dominant in $\mathcal {Z}$), we only need their ratio with respect to the part of the partition function corresponding to the vacuum saddle  $\mathcal {Z}_N/\mathcal {Z}_0$. For the $1$-instanton saddle, this means that what we really need to compute is the following expression 
\begin{equation}
\label{eq:Z1_Z0}
\frac{\mathcal{Z}_1}{\mathcal{Z}_0} =2 N_S  L^{(1)} e^{-\tilde S^{(1)}}  \left( 2\pi \frac{\det \mathcal {H}^{(1)}_\perp}{\det \mathcal {H}^{(0)}  } \right)^{-1/2},
\end{equation}
where $\tilde S^{(i)} = S^{(i)} - S_{vac.}$. In this expression, $L^{(1)}$ is $\Delta \tau$ dependent and thus the Gaussian fluctuations in the perpendicular directions must be taken into account to achieve the $\Delta \tau$-independent results in the continuum limit. In this case, the $\Delta \tau$-dependencies in $L^{(1)}$ and in the Hessian matrices compensate each other.  The numerical results for the expression in Eq.~(\ref{eq:Z1_Z0}) are shown in Table~\ref{table:1}. The independence of $\tilde S^{(i)} $ on the step size in Euclidean time is shown in Appendix \ref{sec:AppendixA}.  Evidently, our simulations are already close to the continuum limit, as the final result for the ratio $\mathcal{Z}_1/\mathcal{Z}_0$ is practically $\Delta \tau$-independent.

\begin{figure}[]
    \centering
             {\includegraphics[width=0.3\textwidth, angle=270,clip]{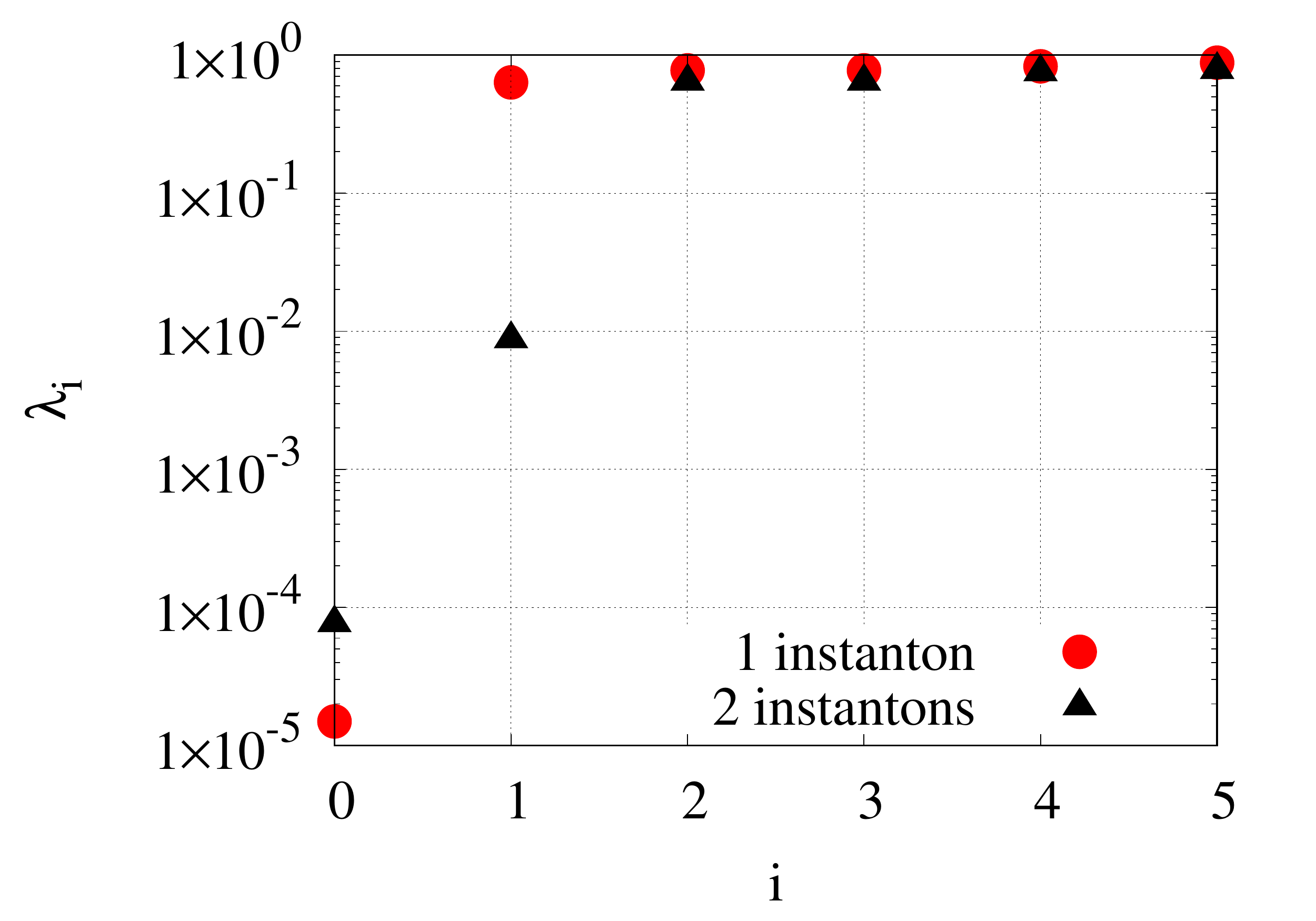}}
      \caption{The five lowest eigenvalues of the Hessian matrices computed at the one- and two-instanton saddle points. The calculations were performed on a $6\times6$ lattice with interaction strength $U=6.0\kappa$, $N _{\tau}=512$ and $\beta\kappa=20$.}
   \label{fig:hessian_eigenvalues}
\end{figure}

\begin{table}[]
\centering
\begin{tabular}{||c c c c||} 
 \hline
 $N_\tau$ & $L^{(1)}$ &  $ \left(\frac{\det \mathcal {H}^{(1)}_\perp}{\det \mathcal {H}^{(0)}  } \right)$  &  $\frac{\mathcal{Z}_1}{2 N_S \mathcal{Z}_0}$ \\ [0.5ex] 
 \hline\hline
 256 & 62.699 & 0.124533  & 445.398 \\ 
 512 & 44.272 & 0.06208   & 445.355 \\ [1ex] 
 \hline
\end{tabular}
\caption{Here we display the valley lengths and Hessians for the one-instanton saddles on lattices of two different sizes in Euclidean time: $N_\tau=256$ and $N_\tau=512$. The remaining parameters are fixed with the spatial size given by $6\times 6$ with $U=2.0 \kappa$, $\kappa \beta=20$. }
\label{table:1}
\end{table}

\subsection{\label{subsec:Interaction}Interaction of instantons}

In the previous section, we have described in detail the one-instanton saddle and its partition function. As the interaction $U$ becomes large, it becomes increasingly likely for multi-instanton configurations to appear. While it may be appropriate in some regimes to treat these systems as a non-interacting gas of instantons, one would like to understand the interaction between instantons. 
In this section, we consider the two-instanton field configurations $\phi^{((X_1, T_1), (X_2, T_2))}_{{\ve{x}},\tau}$, where the coordinates $(X_i, T_i), ~i=1,2$ define the positions of the centers of the instantons.

The lowest eigenvalues of the Hessian  for the one- and two-instanton saddles are shown in Fig.~\ref{fig:hessian_eigenvalues}.  Here, $\lambda_0$ corresponds to the zero mode (in actual numerical computations it is never exactly equal to zero due to the finite lattice spacing in Euclidean time). However, for the 2-instanton saddle we see that the next eigenvalue $\lambda_1$ is still much smaller then all $\lambda_i$ for $i>1$. This is what we refer to as a quasi-zero mode, which occurs due to the symmetry by which  the instantons are shifted with respect to one another. Imagine two instantons with fixed spatial positions $X_1$ and $X_2$. Then, one can vary their time coordinates $T_1$ and $T_2$. The field configurations generated in this way form a two-dimensional torus in configurations space and the eigenvectors for $\lambda_0$ and $\lambda_1$ define tangent planes to this torus. 

As was done in Eq. (\ref{eq:Z1_final1}), we would like to construct an expression for the partition function of the two-instanton saddle. To faithfully represent the physics of the multi-instanton saddle, we must take into account the change of the action along the previously mentioned torus which defines the symmetry of the two-instanton saddle. At the same time, we must guarantee that our expression has a well-defined continuum limit in Euclidean time, $\Delta \tau \rightarrow 0$. This is encoded in the following expression
\begin{eqnarray}
\label{eq:L_det_H1}
\mathcal{Z}_2=\frac{1}{2}\sum_{X_1, X_2} W^{(2)}(X_1, X_2),
\end{eqnarray}
where
\begin{eqnarray}
\label{eq:L_det_H2}
W^{(2)}(X_1, X_2)=\int d \bar T d \Delta T e^{-S(X_1,X_2, \Delta T)} \times \nonumber \\  \sqrt{g (X_1,X_2,\Delta T) } \left( \prod^{N_S N_\tau-1}_{i=2} \frac{2 \pi}{\lambda_i (X_1,X_2,\Delta T)} \right)^{1/2}.
\end{eqnarray}
As in the case of the one-instanton saddle, both coordinates $X_i$ include the spatial part with sublattice index and the binary instanton-anti-instanton index. The action of the field configuration, $S(X_1,X_2, \Delta T)$, is characterized by the two spatial locations and their separation in Euclidean time. The quantity $\sqrt{g}$ is the first fundamental form of the mapping of the surface of the two-dimensional torus to the ``center of mass" and relative Euclidean time coordinates, $(\bar T, \Delta T)$, where 
\begin{eqnarray}
\label{eq:center_delta_time}
\bar T= \frac{T_1+T_2}{2}, \\
\Delta T= T_1-T_2.
\end{eqnarray}
Finally, the factor of $\frac{1}{2}$ in Eq. (\ref{eq:L_det_H1}) compensates for the double counting, which appears due to the fact that the instantons are indistinguishable. Thus after the exchange $X_1 \leftrightarrow X_2$ and $T_1 \leftrightarrow T_2$ we still arrive at the same saddle.

One can immediately notice that the integrand in (\ref{eq:L_det_H2}) is independent of $\bar T$, as this direction corresponds to the true zero mode, where  both instantons are simultaneously translated in the Euclidean time direction. 
We now can rewrite the integrand of the above expression in a suggestive way 
\begin{eqnarray}
\label{eq:U_eff_2inst}
U^{(2)}(X_1, X_2, \Delta T) \equiv U^{(2)}_S(X_1, X_2, \Delta T) +\nonumber \\ U^{(2)}_g(X_1, X_2, \Delta T)+ \\ U^{(2)}_\lambda(X_1, X_2, \Delta T), \nonumber 
\end{eqnarray}
which we can identify as the two-body instanton interaction. The individual terms in (\ref{eq:U_eff_2inst}) each have a clear interpretation and meaning. The first term represents the change in the action with respect to two, infinitely separated instantons
\begin{eqnarray}
\label{eq:U_eff_2inst_S}
U^{(2)}_S(X_1, X_2, \Delta T)=S(X_1,X_2, \Delta T) -S^{(2)}. \end{eqnarray}
The next two terms come from re-exponentiating both the first fundamental form and the zero-mode regulated determinant of the Hessian matrix 
\begin{eqnarray}
\label{eq:U_eff_2inst_g}
U^{(2)}_g(X_1, X_2, \Delta T)= - \frac{1}{2} \left(  \ln {g (X_1,X_2,\Delta T) } \right. \\ \nonumber- \left. \ln {g (X_1,X_2,\infty) }  \right), \\
\label{eq:U_eff_2inst_l}
U^{(2)}_\lambda(X_1, X_2, \Delta T)= \frac{1}{2} \sum^{N_S N_\tau-1}_{i=2} \left( \ln \lambda_i (X_1,X_2,\Delta T) \right. \\ \nonumber- \left.  \ln \lambda_i (X_1,X_2,\infty)    \right).
\end{eqnarray}
$U^{(2)}_g$ is computed using the triangulation of the surface of a 2D torus formed by the field configurations of the two instanton solutions. All potentials are normalized by their values at infinitely large time separation between instantons. Together, these can be taken as the starting point for a many-body theory of pairwise-interacting semiclassical objects.

\begin{center}
  \begin{figure}[t!]
   \subfigure[]{\label{fig:interactionTime1}\includegraphics[width=0.23\textwidth,clip]{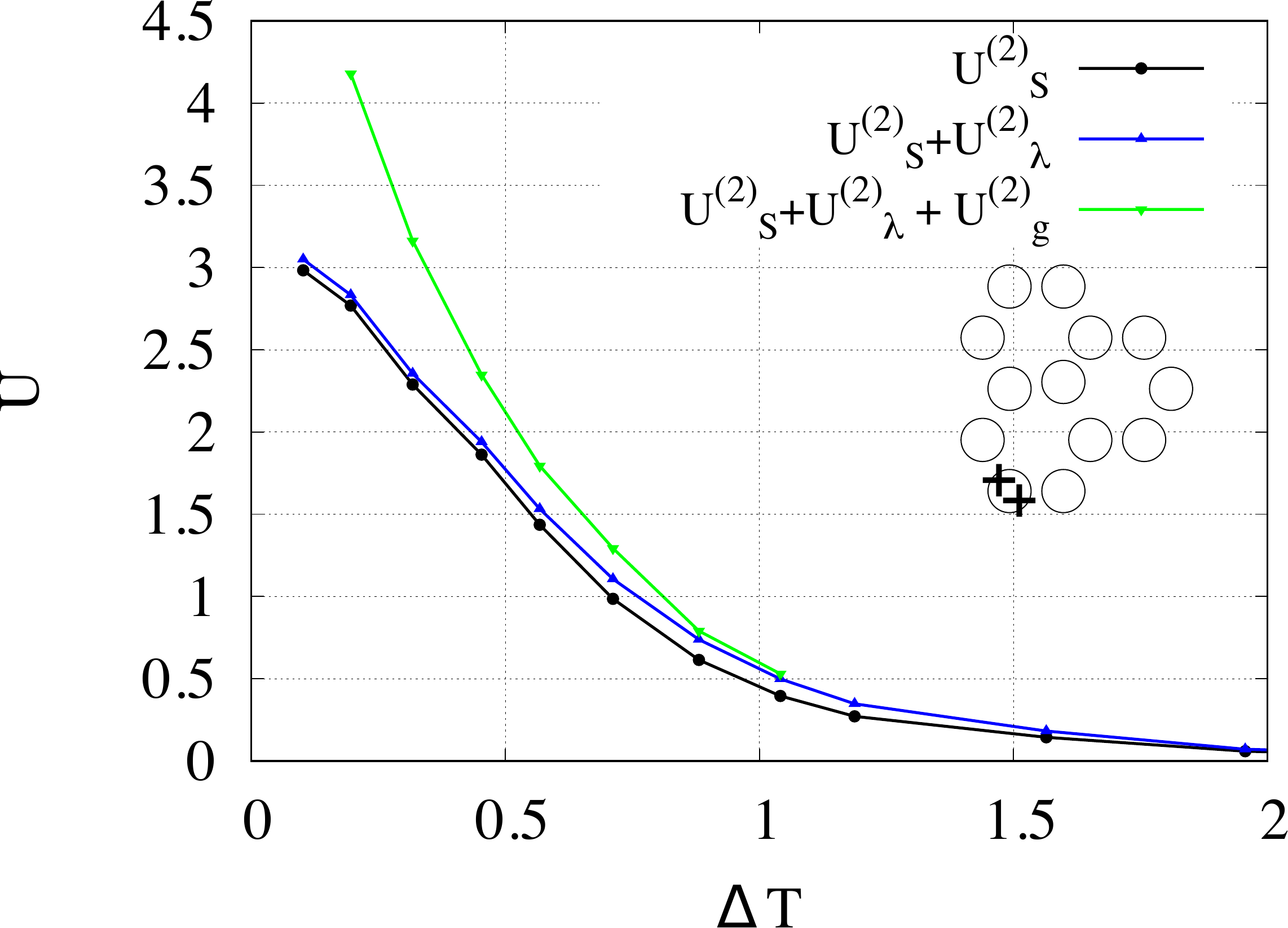}}
   \subfigure[]{\label{fig:interactionTime2}\includegraphics[width=0.23\textwidth,clip]{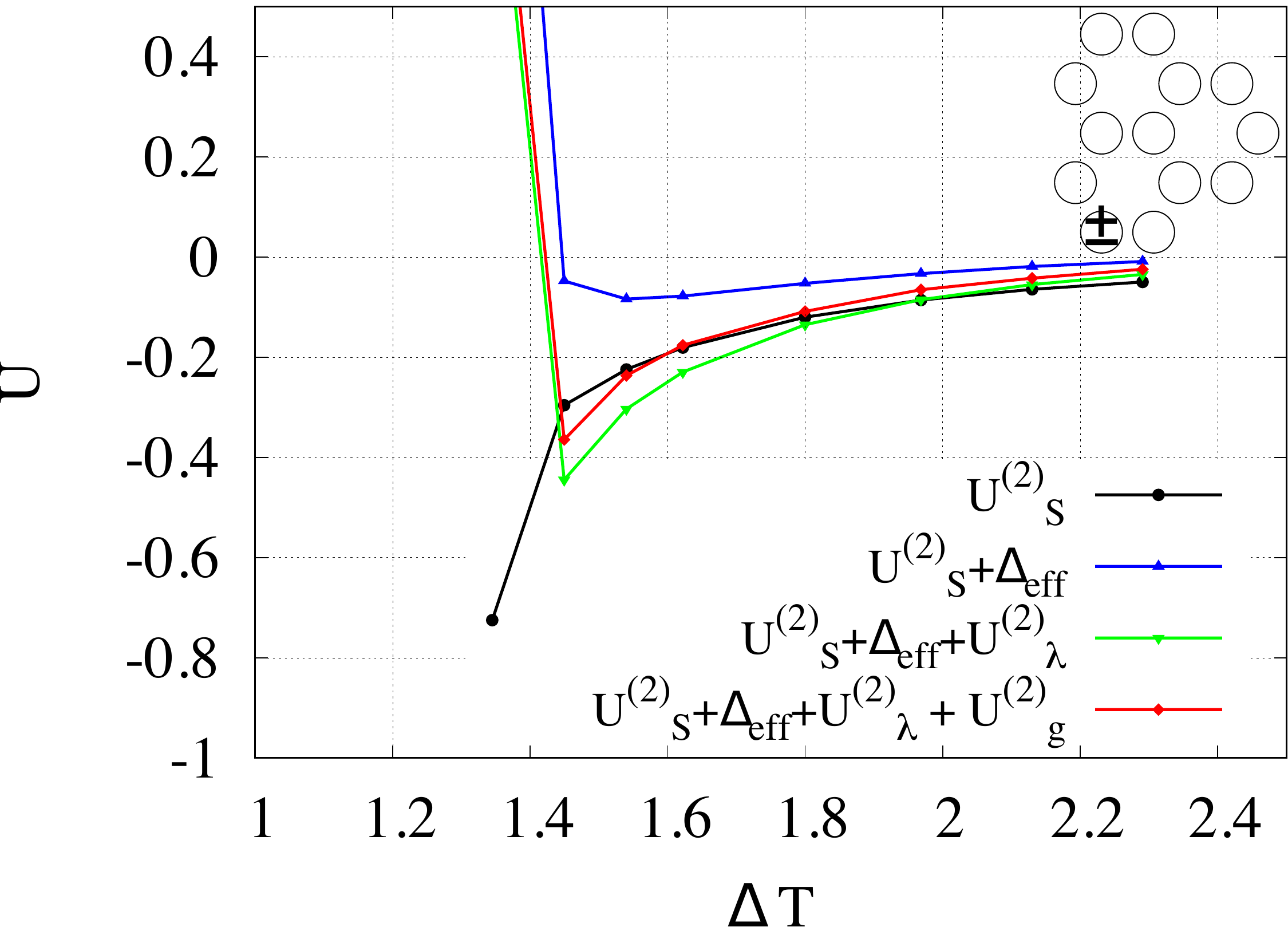}}
   \subfigure[]{\label{fig:interactionTime3}\includegraphics[width=0.23\textwidth,clip]{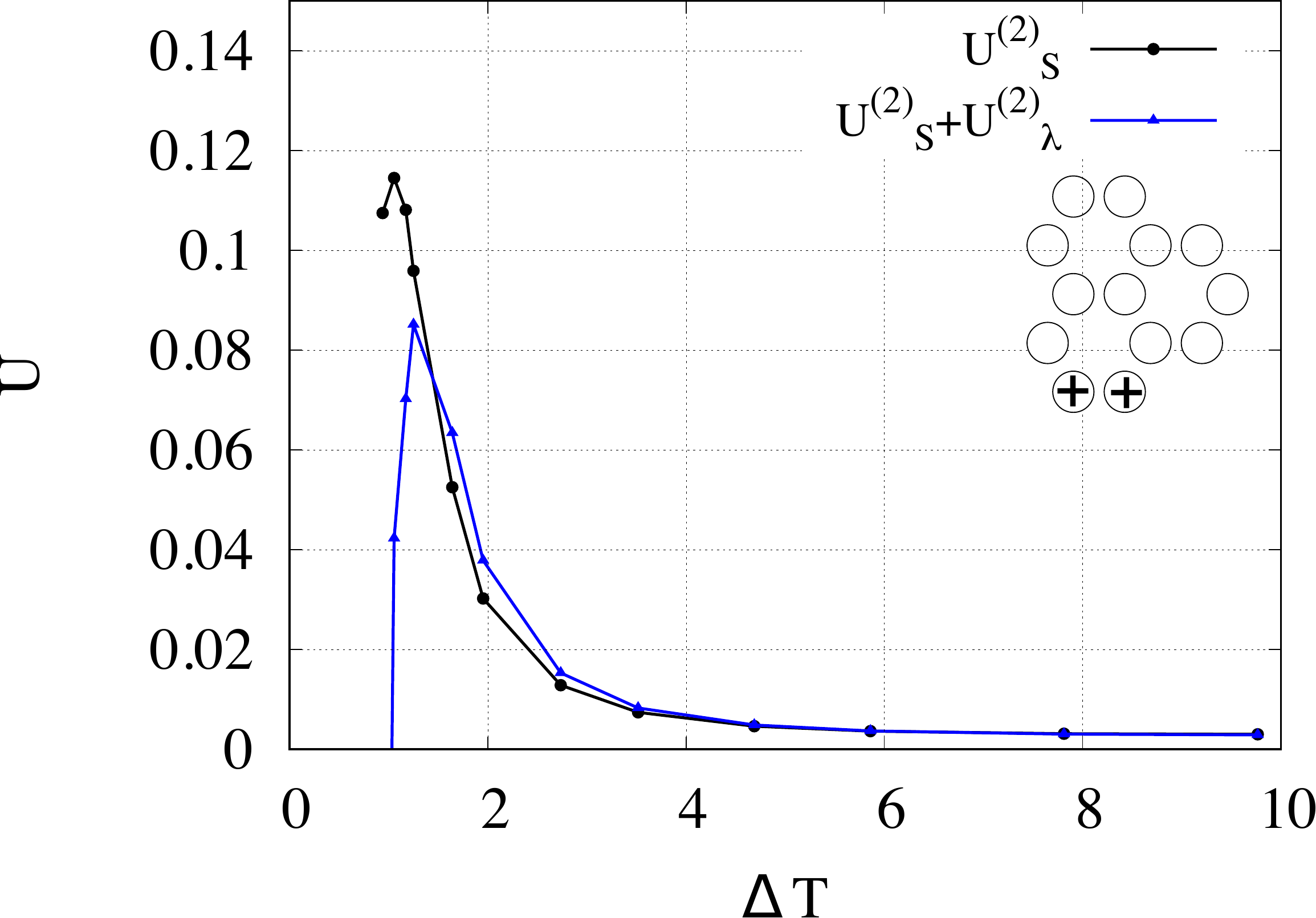}}
    \subfigure[]{\label{fig:interactionTime4}\includegraphics[width=0.23\textwidth,clip]{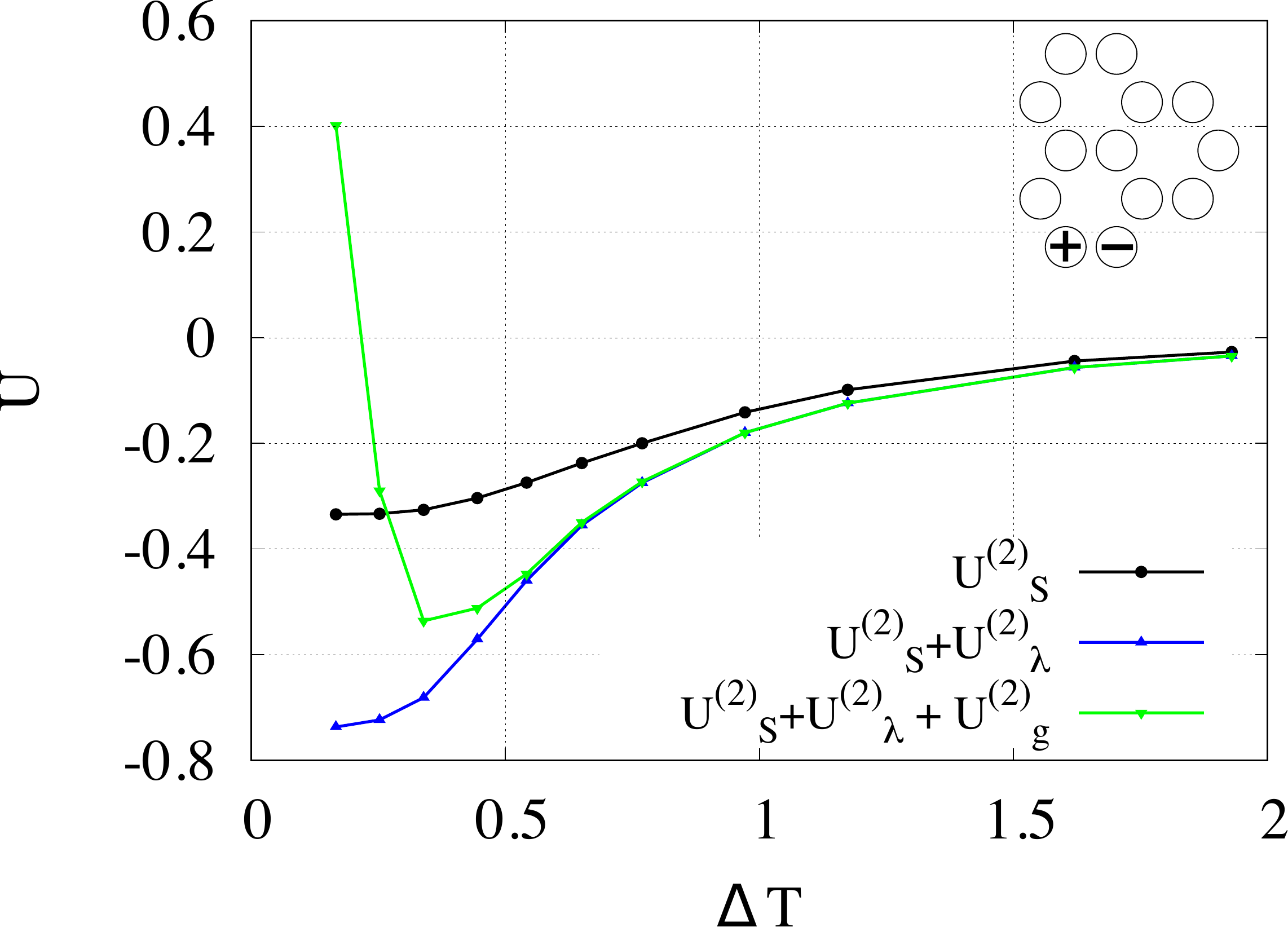}}
   \subfigure[]{\label{fig:interactionTime5}\includegraphics[width=0.23\textwidth,clip]{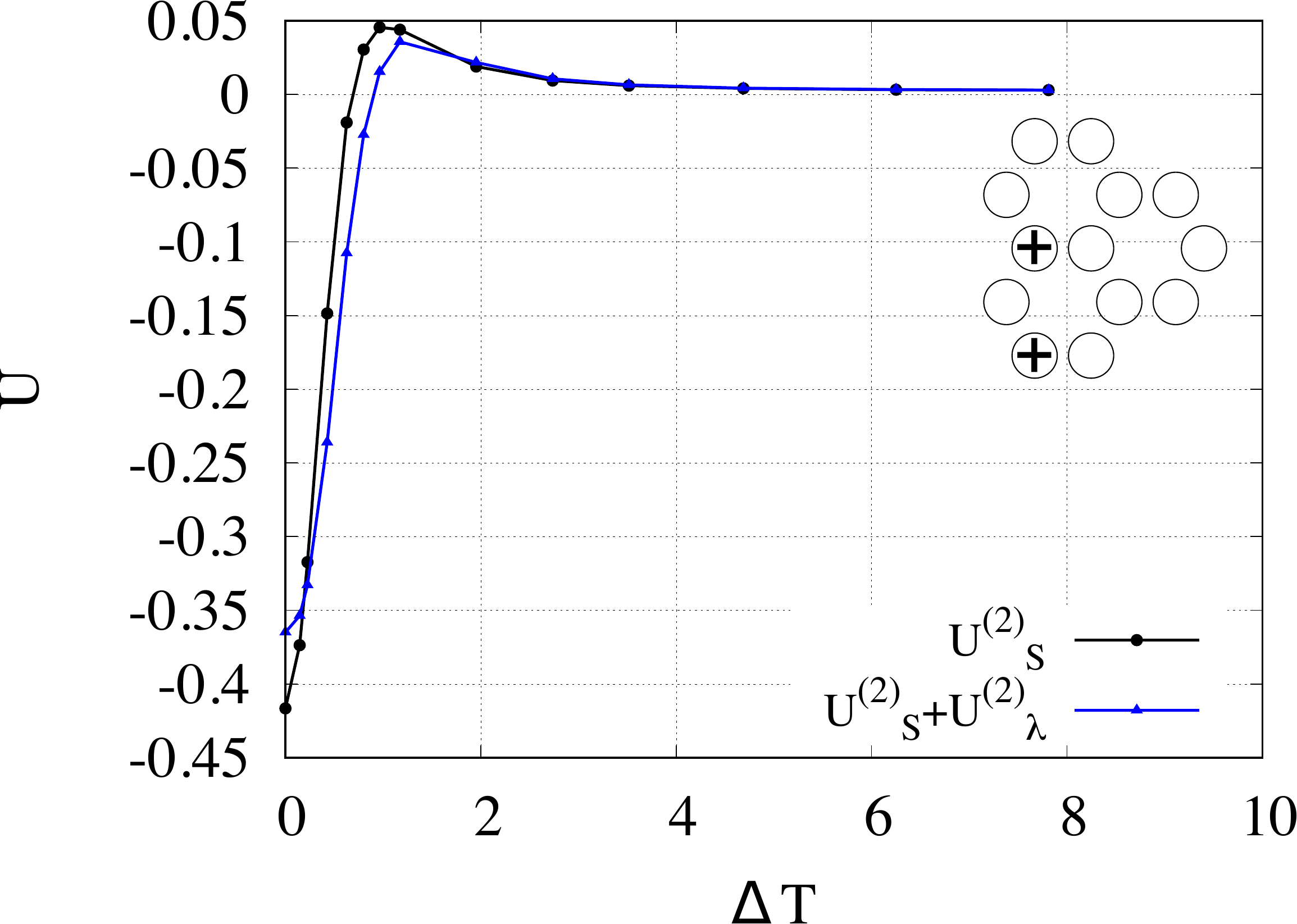}}
   \subfigure[]{\label{fig:interactionTime6}\includegraphics[width=0.23\textwidth,clip]{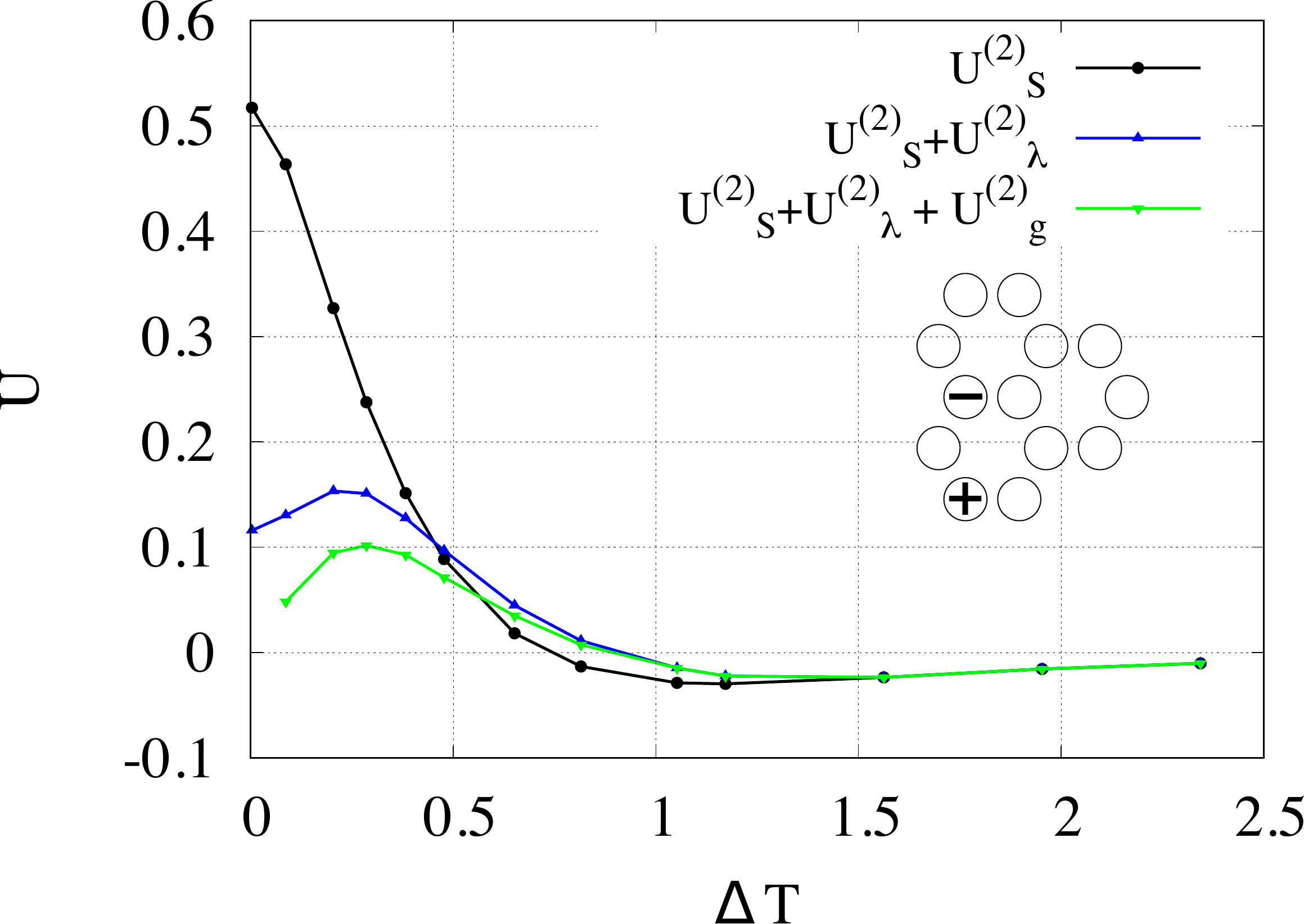}}
     \subfigure[]{\label{fig:interactionTime7}\includegraphics[width=0.23\textwidth,clip]{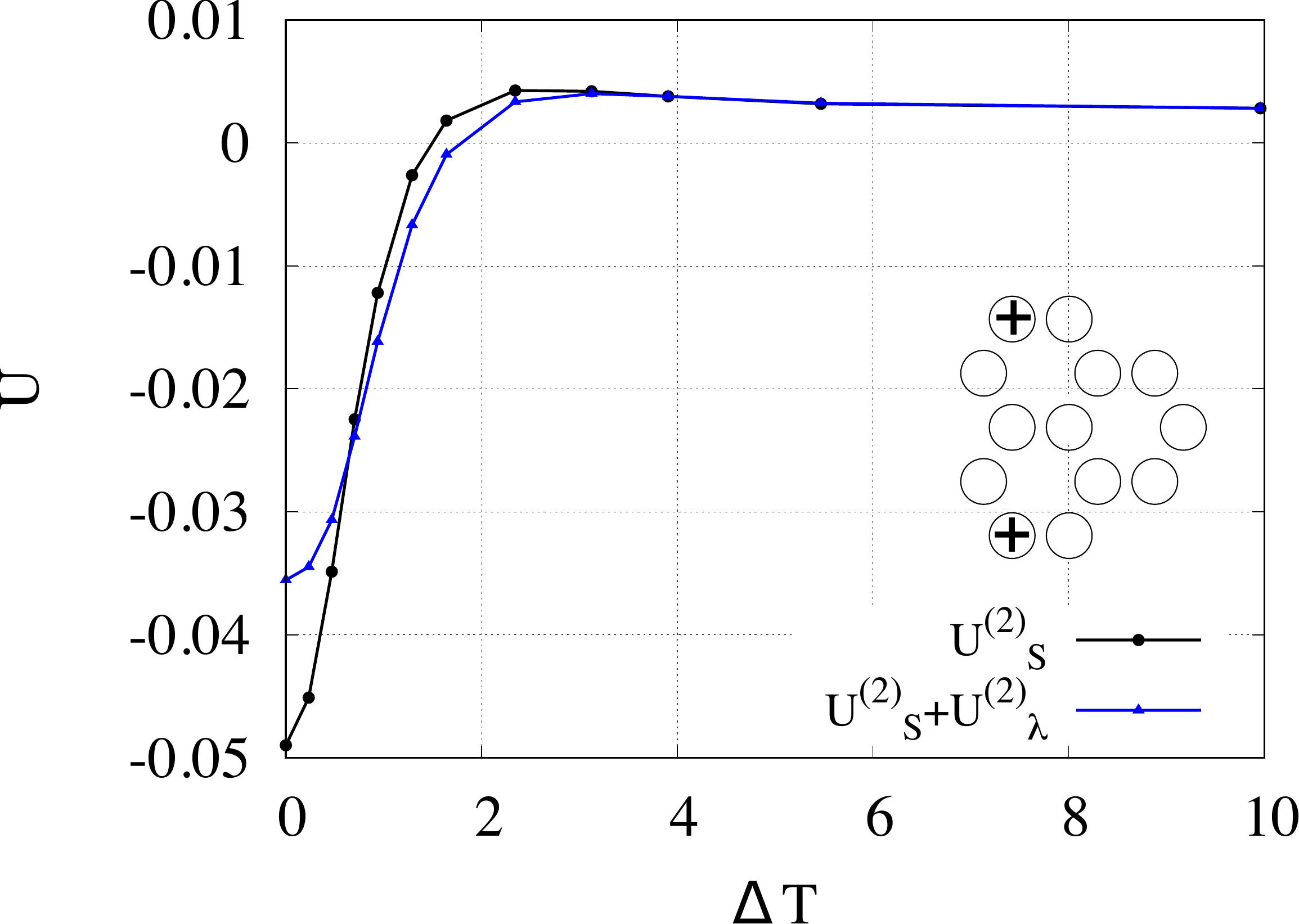}}
   \subfigure[]{\label{fig:interactionTime8}\includegraphics[width=0.23\textwidth,clip]{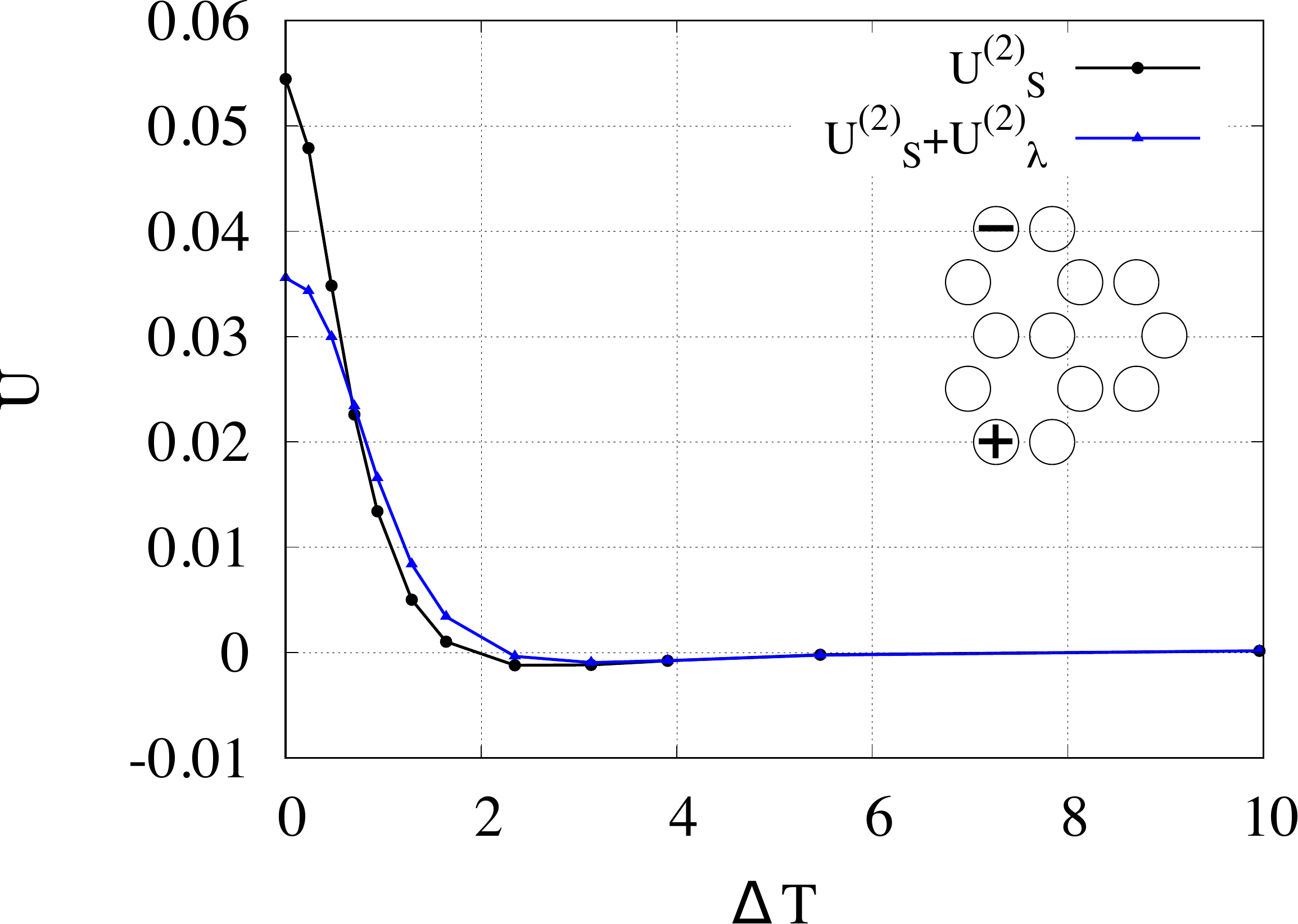}}
         \caption{The pairwise interaction of instantons and anti-instantons as a function of their separation in Euclidean time for a fixed spatial location. The plots in the left column (a,c,e,g) correspond to interaction between a pair of instantons, while the plots in the right column (b,d,f,h) correspond to the interaction between an instanton and an anti-instanton. The first row (a and b) shows (anti)instantons at the same spatial lattice site, the second row (c and d) corresponds to (anti)instantons located at nearest neighbours (opposite sublattices), the third row (e and f) corresponds to a spatial separation of next-nearest neighbours, and the last row (g and h) shows the interaction of (anti)instantons located at sites which are separated by two lattice unit vectors. In each case, we include a sketch of the corresponding spatial configuration on the hexagonal lattice in the inset. If $U^{(2)}_g$ is not shown it means that its influence is small and the  $U^{(2)}_S + U^{(2)}_\lambda + U^{(2)}_g$ line coincides with the $ U^{(2)}_S + U^{(2)}_\lambda$ line. These calculations were performed on a $12\times12$ lattice with $\beta \kappa =20$ and $N_{\tau}=512$, with interaction strength $U=4.6 \kappa$. }
   \label{fig:interactionsTime}
\end{figure}
\end{center}

\begin{figure}[]
   \centering
   \subfigure[]{\label{fig:interactionSpacePlus}\includegraphics[width=0.35\textwidth,clip]{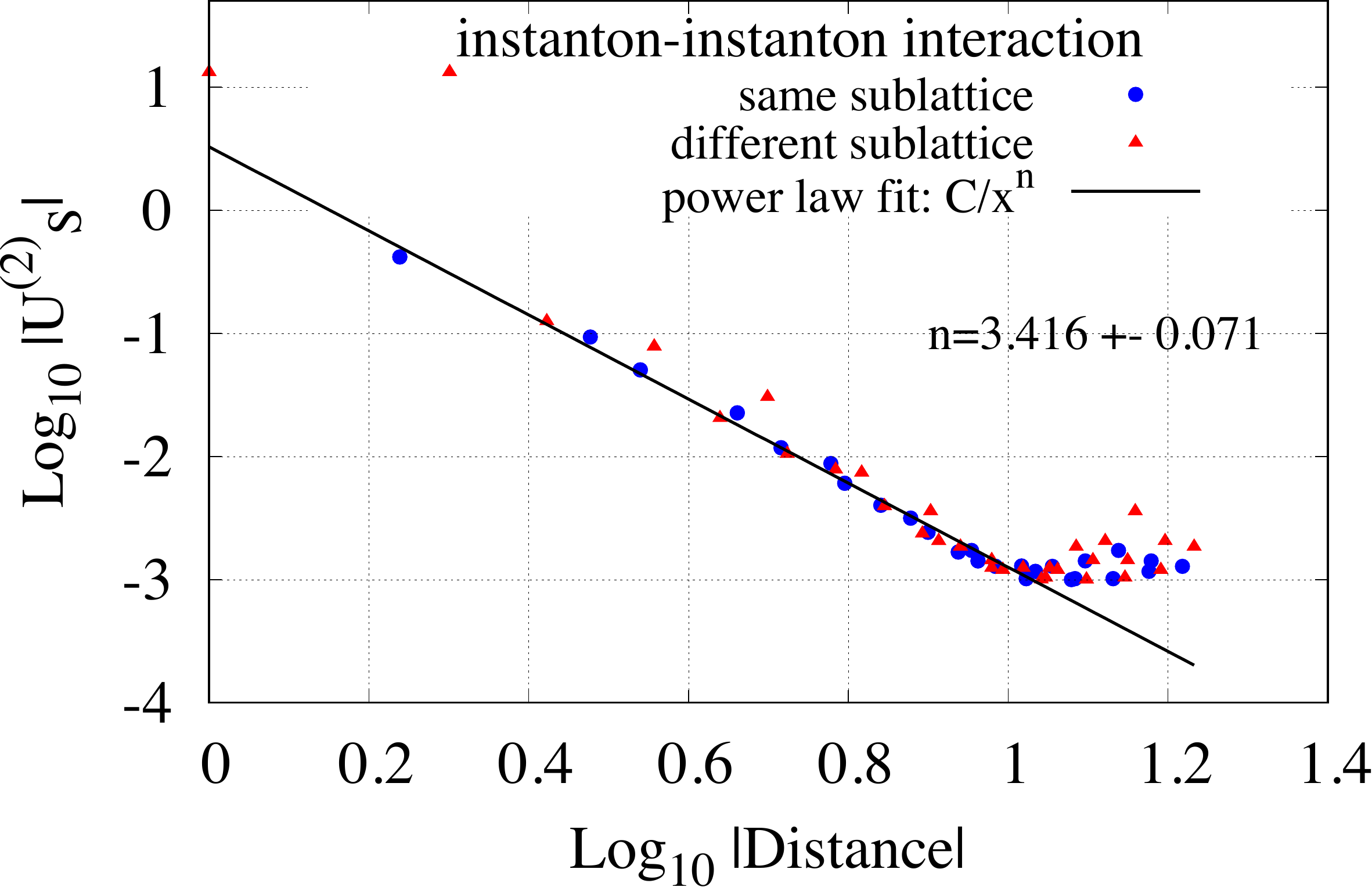}}
   \subfigure[]{\label{fig:interactionSpaceMinus}\includegraphics[width=0.35\textwidth,clip]{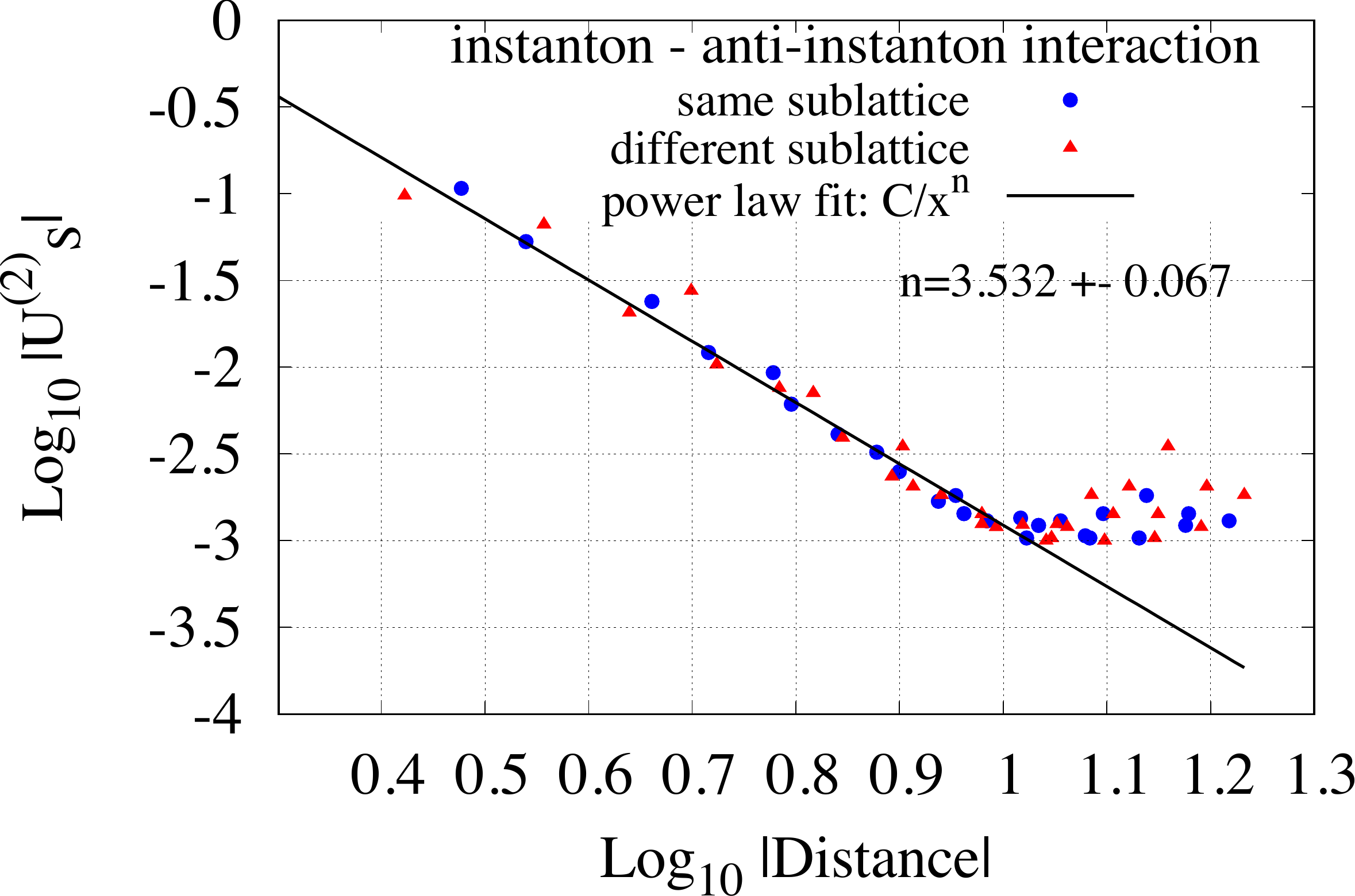}}
          \caption{(a) The action of an instanton-instanton pair as a function of the spatial separation between the instantons. In this case, zero action corresponds to the instantons being infinitely separated. (b) The same situation but for an instanton-anti-instanton pair. These calculations were performed on a $12\times12$ lattice with $\beta \kappa=20$ and $N_{\tau}=512$, and interaction strength $U=4.6 \kappa$. The (anti) instantons are placed at the same Euclidean time slice. The spatial separation is in units of the distance between nearest neighbours. In both cases, a power law fit has been added.}
   \label{fig:interactionsSpace}
\end{figure}

The two-body instanton interaction can be investigated numerically. This is simply done, by hand, by combining two separate one-instanton configurations of the auxiliary bosonic field, where the instantons are located at two different lattice sites and separated by a fixed distance in Euclidean time. The bosonic term in the action, being Gaussian, is trivial, whereas the fermion determinant on a fixed background can be computed using the Schur complement solver \cite{ULYBYSHEV2019118}. Our findings are illustrated in Fig.~\ref{fig:interactionsTime}. 
Here, several profiles of $U^{(2)}(X_1, X_2, \Delta T)$ for the instanton-instanton and instanton-anti-instanton pairs are plotted. As the instantons and anti-instantons are ``ultra-local" in space (almost delta-function-like), only separations up to the distance of fourth-nearest-neighbors on the hexagonal lattice are displayed. The two-body interaction rapidly decreases with increasing separation in Euclidean time as is visible from each of the plots in Fig.~ \ref{fig:interactionsTime}. For reference, the difference between the action of one instanton and the vacuum is equal to $\tilde S^{(1)}=6.5658$ in this case. This means that, in general, the interaction strength is at least one order of magnitude smaller than the difference in action between the  vacuum and a single instanton. Thus, we can treat instantons as non-interacting classical particles in 3D space, except for the case when they occupy the same spatial site. This is the so-called ``hard-core" repulsion between instantons and anti-instantons which also appears in semiclassical models for the vacuum in quantum chromodynamics (QCD) \cite{doi:10.1142/5367}. In addition to this,  noticeable effect is also a local, attractive interaction between an instanton and an anti-instanton on nearest-neighbour sites. The conclusion about the locality of the instanton-instanton interactions is further supported by the spatial profiles plotted in Fig. \ref{fig:interactionsSpace}. In this case, we plot only the variation of the action $U^{(2)}_S(X_1, X_2, \Delta T=0)$ and omit the other two terms. It is clearly shown to rapidly decrease with increased spatial separation. This implies that an ultra-local interaction accurately captures the physics of the saddle points.

However there is a small caveat which we here note. Special treatment is needed when we consider the instanton and anti-instanton occupying the same spatial lattice site. In this case, they can actually annihilate, which means that the ``valley" for the instanton-anti-instanton configuration is smoothly connected to the vacuum saddle
\begin{eqnarray}
\left\{ { {\tilde X_1=(\nu, \ve{r})}\atop{\tilde X_2=(-\nu, \ve{r})} } \right.  \Rightarrow S(\tilde X_1,\tilde X_2, \Delta T=0)=S_{vac.}. 
\label{eq:instanton_anti_instanton_collapse}
\end{eqnarray}

When constructing a semiclassical model of instantons and anti-instantons, a question naturally arises as to double counting. This is due to the fact that the vacuum saddle and Gaussian fluctuations around it were already taken into account by the factor $e^{-S_{vac.}}\det \mathcal{H}^{(0)}$ in the approximate partition function. One way to address this issue of double counting is to consider the profile of the action and the corresponding probability distribution $e^{-S(\tilde X_1,\tilde X_2, \Delta T)}$ along the coordinate $\Delta T$.  If $\Delta T$ is small, we are close to the vacuum and the probability distribution for the field configurations can be written in Gaussian approximation as
\begin{eqnarray}
\label{eq:P_vacuum}
P^{\mathcal{H}}_{vac}(\Delta T)=e^{-S_{vac}-\frac{1}{2}\mathcal{H}^{(0)}_{ij}\phi(\Delta T)_i \phi(\Delta T)_j},
\end{eqnarray}
where $\phi(\Delta T)_i$ is the field configuration for the instanton-anti-instanton pair at the same spatial lattice site $x$ separated in Euclidean time by $\Delta T$. For convenience, one can transform the lattice coordinates $(\ve{x}, \tau)$ into one-dimensional indices $i$ and $j$ via Eq. (\ref{eq:conversion_index}). However, in reality the probability distribution does not sharply vary with $\Delta T$, since the action stabilizes around $S^{(2)}$. Thus, the distinction between this ``real" probability distribution $e^{-S(\tilde X_1,\tilde X_2, \Delta T)}$ and $P^{\mathcal{H}}_{vac}$ is exactly the input needed from the instanton-anti-instanton saddle in the case of equal spatial coordinates. As a result, in the case of $X_1=\tilde X_1$ and $X_2=\tilde X_2$ (see Eq. (\ref{eq:instanton_anti_instanton_collapse}))  we replace $S( X_1,X_2, \Delta T)$ with $S_{eff}(\tilde X_1,\tilde X_2,\Delta T)$ in Eq. (\ref{eq:U_eff_2inst}), where $S_{eff}(\tilde X_1,\tilde X_2,\Delta T)$ is defined by the relation
\begin{eqnarray}
\label{eq:S_eff_definition}
e^{-S_{eff.}}=e^{-S(\tilde X_1,\tilde X_2,\Delta T)} - P^{\mathcal{H}}_{vac}(\Delta T).
\end{eqnarray}
It is convenient to write this in the form 
\begin{eqnarray}
\label{eq:S_eff_inst_anti_inst}
S_{eff}(\tilde X_1,\tilde X_2, \Delta T)=S(\tilde X_1,\tilde X_2, \Delta T) + \Delta_{eff.}(\Delta T),
\end{eqnarray}
where we have introduced the following quantity
\begin{eqnarray}
\label{eq:S_eff_inst_anti_inst1}
\Delta_{eff.}(\Delta T)= \nonumber \\ - \ln \left( 1- e^{-S(\tilde X_1,\tilde X_2, \Delta T) + S_{vac}+ \frac{1}{2}\mathcal{H}_{ij}\phi(\Delta T)_i \phi(\Delta T)_j}   \right).
\end{eqnarray}
Results for this correction term are shown in the corresponding plot in Fig.~\ref{fig:interactionTime2}, where the instanton and the anti-instanton reside at the same site of the hexagonal lattice (illustrated in the inset). As one would expect, its role rapidly decreases with increasing $\Delta T$, since $S_{eff}(\tilde X_1,\tilde X_2, \Delta T)$ is almost indistinguishable from $S(\tilde X_1,\tilde X_2, \Delta T)$ in this limit. However, in the limit of small $\Delta T$, the correction is extremely important. It forms a sharp repulsive barrier, which prevents the instanton and anti-instanton from annihilation, thus preventing the double counting of the vacuum saddle in the saddle point decomposition. 

In closing, we summarize the results of this section. The saddle points for the Hubbard model on the hexagonal lattice in the charge-density channel consist of individual localized field configurations: instantons and anti-instantons. These semi-classical objects form a weakly-interacting gas in 3D (also taking into account the localization in Euclidean time).  The only noticeable instanton-instanton or instanton-anti-instanton interaction is a strong repulsion when they occupy the same spatial lattice site and are closely separated in Euclidean time. This notion will be used in the next section for the construction of an analytical saddle point approximation which will be used to reproduce the physics of the full theory as elucidated by our QMC calculations.

\section{\label{sec:GasModel}Instanton gas model}
Using analytical insights from Appendix \ref{sec:AppendixB} and \ref{sec:AppendixC}, as well as the numerical data described in the previous section, we can now switch to the construction of a weakly-interacting instanton gas model. 
First, we derive the approximate analytic expression for the free energy only taking into account the hardcore repulsion in the instanton-instanton and instanton-anti-instanton pairs located at the same lattice site. As was mentioned before, the minimal approximation, which supports  the correct continuum limit $\Delta \tau=0$ is the one which includes the Gaussian fluctuations around the saddle points. We propose that  the partition function can be written as 
\beq
\mathcal{Z}=\mathcal{Z}_0 \left( 1 + \sum_{k=1}^\infty \frac{\mathcal{Z}_k}{\mathcal{Z}_0} \right),
\label{eq:PartFuncStart}
\eeq
where $\mathcal{Z}_0$ is the "vacuum" partition function, which corresponds to the Gaussian integral around the vacuum saddle point $\phi_{{\ve{x}},\tau}=0$ and $\mathcal{Z}_k$ corresponds to the Gaussian integral around the $k-$instanton saddle point. 

For the one-instanton saddle point, we take the ratio $\frac{\mathcal{Z}_1}{\mathcal{Z}_0}$ from Eq. (\ref{eq:Z1_Z0}). 
For the $k-$instanton saddle point, the weight within the Gaussian approximation can be computed using the results from Appendix \ref{sec:AppendixC}. Namely, we use Eq. (\ref{eq:general_hessian}) and neglect the variation of the first fundamental form $\sqrt{g}$ along the surface of the $k-$torus formed by the saddle point field configurations. One should also exclude the volume in Euclidean time and space, $\Delta \beta \mathcal{X}$, which is occupied by each instanton. Putting this all together, the final expression for the ratio  $\frac{\mathcal{Z}_k}{\mathcal{Z}_0}$ takes the following form
\begin{widetext}
\begin{eqnarray}
\frac{\mathcal{Z}_k}{\mathcal{Z}_0}=\frac{1}{k!} \left[ \prod_{m=1}^k (\beta V - (m-1) \Delta \beta \mathcal{X}) \right]  2^{2k} e^{- k \tilde S^{(1)}} \left( \frac{L^{(1)}}{\beta}\right)^k \left( \frac {\det \mathcal{H}^{(1)}_\perp}{\det \mathcal{H}^{(0)} } \right)^{-k/2}  \frac {1}{(2 \pi)^{k/2}},
\label{eq:ZRatioK}
\end{eqnarray}
\end{widetext}
where $V=N_S/2$ is the spatial volume of the lattice, and the sublattice index is taken into account in the $2^{2k}$ multiplier alongside the instanton-anti-instanton degeneracy. The multiplier $k!$ comes from the fact that the instantons are indistinguishable. 

Formally, the sum over instanton sectors in Eq. (\ref{eq:PartFuncStart}) runs to infinity. However, in practice, we should truncate it at $k_{max}= \lfloor \beta V /(\Delta \beta \mathcal{X}) \rfloor$, when all free ``slots" for instantons are taken. In other words, we stop at the point where the entire spacetime volume is packed full of instantons.  Under this assumption, the expression for the ratio $\frac{\mathcal{Z}}{\mathcal{Z}_0}$ can be summed exactly which yields the following expression:
\beq
\frac{\mathcal{Z}}{\mathcal{Z}_0}=1+\sum_{k=1}^{k_{max}} \frac{k_{max}!}{k! (k_{max}-k)!} \gamma^k = (1+\gamma)^{k_{max}},   
\label{eq:ZRatioFinal}
\eeq
where we have introduced the quantity
\beq
\gamma \equiv \frac{4}{\sqrt{2\pi}} e^{- \tilde S^{(1)}} \left( \frac{\mathcal{X} \Delta \beta}{\beta}\right) L^{(1)} \left( \frac {\det \mathcal{H}^{(1)}_\perp}{\det \mathcal{H}^{(0)} } \right)^{-1/2}.
\label{eq:ZRatioGamma}
\eeq
For practical calculations, we take $\mathcal{X}=1$ and $\Delta \beta$ equal to the width of the instanton's profile (Fig.~ \ref{fig:single_instanton}) at half height.  Both the instanton profiles as well as the product  $L^{(1)} \left( \frac {\det \mathcal{H}^{(1)}_\perp}{\det \mathcal{H}^{(0)} }\right)^{1/2}$ are taken from the exact one instanton saddle point.

Once we have the partition function for our gas of instantons, the free energy density can be computed as follows
\beq
f-f_0=-\frac{1}{\beta V} \ln \frac{\mathcal{Z}}{\mathcal{Z}_0},
\label{eq:FreeEnergyAnalyticStart}
\eeq
where 
\beq
f_0=-\frac{1}{\beta V} \ln \mathcal{Z}_0
\label{eq:FreeEnergyAnalytic0}
\eeq
is the contribution to the free energy from the Gaussian integral around the vacuum saddle point. Using our previous results we find
\beq
f=f_0-\frac{1}{\Delta \beta \mathcal{X}} \ln (1+\gamma).
\label{eq:FreeEnergyAnalyticFin}
\eeq
The physics described by the model has been reduced to the single parameter $\gamma$, which has temperature as well as coupling dependence. 

\begin{figure}[]
   \centering
   \subfigure[]{\label{fig:instantonsNumberQMCLat6}\includegraphics[width=0.3\textwidth,clip]{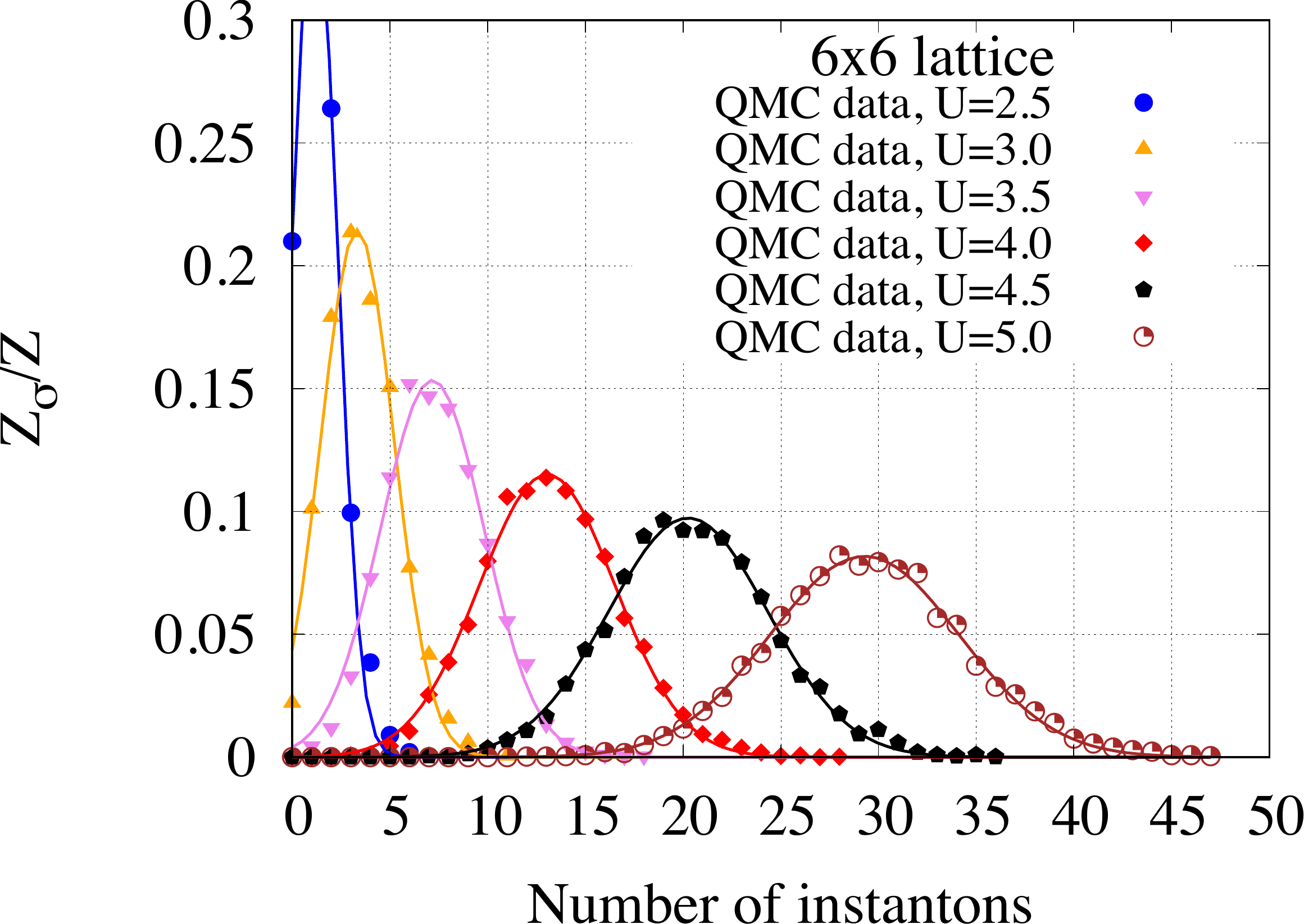}}
   \subfigure[]{\label{fig:einstantonsNumberQMCLat12}\includegraphics[width=0.3\textwidth,clip]{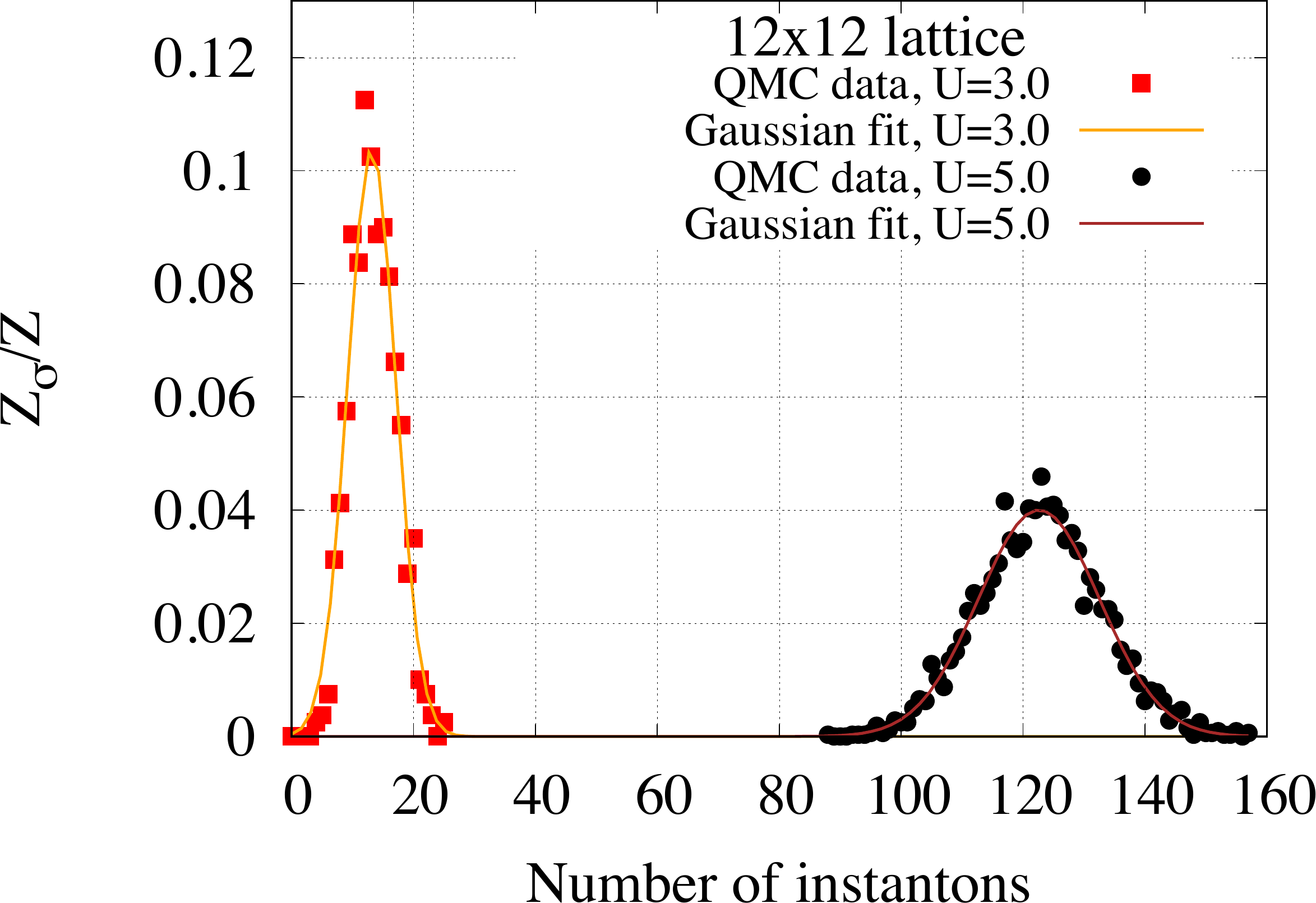}}
    \subfigure[]{\label{fig:instantonsNumberQMCComparison}\includegraphics[width=0.3\textwidth,clip]{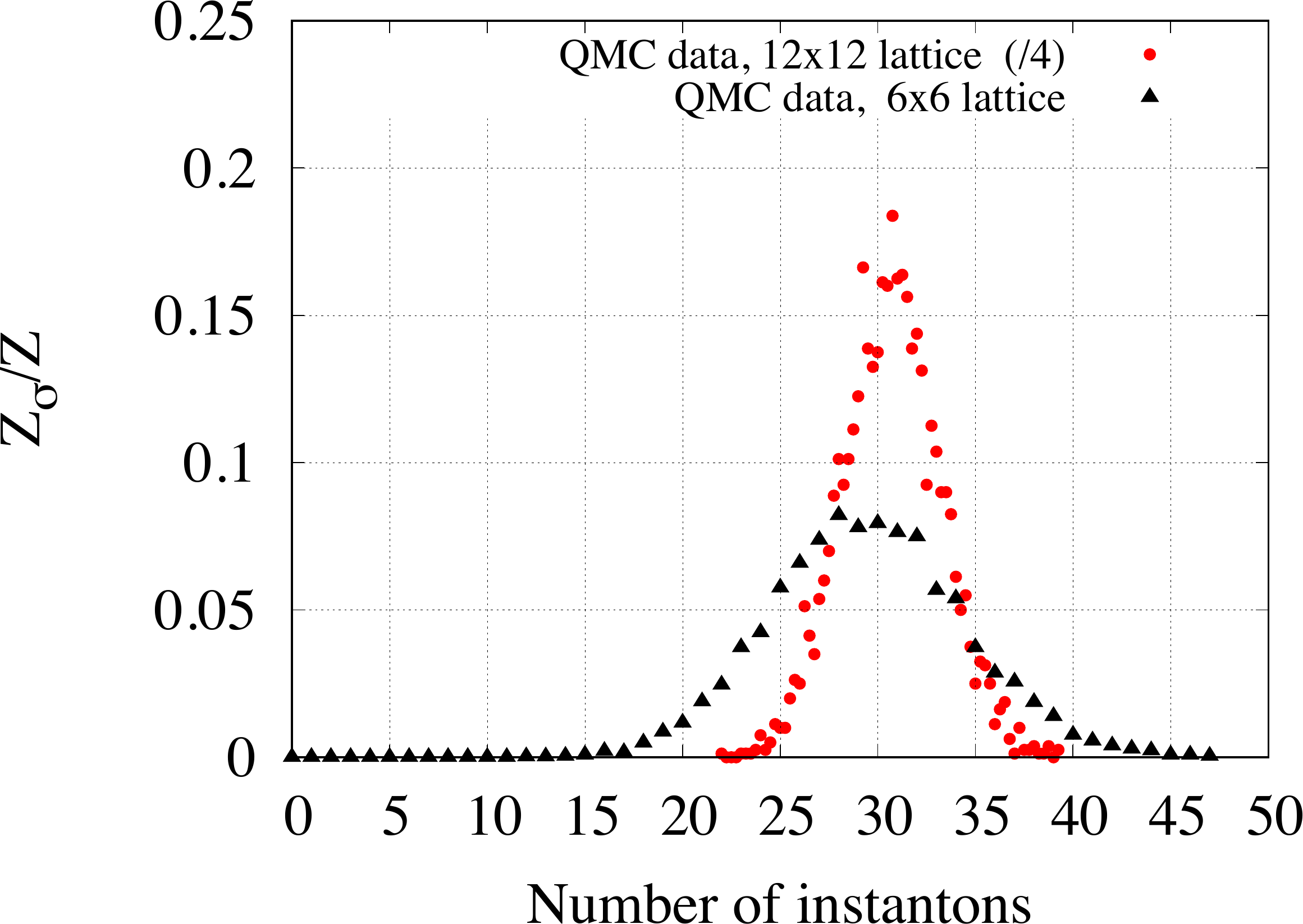}}
          \caption{The distribution of the number of instantons at various interaction strengths obtained from QMC configurations. The top two plots (a and b) show the distributions obtained on $6\times6$ and $12\times12$ lattices respectively and the bottom plot (c) compares these data in the case of large interaction strength $U=5.0 \kappa$. We note that the x-axis of the $12\times12$ lattice data is rescaled by a factor of four in order to have the curves lie on top of each other and to make the comparison more straightforward. All of these calculations were performed at $\beta \kappa=20$ and $N _{\tau}=512$. For each data set, a Gaussian fit of the lattice data has also been performed, it is shown with the lines of the same colors as corresponding data sets.}
   \label{fig:instantonsNumberQMC}
\end{figure}

Before further addressing the physics encoded in these expressions, we describe the predictions which can be made for the structure of the thimbles decomposition from the analytical partition function. We are particularly interested in the possibility to predict the structure of the dominant saddles which form the peaks in the distributions displayed in Fig. ~\ref{fig:histogramsU}. Since each saddle can be characterised by the number of instantons, we can replot the distributions from Fig. ~\ref{fig:histogramsU} in terms of the number of instantons.   This is done in  
Fig.~\ref{fig:instantonsNumberQMC} with an additional  fit of the data to  a  Gaussian form.  As one can see, the distributions can be quite precisely described by these curves whereby only two parameters (the mean $\mathcal{C}$ and variance $\mathcal{D}$) are needed to characterise them. Another important effect which we observe is that the relative width of the distribution goes down with the increased system size. As one can see from the last plot in  Fig.~ \ref{fig:instantonsNumberQMC}, the mean of the distribution scales as $\sim V$, but the width scales much more slowly. In fact, we will further show that the precise form of the scaling of the width with the volume is $\sim\sqrt{V}$. Thus, the distribution for the density of the instantons will   approach  a Dirac $\delta$-function in the thermodynamic limit. Therefore, in  this  limit,  it   suffices  to  consider  only saddles where  the number of instantons  matches  the mean value of the distribution.

The distribution of the number of instantons can be obtained from Eqs. (\ref{eq:PartFuncStart}-\ref{eq:ZRatioGamma}) as $\mathcal{Z}_k/\mathcal{Z}$ in the limit $V\rightarrow \infty$ after we apply Stirling's approximation to the factorials in Eq. (\ref{eq:ZRatioFinal}).  
The final expressions for the mean of the distribution  $\mathcal{C}$ and the variance $\mathcal{D}$ which follow from the instanton gas model are
\begin{eqnarray}
\mathcal{C}_f=k_{max} \frac{\gamma}{\gamma+1}, \label{eq:center}
\\
\mathcal{D}_f= \sqrt{\frac{1}{\gamma+1}}. \label{eq:dispersion}
\end{eqnarray}
From this one sees that the physics of the saddles in the charge-density-coupled channel can be accurately described by the single parameter $\gamma$ of our instanton gas model. As we will demonstrate further, despite its apparent simplicity, the model reproduces key features of the physics of the saddle decomposition of the Hubbard model on the hexagonal lattice.

Going further, it would be crucial to quantify the importance of interactions. Despite the relative smallness of interactions with respect to the action of a single semiclassical object, it is still worth checking whether they influence the characteristics of the final distribution of the instanton number. For this purpose, we construct a model of interacting instanton gas, where the pair-wise interaction profiles are taken directly from our QMC data as described in Section \ref{subsec:Interaction}. As the number of instantons is one of the thermodynamic variables of the system, it is necessary to work in the grand canonical ensemble. The microscopic state of the system is defined by the set of $N$ coordinates for the individual instantons: $\{X_i, T_i \}$, $i=1...N$.  By generalizing Eqs. (\ref{eq:L_det_H1}) and (\ref{eq:L_det_H2}) to the case of an $N$-instanton saddle and taking into account only pairwise interactions, we arrive at the following grand canonical partition function for the interacting instanton gas
\beq
 \zeta=\sum_N\frac{1}{N!} \sum_{\{X_i\}} \int \prod_{i=1}^N dT_i \nonumber \\ \times e^{-\frac{1}{2}\sum_{i \neq j}^{N} U^{(2)} (X_i, X_j, T_{i}-T_{j}) + N \ln \tilde \gamma},
 \label{eq:ClassicalModelInt}
\eeq
where 
\beq
\tilde \gamma \equiv \frac{1}{\sqrt{2\pi}} e^{- \tilde S^{(1)}} \frac{L^{(1)}}{\beta} \left( \frac {\det \mathcal{H}^{(1)}_\perp}{\det \mathcal{H}^{(0)} } \right)^{-1/2}.
\label{eq:ZRatioGammaTilde}
\eeq
As we did for the non-interacting instanton gas in (\ref{eq:ZRatioK}), we only consider the ratios ${\mathcal{Z}_k}/{\mathcal{Z}_0}$.
We recall that the pairwise interaction $U^{(2)} (X_i, X_j, T_{i}-T_{j})$ is defined in Eq. (\ref{eq:U_eff_2inst}). This model can be simulated using a classical grand canonical Monte Carlo (GCMC), where the number of instantons $N$ can be changed in the updates alongside with the coordinates $(X_i, T_i)$ of the instantons already present in the system. The details of these calculations are described in Appendix \ref{sec:AppendixD} while below we discuss the important results of these classical simulations of the interacting instanton gas model. 

\begin{figure}[]
   \centering
   \subfigure[] {\label{fig:DistributionComparisonCenter}\includegraphics[width=0.35\textwidth,clip]{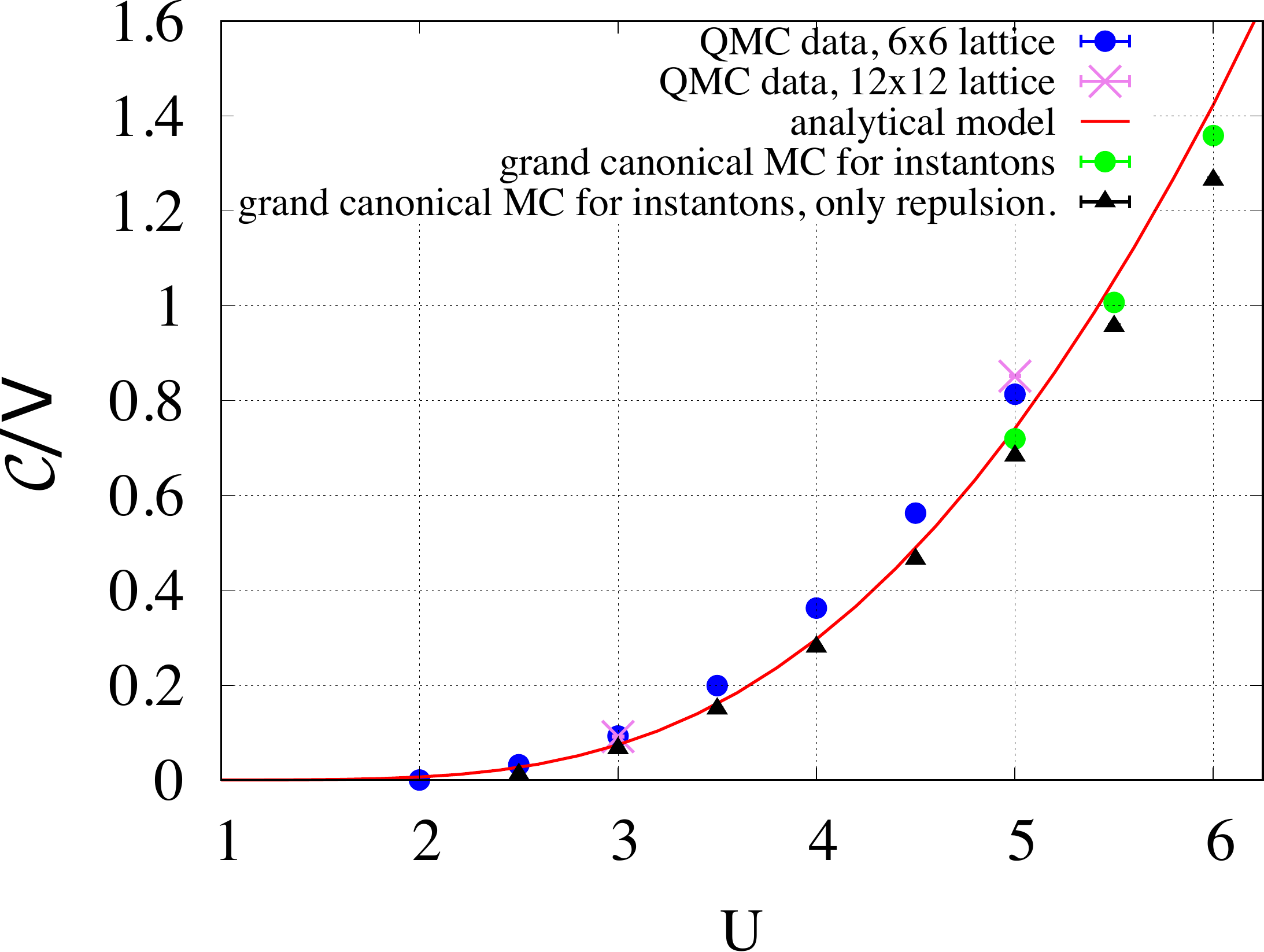}}
   \subfigure[] {\label{fig:DistributionComparisonWidth}\includegraphics[width=0.35\textwidth,clip]{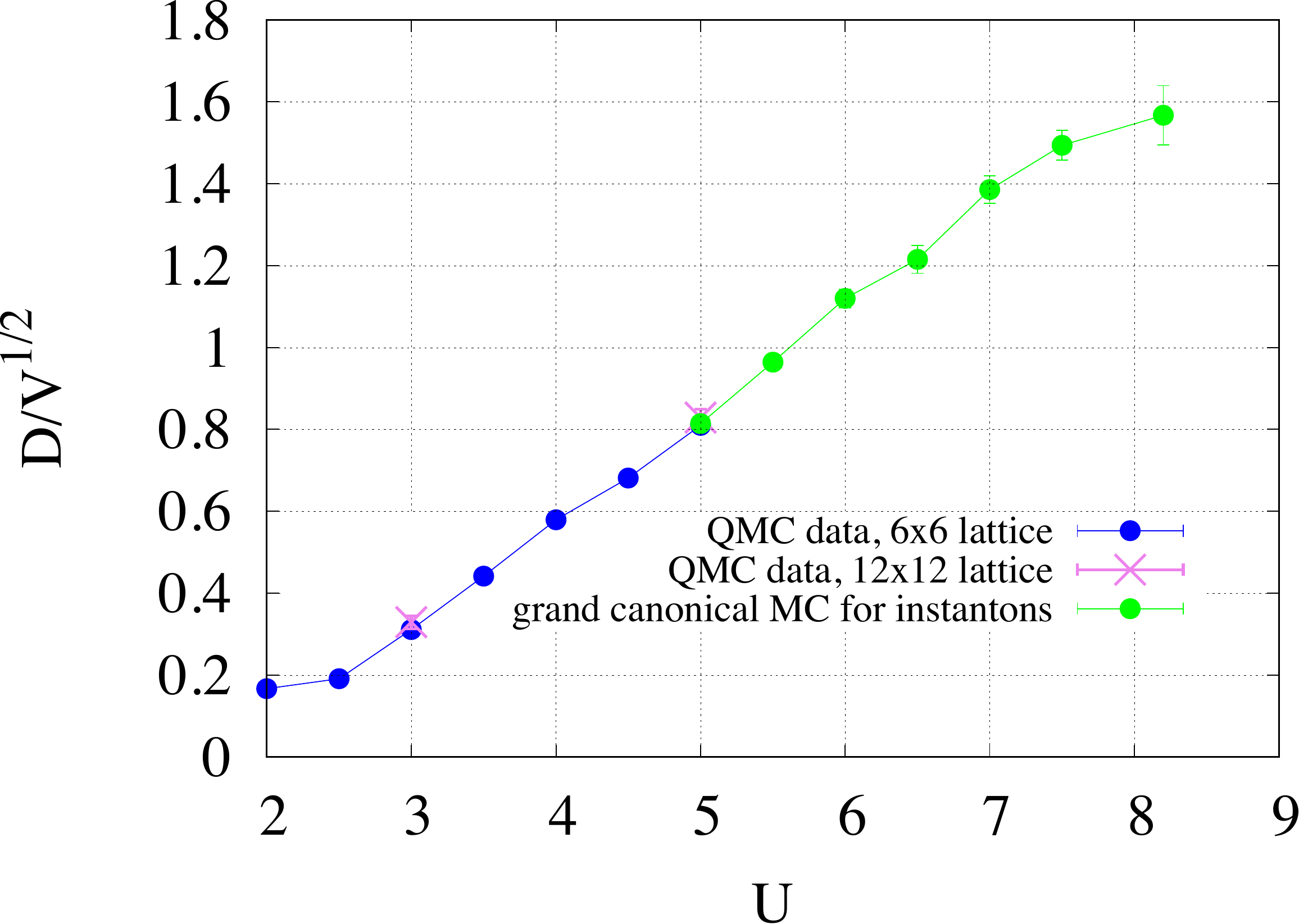}}
            \caption{(a)  A comparison of the instanton density obtained from real QMC data with various instanton gas models. (b) The same comparison but for the variance. The QMC data corresponds to $N _{\tau}=512$ and $\beta \kappa=20 $. } 
   \label{fig:DistributionComparison}
\end{figure}

A comparison of the predictions from the instanton gas model with the results from QMC for the structure of the thimbles decomposition is shown in
 Fig.~\ref{fig:DistributionComparison}. In order to compare the data for different lattice sizes in a uniform way, we plot the density of the instantons $\mathcal{C}/V$. In order to show that the variance scales as $\sqrt{V}$, we plot $\mathcal{D}/\sqrt{V}$ to demonstrate the collapse of the data obtained on different lattice volumes onto one curve. For the mean of the distribution, $\mathcal{C}$ (Fig.\ref{fig:instantonsNumberQMC}), we used the QMC data on $6\times6$ and $12\times12$ lattices to check that indeed, the data from the full theory scales linearly with the volume. As we can see, both the analytical model (\ref{eq:center}) and the classical GCMC simulations which include pairwise interactions (\ref{eq:ClassicalModelInt}) yield a prediction for the mean of the distribution which is consistent with the one obtained in our QMC calculations. Thus, one could in principle predict the dominant thimble for a given set of lattice parameters (including lattice size, temperature and interaction strength) even without doing actual QMC simulations, which are much more expensive.

Furthermore, the classical GCMC simulations of the instanton gas model (\ref{eq:ClassicalModelInt}) also provide an accurate prediction for the variance, as shown in   Fig.~\ref{fig:DistributionComparisonWidth}. In particular, we obtain exactly the same results as QMC on a $6\times6$ lattice. In addition to that, the QMC data for a $12\times12$ lattice which has been rescaled by a factor of $2$ exactly coincides with the data for the $6\times6$ lattice.  This implies  that unlike the mean $\mathcal{C}$, the variance $\mathcal{D}$ scales only as $\sim \sqrt{V}$. These two facts together show that, indeed, the distribution for the density of instantons, $\mathcal{C}/V$, tends to the $\delta-$function in thermodynamic limit.  As the prediction from the analytical model in Eq. (\ref{eq:dispersion}) was obtained exactly in the thermodynamic limit, $V\rightarrow\infty$, this model does not provide a good estimate for the variance on a finite lattice volume.

 \section{\label{sec:GasPhysics}Physics from the instanton gas approximation}
 
 In this section we will concentrate on further physical predictions of the instanton gas model. First, we will consider the possibility to describe the  semi-metal (SM) to AFM phase transition, which is one of the most prominent features of the Hubbard model on the hexagonal lattice. Second, we consider the evolution of the electron density of states away from the  Dirac point with increasing interaction strength.

  \begin{figure}[]
   \centering
   \subfigure[] {\label{fig:AnalyticalModelDF}\includegraphics[width=0.35\textwidth,clip]{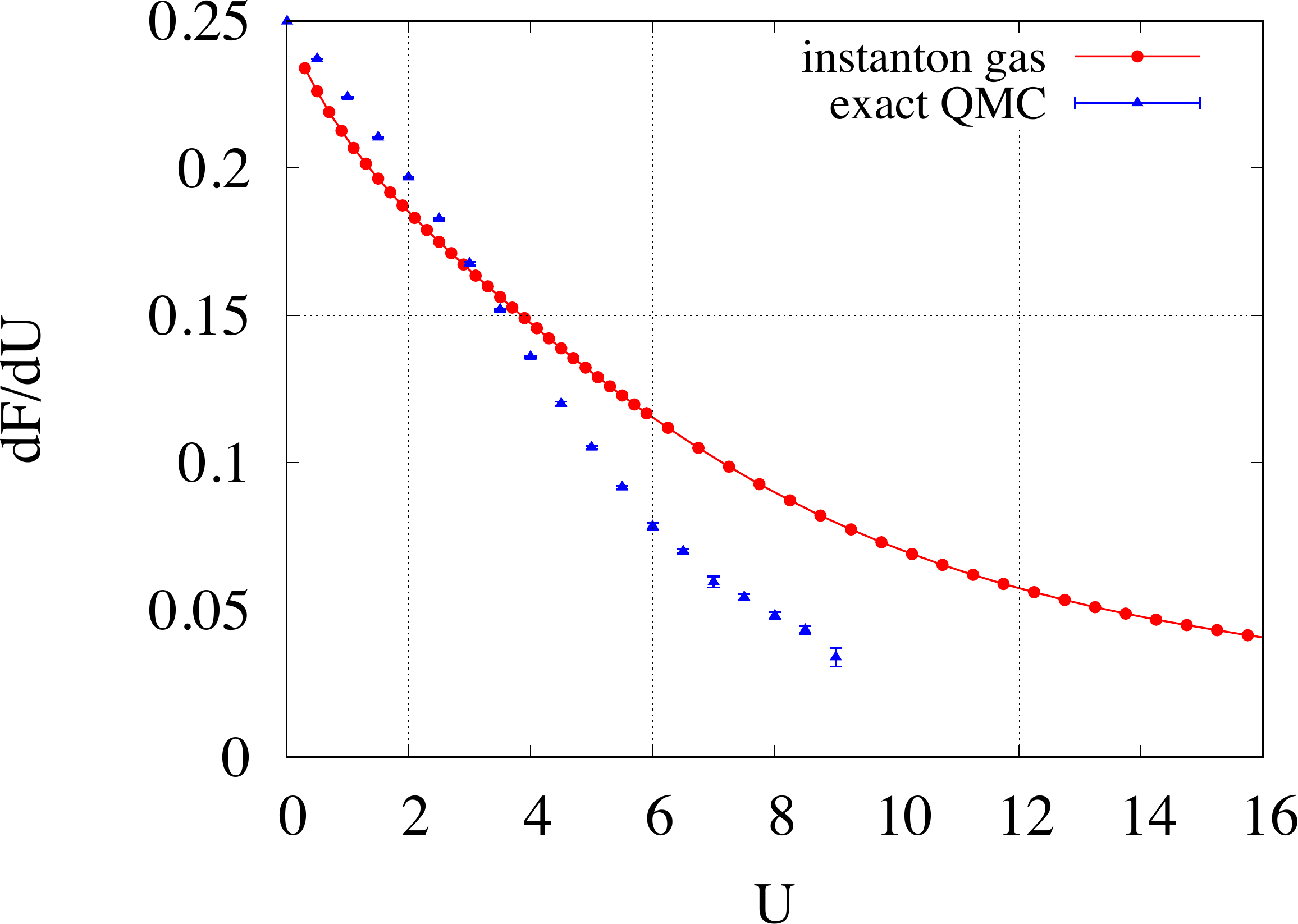}}
   \subfigure[]  {\label{fig:AnalyticalModelChi}\includegraphics[width=0.35\textwidth,clip]{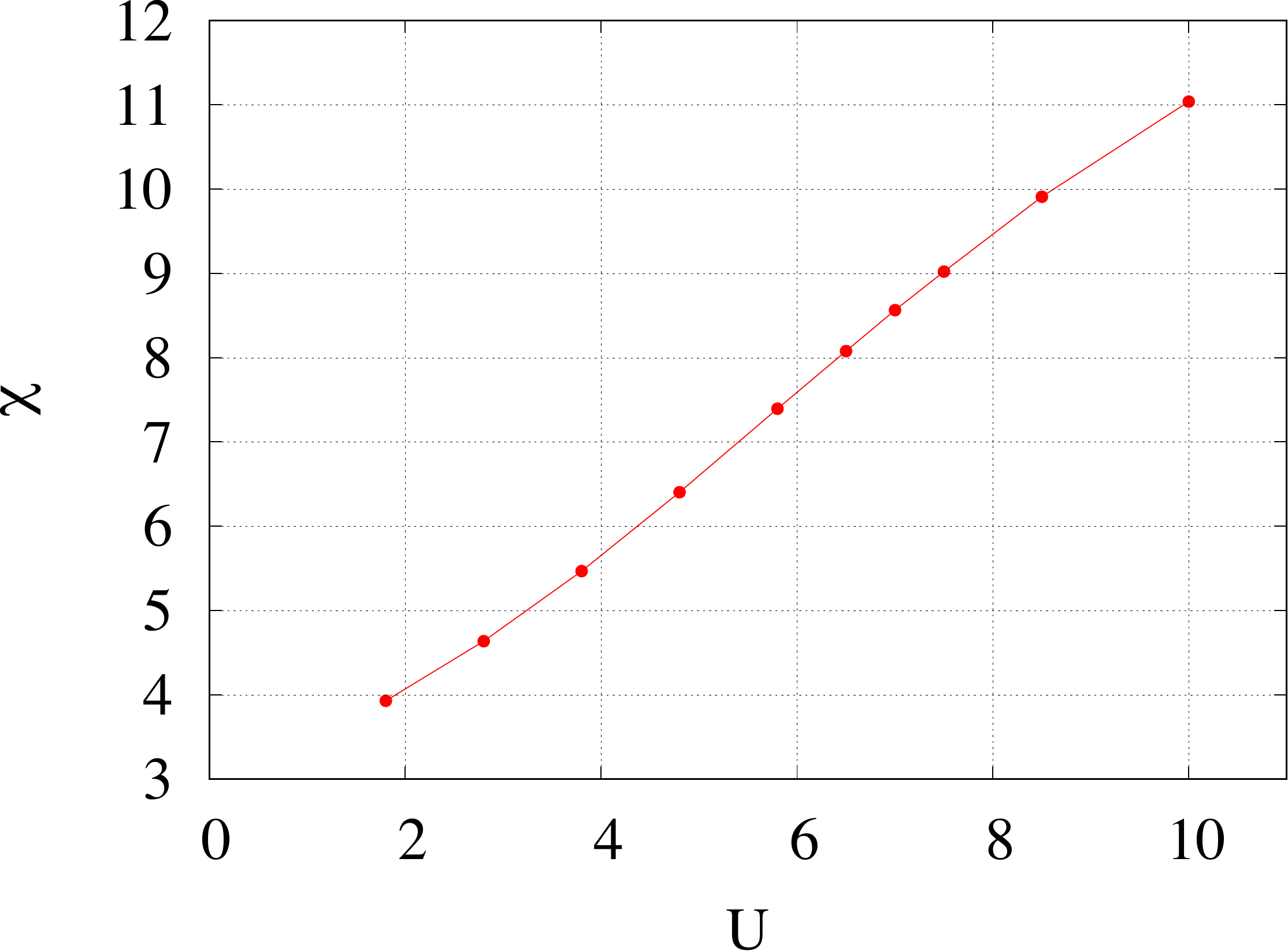}}
       \caption{(a) Double occupancy obtained from the instanton gas model and from QMC data. (b) AFM susceptibility obtained from the analytical model. In both cases, the instanton profiles and actions are obtained on the $6\times6$ lattice with $N _{\tau}=512$ and $\beta \kappa =20$ were used as input. QMC data were obtained on the same lattice.}
   \label{fig:AnalyticalModel}
\end{figure}

 \subsection{\label{subsec:GasPhysicsK} Local  magnetic  fluctuations  and  long  ranged  order}
 Starting from our simple expression for the free energy of the ensemble of instantons in (\ref{eq:FreeEnergyAnalyticFin}), one can obtain further thermodynamic quantities by taking appropriate derivatives. In particular, the derivative of   the  free  energy  density, $f$,  with respect to the Hubbard interaction gives us  the  double occupancy, which is defined as
\beq
\langle  \hat{n}_{\ve{x},\uparrow} \hat{n}_{\ve{x}, \downarrow}  \rangle =   \frac{\partial f}{\partial U}. 
\label{eq:double_occupancy}
\eeq
In practice, the derivative of the free energy over the interaction is computed by noting that $\gamma$ is a function of $U$ and taking the appropriate partials, $\partial_U \gamma$ and $\partial_U  f_0$. The latter quantity can be directly obtained in the Gaussian approximation for $S_{vac}(U)$, whereas $ \det \mathcal{H}^{(0)}(U)$ can be computed numerically for a fixed spatial lattice size. The profiles for $\Delta \beta(U)$, $\tilde S^{(1)}(U)$, $L^{(1)}(U)$ and $\det \mathcal{H}^{(1)}_\perp(U)$  are obtained from the exact one-instanton field configurations we have obtained from our GF procedure discussed previously.

  \begin{figure}[]
   \centering
   \subfigure[] {\label{fig:Spin1Instanton}\includegraphics[width=0.30\textwidth,clip]{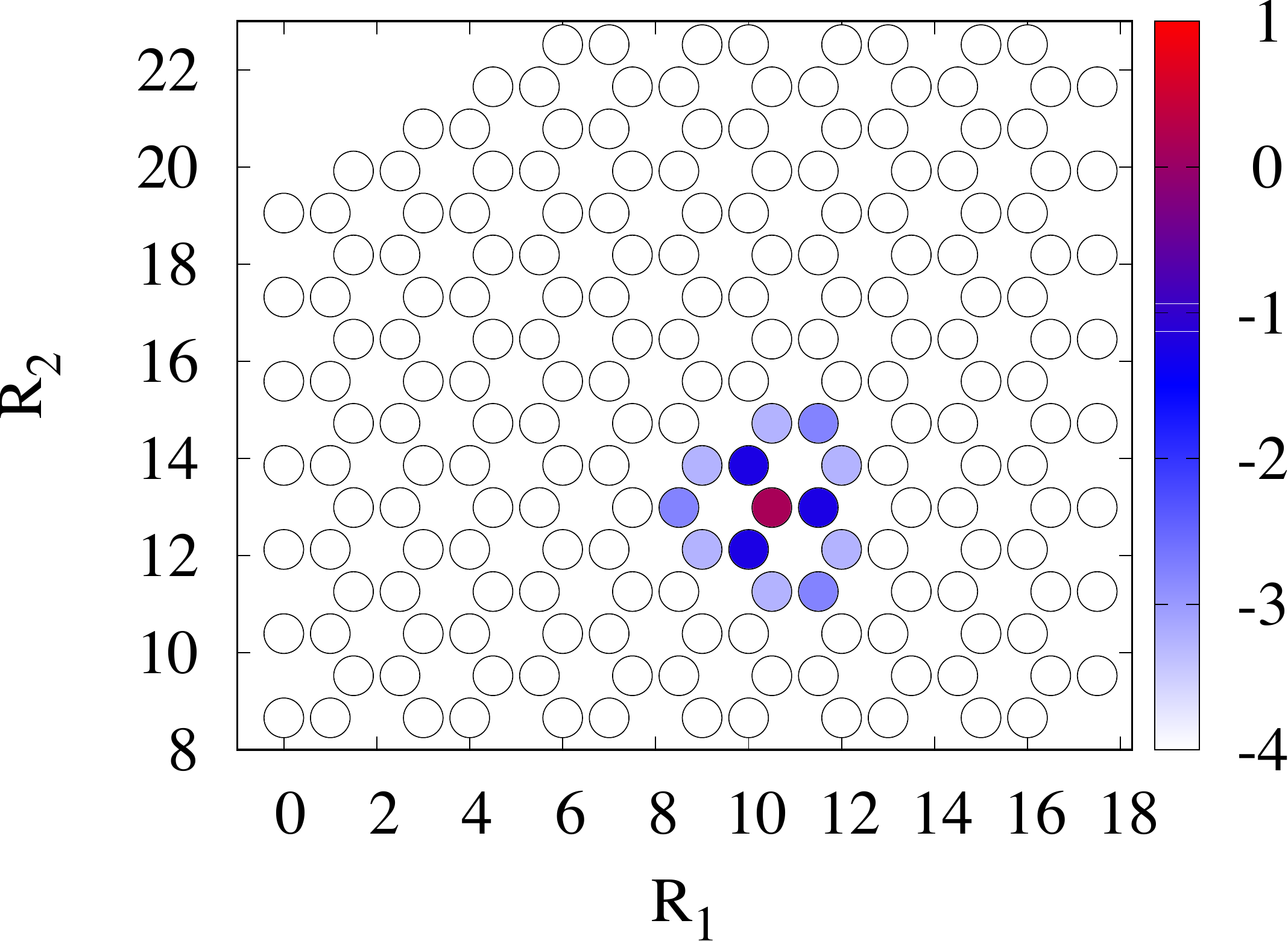}}
   \subfigure[]   {\label{fig:Spin2Instantons}\includegraphics[width=0.30\textwidth,clip]{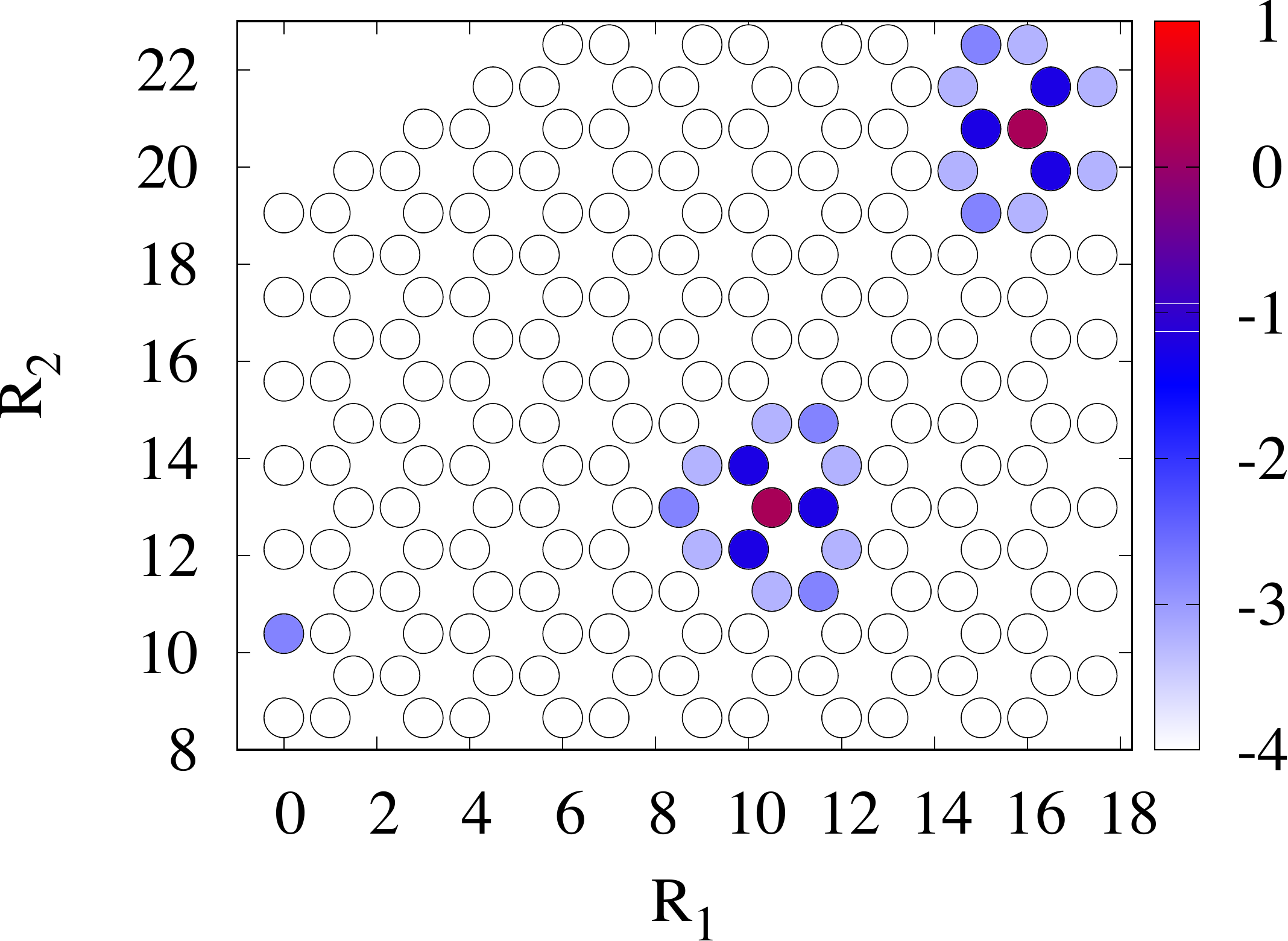}}
      \caption{The distribution of the squared spin across the hexagonal lattice for background field configuration consisting of a single instanton (a) or two instantons (b). What is plotted is the difference between the squared spin at a given lattice site $x$ in the presence of semiclassical objects and its value for the vacuum configuration: $\langle \hat S^2_x\rangle|_{N_{inst.}} - \langle \hat S^2_x\rangle|_{vac}$. A base-10 logarithmic scale is used for the color scale. Both instantons are located at the same Euclidean timeslice where the spin operator is measured. This calculation refers to a $12\times12$ lattice with $\beta \kappa=20$ and $N _{\tau}=512$, with interaction strength $U=2.0 \kappa$. $R_1$ and $R_2$ are Cartesian coordinates of the lattice sites, displayed in the units of the distance between nearest neighbours.}
   \label{fig:SpinInstantons}
\end{figure}

  \begin{figure}[]
   \centering
   \subfigure[] {\label{fig:SpinChargeCorrelations}\includegraphics[width=0.30\textwidth,clip]{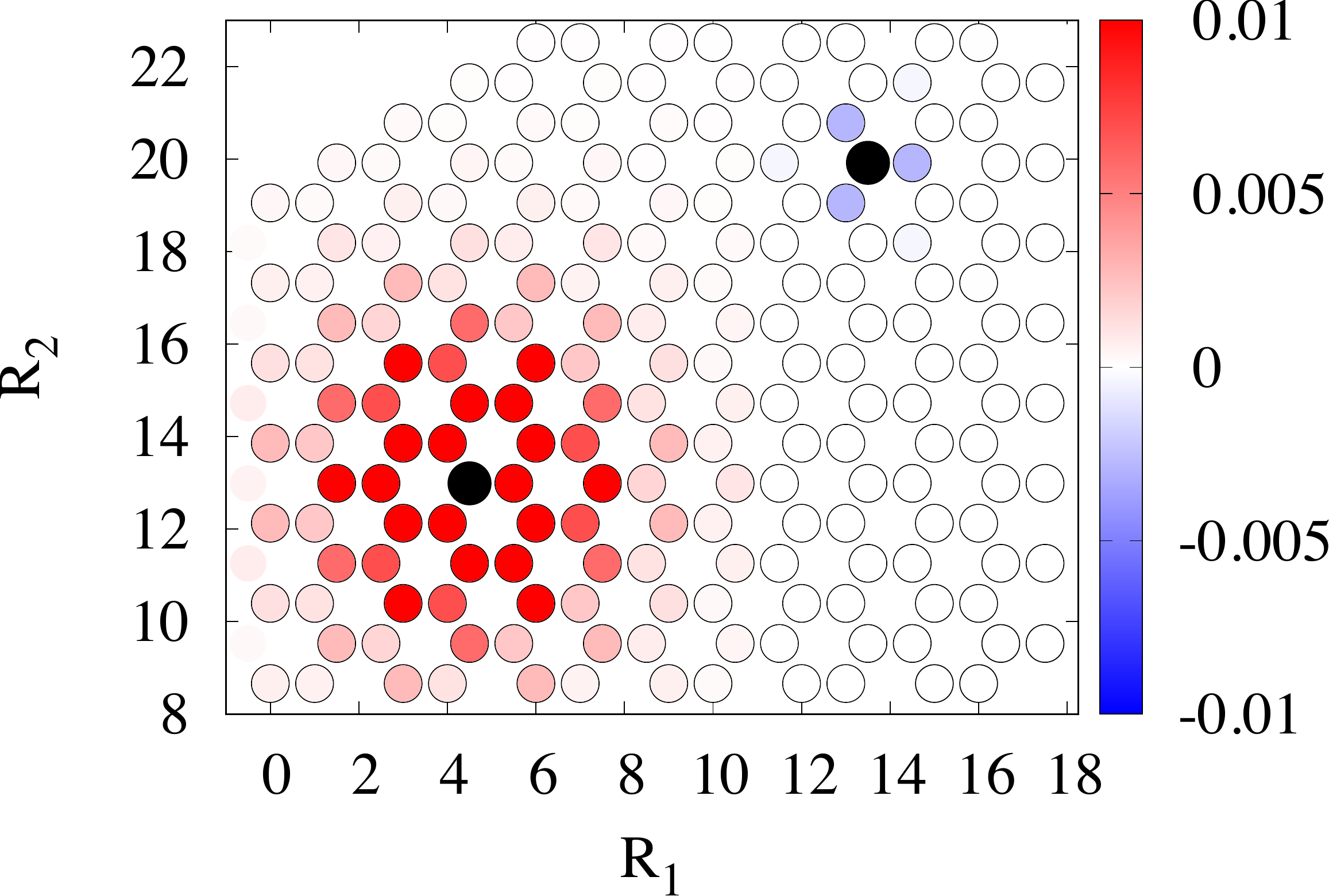}}
   \subfigure[]   {\label{fig:InstantonVacuumComparison}\includegraphics[width=0.30\textwidth,clip]{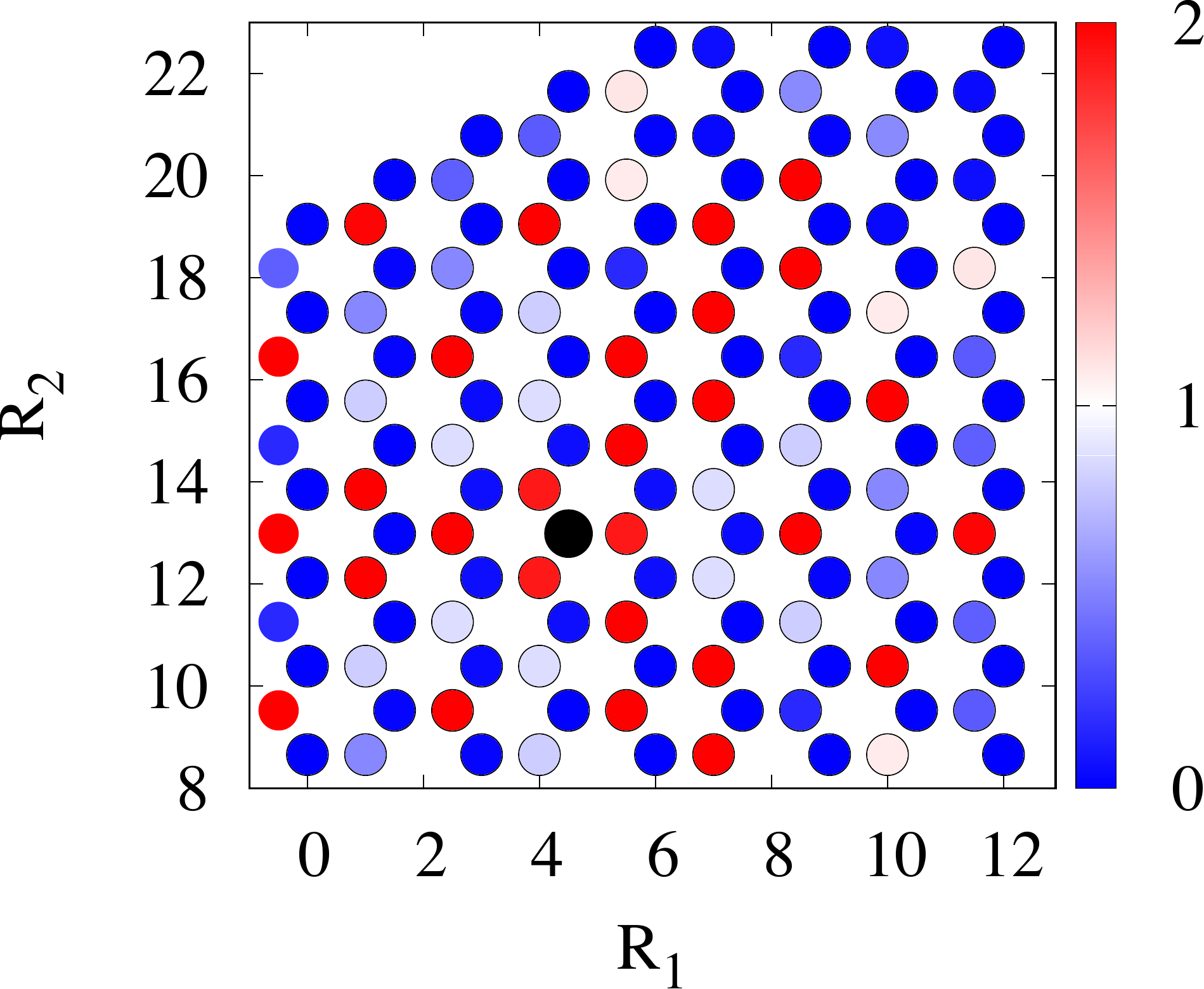}}
      \caption{(a) Charge-charge correlations, $\frac{1}{4}\langle \hat{{q}}_{\ve{x}_0}(T) \cdot \hat{{q}}_{\ve{x}_0  +  \ve{x}} (T) \rangle$,  for a field  configuration  with one instanton at space time   $(X,T)$.   We  consider two  values of   $\ve{x}_0$.   The left  black  circle  corresponds  to  $\ve{x}_0 = X$. The other value of  $\ve{x}_0$  (right black  circle)  is  \text{far}   from the  instanton.  $R_1$ and $R_2$ are two Cartesian coordinates of the lattice sites, displayed in the units of the distance between nearest neighbours. The corresponding plot for the spin-spin fluctuations can be found in the Fig.   \ref{fig:SpinCorrInstantonVacuumMap}. (b) Ratio of spin-spin and charge-charge correlators on the same one-instanton saddle $4 \langle \hat S^{(3)}_{\ve{x}_0}(T)  \hat S^{(3)}_{\ve{x}_0+\ve{x}} (T) \rangle / \langle \hat q_{\ve{x}_0}(T)  \hat q_{\ve{x}_0+\ve{x}} (T) \rangle$ centered at the location of the instanton $\ve{x}_0=X$. The calculation was done on the same lattice as for the Fig.~\ref{fig:Spin1Instanton}.}
   \label{fig:InstantonCorrelations}
\end{figure}

The   double occupancy as a function of interaction strength is plotted in Fig.~\ref{fig:AnalyticalModelDF}. As one can see, we can successfully describe the increasing localization of electrons with increasing interaction strength. 
To  understand  how  the reduction of the   double  occupancy  comes  about,  we   consider  the local  moment  
\begin{equation}
	\langle \hat S_{\ve{x}}^2 \rangle   -  \langle \hat S_{\ve{x}}^2 \rangle_{U=0}   \equiv  \frac{1}{4}  \left[ \frac{1}{2} - 2 \langle \hat{n}_{\ve{x},\uparrow}  
	\hat{n}_{\ve{x},\downarrow} \rangle     \right] 
\end{equation}
 at the saddle point field configurations (Fig. ~\ref{fig:SpinInstantons}).  This  quantity  is computed using the fermionic propagator calculated at the saddle point configuration of the auxiliary field. The distribution is shown both for the one- and two-instantons saddle points, where the centers of the instantons are  located at the same time slice.  As one can see, each instanton generates a localised region of excess spin,   or  reduced  double occupancy, around the instanton center.

Fig.~\ref{fig:AnalyticalModelDF}  equally plots  the  double  occupancy   obtained    with  the  Algorithm for Lattice Fermions \cite{ALF_v2}  implementation of the  finite-temperature auxiliary  field  QMC  \cite{Blankenbecler81}.    As in the instanton  approach,    local moment formation  leads  to  a  decrease of double  occupancy.  Owing to  Eq.~(\ref{eq:double_occupancy}),  we  expect double  occupancy  to  show  non-analytical  behaviour at the  Gross-Neveu  transition  located  at  $U_c/\kappa \simeq  3.8$.   The QMC  data  hints to  non-analytical behaviour, whereas the instanton gas solution  exhibits  a very  smooth  curve.  We will see below that this stems  from the fact that the instanton approach  does not capture  the onset of the magnetic ordering and  the resulting  mass generation.       We again note  that the  reduction of  double occupancy is a  dynamical  effect  that  cannot be obtained  at  the  mean-field  level   without  breaking     time-reversal   symmetry  \cite{Anderson61,Raczkowski2019}.

 \begin{figure*}
        \centering
        \includegraphics[scale=0.9, angle=0]{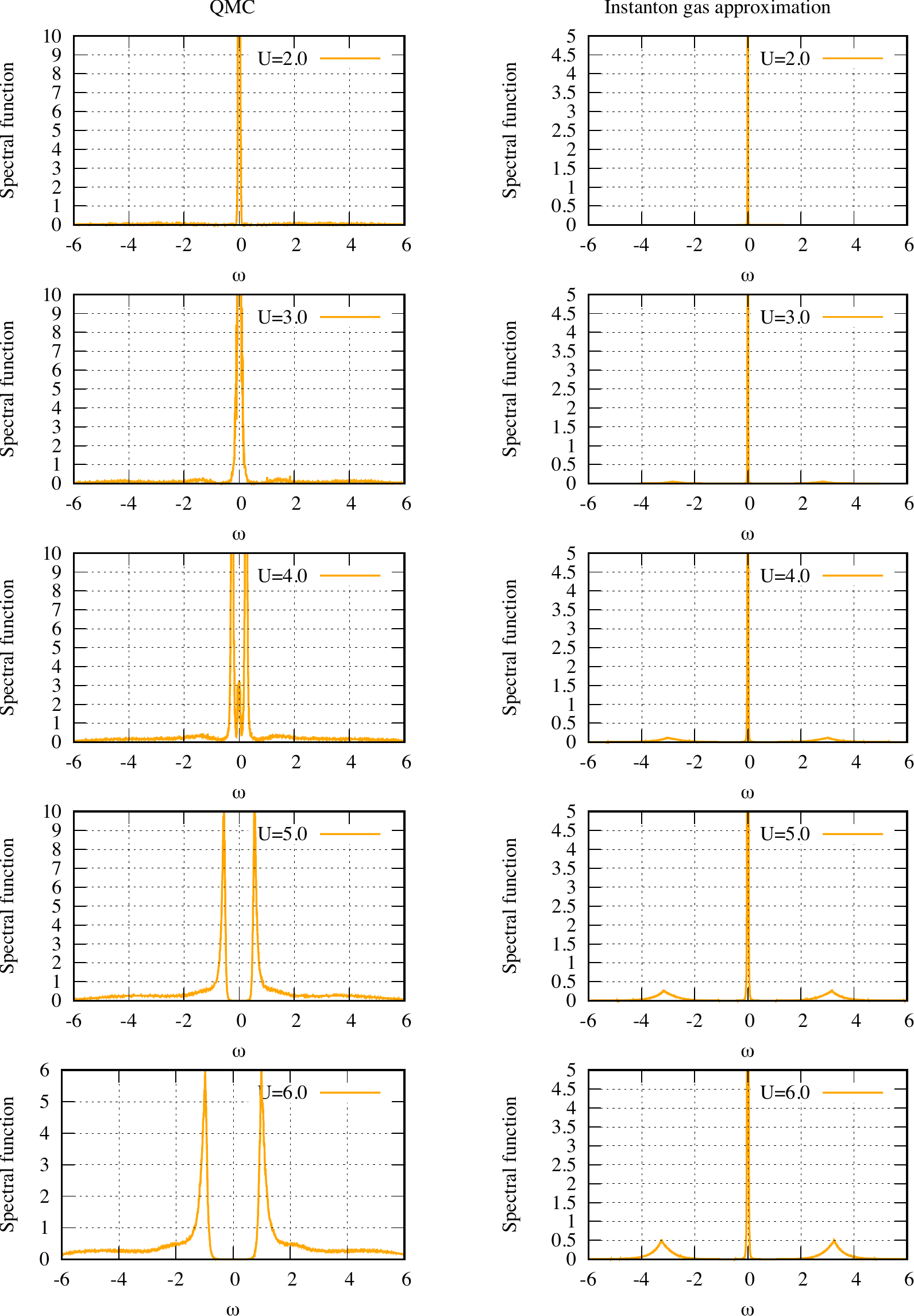}
        \caption{A comparison of the density of states at the Dirac point obtained from QMC data (left column) with that obtained from the instanton gas model (right column). All calculations correspond to a spatial volume of $12\times12$ with $\beta \kappa=20 $ and $N _{\tau}=256$, while the corresponding interaction strength in units of $\kappa$ is displayed on each plot. The analytical continuation has been performed with stochastic MEM.}
        \label{fig:SpectralFunctionsK}
\end{figure*}

Fig. \ref{fig:SpinChargeCorrelations} shows the charge-charge ($\frac{1}{4}\langle \hat q_{\ve{x}}  \hat q_{\ve{y}} \rangle$) correlations at  the one instanton saddle, in the vicinity of the instanton and for reference away from it, where the correlator coincides with its vacuum values.  The plot should be compared with the corresponding spin-spin correlator $\langle \hat S^{(3)}_{\ve{x}} \hat S^{(3)}_{\ve{y}} \rangle $  displayed in Fig. \ref{fig:SpinCorrInstantonVacuumMap}.  The spin and charge correlators are equal in the  vacuum. Both are zero if $\ve{x}$ and $\ve{y}$ are on the same sublattice and negative if $\ve{x}$ and $\ve{y}$ are at different sublattices. Around the instanton, both correlations are substantially enhanced, spin correlations remain anti-ferromagnetic with the largest correlations still between points on different sublattices. The charge correlators change sign. The ratio $4 \langle \hat S^{(3)}_{\ve{x}} \hat S^{(3)}_{\ve{y}} \rangle / \langle \hat q_{\ve{x}}  \hat q_{\ve{y}} \rangle$ plotted in the figure  \ref{fig:InstantonVacuumComparison} shows that the spin-spin correlations dominate over charge-charge correlations in the vicinity of the instanton. 
In particular,   at  lattice  sites  $\ve{x}$ and $\ve{y}$   that belong to different  sublattices and   where both spin-spin and charge-charge correlators acquire their  largest values,  spin  correlations   dominate.   In summary, these figures demonstrate that the increased spin localization at the instanton core is surrounded by local AFM correlations.   This  key  result  is  also shown in Fig.~\ref{fig:SpinCorrInstantonVacuumMap}.

We next investigate   long-range spin order    characteristic of the AFM phase. The most obvious quantity to check is the spin susceptibility. It can be computed via the second derivative of the free energy with respect to an external, alternating magnetic field
\beq
\chi = \left. \frac{\partial^2 f}{\partial m^2}\right|_{m\rightarrow0}, \eeq
where we introduced the explicit  \textit{ staggered }  mass term in the Hamiltonian (\ref{eq:Hamiltonian}):
\begin{eqnarray}
\hat{{H}}_m=m \left( \sum_{x \in \text{1st sublat. }} (\hat n_{x, \text{el.}} + \hat n_{x, \text{h.}})  \right. \nonumber \\ -  \left.  \sum_{x \in \text{2nd sublat. }} (\hat n_{x, \text{el.}} + \hat n_{x, \text{h.}}) \right) .
\label{eq:susceptibility}
\end{eqnarray}
This mass term contains the symmetry of the AFM ordering on the hexagonal lattice rewritten in terms of electrons and holes.   Similarly  to  the  calculation of the  double occupancy (\ref{eq:double_occupancy}), we compute the susceptibility using numerical derivatives of  $S_{vac}(U,m)$,  $ \det \mathcal{H}^{(0)}(U,m)$, $\det \mathcal{H}^{(1)}_\perp(U,m)$, etc with respect to the mass parameter $m$. These derivatives can be obtained from the corresponding one-instanton profiles at finite $m$ after the application of the GF equations to the configurations generated in QMC simulations, in exactly the same manner as was done for $m=0$. This is possible due to the fact that the small staggered mass term  does  not  alter the structure  of the  saddle points in the charge-coupled representation of the functional integral. 
The dependence of the susceptibility on the interaction strength is shown in Fig.~\ref{fig:AnalyticalModelChi}. As apparent,  the susceptibility does 
not form a peak, signaling a phase transition but rather  increases monotonically.  
Hence,  we  conclude  that the instanton gas approach  captures local moment  formation,  but not  the onset of long-ranged anti-ferromagnetic ordering. 

Additional argument in favor of the instanton as a path integral representation of a localized spin is presented in Appendix \ref{sec:AppendixE}, where we describe the technique which allows us to connect certain paths in the path integral to the properties of the ground state wave function. Using this technique, we show that the ground state corresponding to the instanton is local Gutzwiller projection which suppresses the double occupancy at the site occupied by the instanton.

\subsection{\label{subsec:GasPhysicsK}Single  particle spectral function}
  \begin{figure}[]
   \centering
   \subfigure[]{\label{fig:U6ModelCorr}\includegraphics[width=0.35\textwidth,clip]{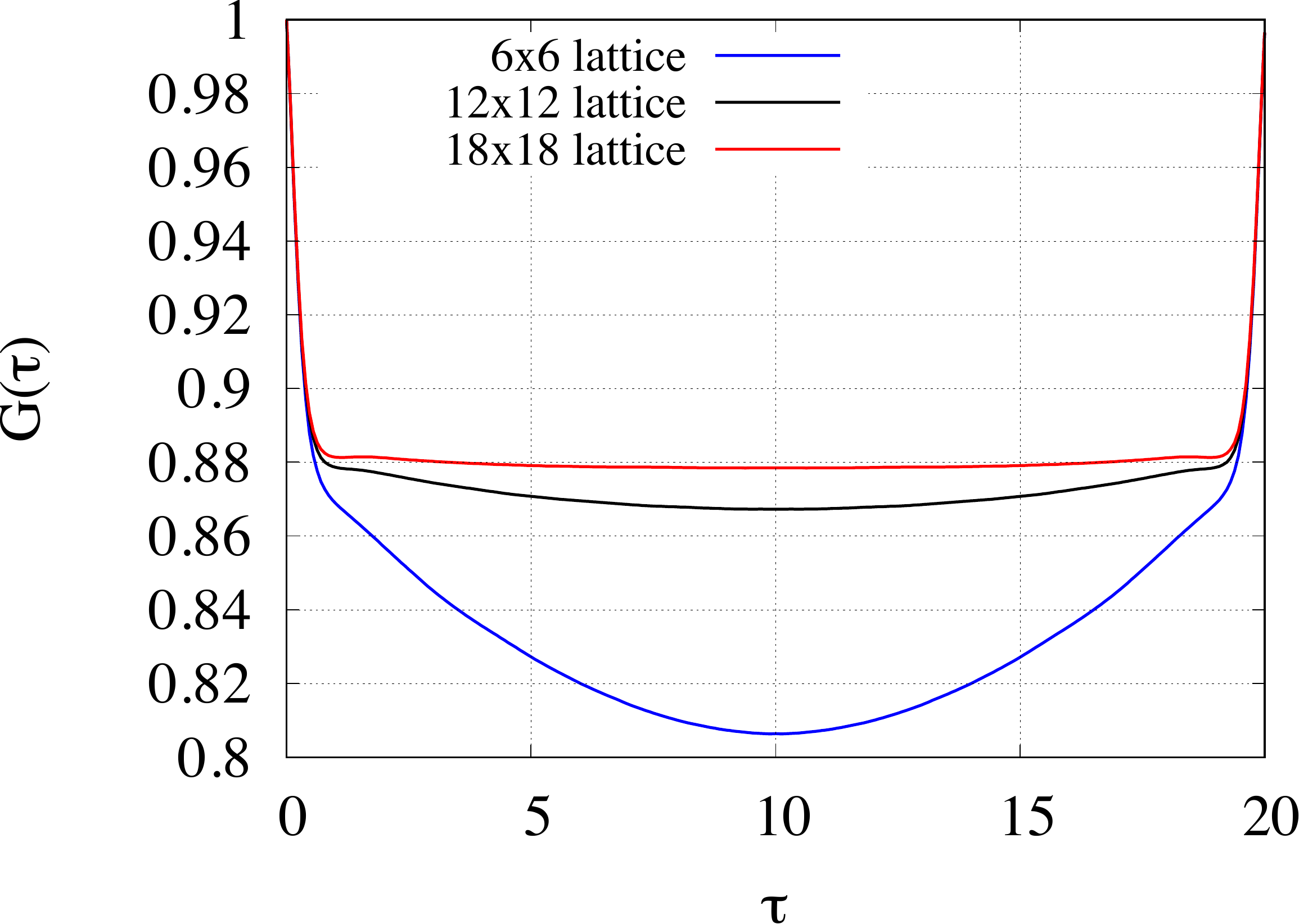}}
   \subfigure[]   {\label{fig:U6ModelSpectral}\includegraphics[width=0.35\textwidth,clip]{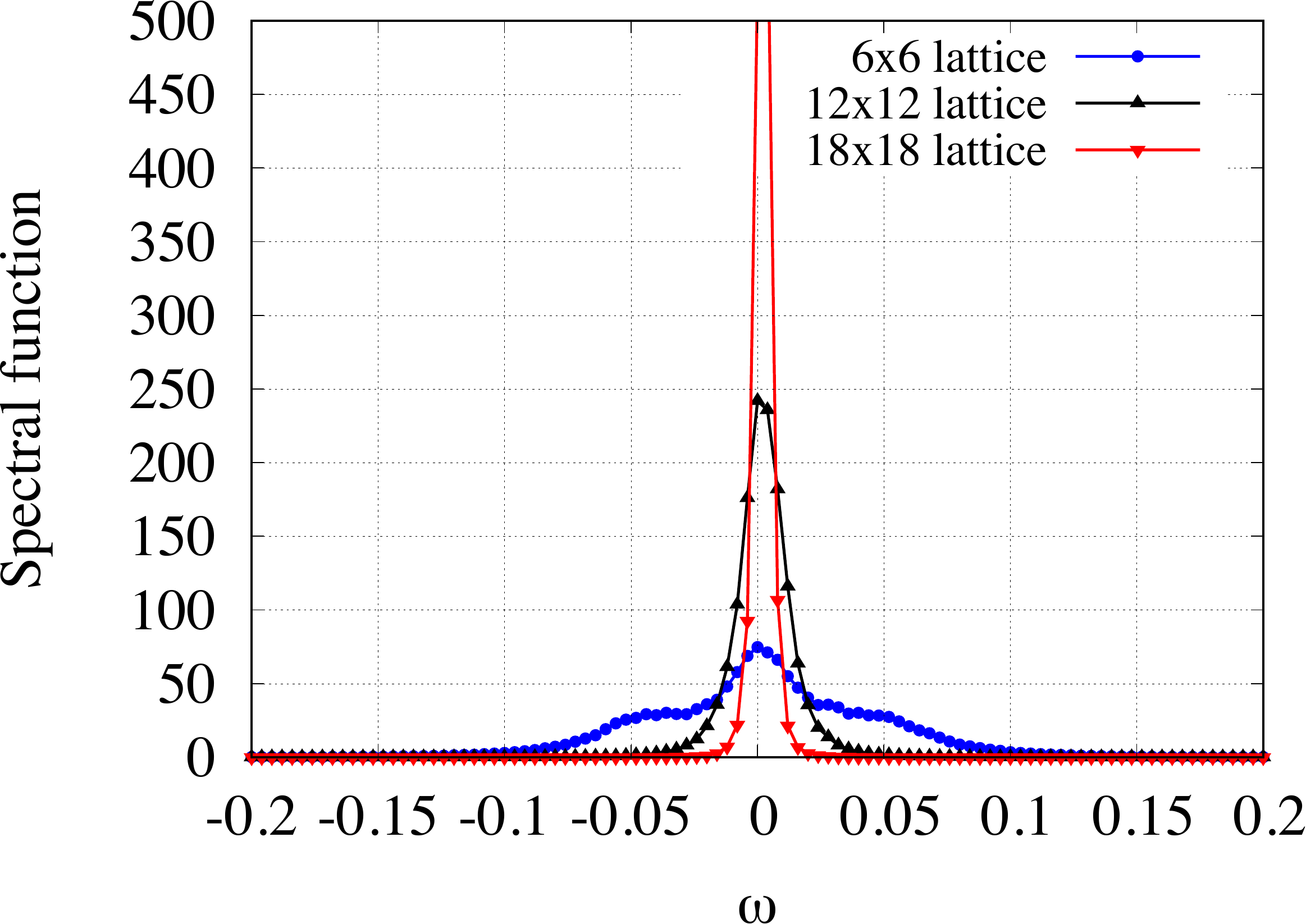}}
       \caption{(a) The Euclidean-time fermion propagator at the Dirac point in momentum space, obtained from the instanton gas model with hardcore repulsion. Spatial volumes of $6\times6$, $12\times12$ and $18\times18$ are compared at $U=6.0 \kappa$, with $N _{\tau}=256$ in each case. (b) The electron density of states obtained after analytical continuation performed with stochastic MEM.}
   \label{fig:U6Model}
\end{figure} 

The    spectral function of  a  single hole in an  anti-ferromagnetic    defines a  rich  problem that has    been    extensively   investigated   with  various  methods.      One   expects   the low  energy region  to be  captured  by  a spin-polaron   and  a  broad  high energy  spectral  weight  that is  referred  to as  the upper  Hubbard  band \cite{Martinez91,Preuss95,Brunner00b,Raczkowski2019,Ponsioen22}.     Clearly the   instanton approach will  not  capture   spin-polaron since  there  are no spin waves -- Goldstone  modes of  the broken global  symmetry --    that will  dress the  doped  hole.    However,  local   magnetic  fluctuations      and   the  reduction of the double 
occupancy  have the potential to   account  for  the upper Hubbard  band.    It  is  very appealing  to   adopt   a  parton  construction  to account  for the  rich structure of   the  single-particle spectral function \cite{Beran96,Grusdt18}.   As an  example,  one can   consider   the orthogonal    fermion  representation of the Hubbard model  \cite{Nandkishore12,Hohenadler18}.     Consider,   
   \begin{equation}
   	  \hat{H}   = -t  \sum_{\left<\ve{i},\ve{j} \right>}    \hat{f}^{\dagger}_{\ve{i},\sigma}  \hat{f}^{\phantom\dagger}_{\ve{j},\sigma} \hat{s}^{z}_{\ve{i}} \hat{s}^{z}_{\ve{j}}     +  \frac{U}{4} \sum_{\ve{i}} \hat{s}^{x}_{\ve{i}} 
\label{eq:orthogonal}
   \end{equation}
 Where  $ \hat{f}^{\dagger}_{\ve{i},\sigma} $  is  a  fermion operator  and $\hat{s}^{x,z}_{\ve{i}}$,    Pauli spin  matrices acting on an Ising   degree of  freedom  per  site.    The  above  defines a  $\mathbb{Z}_2$  lattice  gauge theory  since 
\begin{equation}
\hat{Q}_{\ve{i}}     =   (-1)^{ \hat{n}_{\ve{i}} }   \hat{s}^{x}_{\ve{i}}
 \end{equation}
 is  a  local  conservation law.   Here $ \hat{n}_{\ve{i}} = \sum_{\sigma}   \hat{f}^{\dagger}_{\ve{i},\sigma}  \hat{f}^{\phantom\dagger}_{\ve{i},\sigma} $   and  
 one  will  readily see  that   $\hat{Q}_{\ve{i}}^2 =1$.   Imposing the  Gauss law, 
\begin{equation}
	\hat{Q}_{\ve{i}}       = 1   
\end{equation}
  Eq.~(\ref{eq:orthogonal})  reduces to the  Hubbard  model.    The  physical  electron is  a composite  object,   
\begin{equation}
	   \hat{c}^{\dagger}_{\ve{i},\sigma}    = \hat{f}^{\dagger}_{\ve{i},\sigma}   \hat{s}^{z}_{\ve{i}}, 
\end{equation} 
and using  the  Gauss  law,  we obtain,
\begin{equation}
     \hat{s}^{x}_{\ve{i}} =  (-1)^{ \hat{n}_{\ve{i}} }       =      4  \left( \hat{n}_{\ve{i}, \uparrow}   - 1/2\right)   \left( \hat{n}_{\ve{i}, \downarrow}   - 1/2\right) 
\end{equation}
such that Eq.~(\ref{eq:orthogonal})  maps precisely  onto  the Hubbard model. 
Within this  framework,   the low energy   spectral  function accounts  for  the  electron as  described by  a bound state of the  $f$-electron and an Ising spin.   The  high  energy  is  a  continuum  where  the    composite  object  has  \textit{disintegrated}.      We  note  that  such a  composite  fermion  interpretation of the single particle  spectral  function is equally  appealing  in the realm of  heavy  fermion systems \cite{Danu21,Raczkowski22}.

\begin{figure*}
        \centering
        \includegraphics[scale=0.9, angle=0]{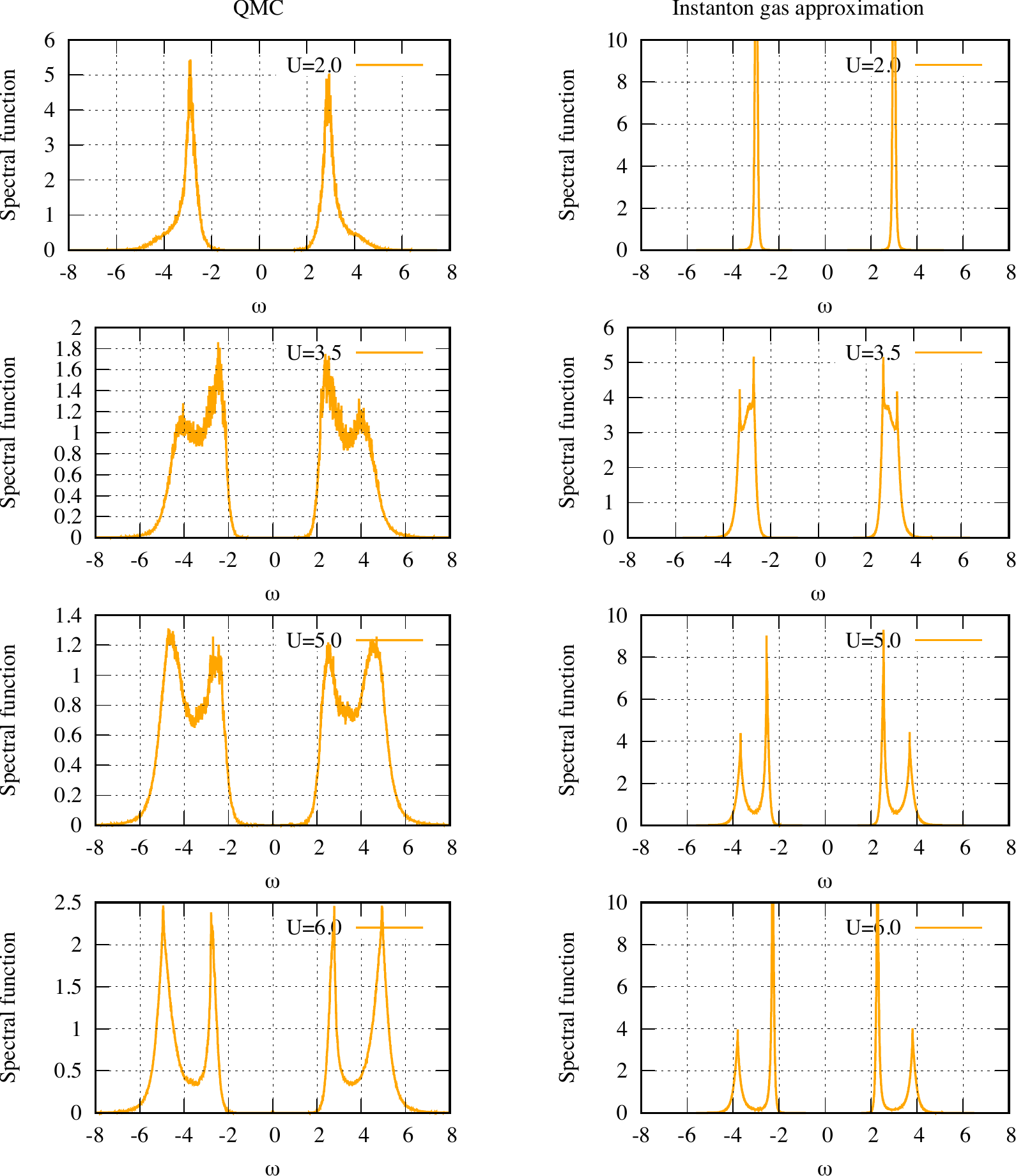}
        \caption{Spectral functions at the $\Gamma$-point obtained from QMC data with Ising fields (left column) and from the instanton gas model with hardcore repulsion (right column). The calculations were performed on a $12\times12$ lattice with $\beta \kappa=20$ and $N _{\tau}=256$, while the corresponding interaction strength is shown on each plot.}
        \label{fig:SpectralFunctionsG}
\end{figure*} 
We  now   compare  the instanton  approach  to  the auxiliary  field  QMC  simulations,  and   will  use  the framework of the  aforementioned  parton  picture  to  interpret the  results. In order to produce the data for the instanton calculation, we first generate $M$ sets of $N^{(r)}$ coordinates $\{X^{(r)}_k, T^{(r)}_k\}$ of instantons, where $r=1...M$ and $k=1...N^{(r)}$ using classical GCMC for the instanton gas model (\ref{eq:ClassicalModelInt}). The details of this 
approach are described in Appendix \ref{sec:AppendixD}. Using these coordinates for the instanton centers, we combine  together the exact profiles of single instantons centered at these coordinates, to get $M$ saddle point field configurations:
\beq
\phi^{(r)}_{{\ve{x}},\tau}=\sum_{k=1}^{N^{(r)}}\phi^{(X^{(r)}_k, T^{(r)}_k)}_{{\ve{x}},\tau}, \, r=1...M. 
\label{eq:InstantonCombination}
\eeq
In this expression, we do not take into account the change of the instanton profiles due to the  overlap between different instantons. This approximation holds due to the relatively small density of the instanton gas for the given range of the interaction strength. Once the configurations have been generated, we average the fermion propagator over these $M$ saddle point field configurations and obtain the spectral function using   the  ALF implementation \cite{ALF_v2} of the stochastic maximum entropy method (SMEM) \cite{PhysRevB.57.10287,Beach04}.

In accordance with the  above  discussion,  we  expect the  biggest  mismatch  between the instanton approach and  auxiliary  field  QMC at  low energies. Namely, in the 
vicinty of the Dirac point (see Fig.~\ref{fig:SpectralFunctionsK}).
The QMC results are plotted in the left row, while the results from the instanton gas model data are shown in the right row. 
As one can see, the QMC results show the appearance of a mass gap starting from $U\approx4.0 \kappa$, while the spectral function obtained within the instanton gas approximation is always concentrated at zero energy. Thus, the formation of a mass gap cannot be described in this simple saddle point approximation. This, along with the failure to reproduce long-range AFM order are two features that are not sensitive to the instantons. A more detailed analysis of the spectral functions within the instanton gas approximation is presented in Fig.~\ref{fig:U6Model}. In this figure, we check the dependence of the spectral functions at the Dirac point on the spatial lattice size. For smaller lattices, we observe noticeable broadening of the spectral function. This broadening is the consequence of the fact that the $6\times6$ lattice is roughly the size of the cluster of local AFM ordering in the vicinity of the centers of instantons (see Figs.~\ref{fig:SpinInstantons}  and \ref{fig:SpinCorrInstantonVacuumMap}). Once the lattice size increases, the simulations start to reflect the absence of long-range AFM ordering and the width of the spectral function decreases.  We thus can conclude that physically, the increase of the instanton density with increasing $U$ corresponds to the increasing local AFM correlations while still not reproducing the long-range ordering of spins.

\begin{figure}
        \centering
        \includegraphics[width=0.35\textwidth, angle=0]{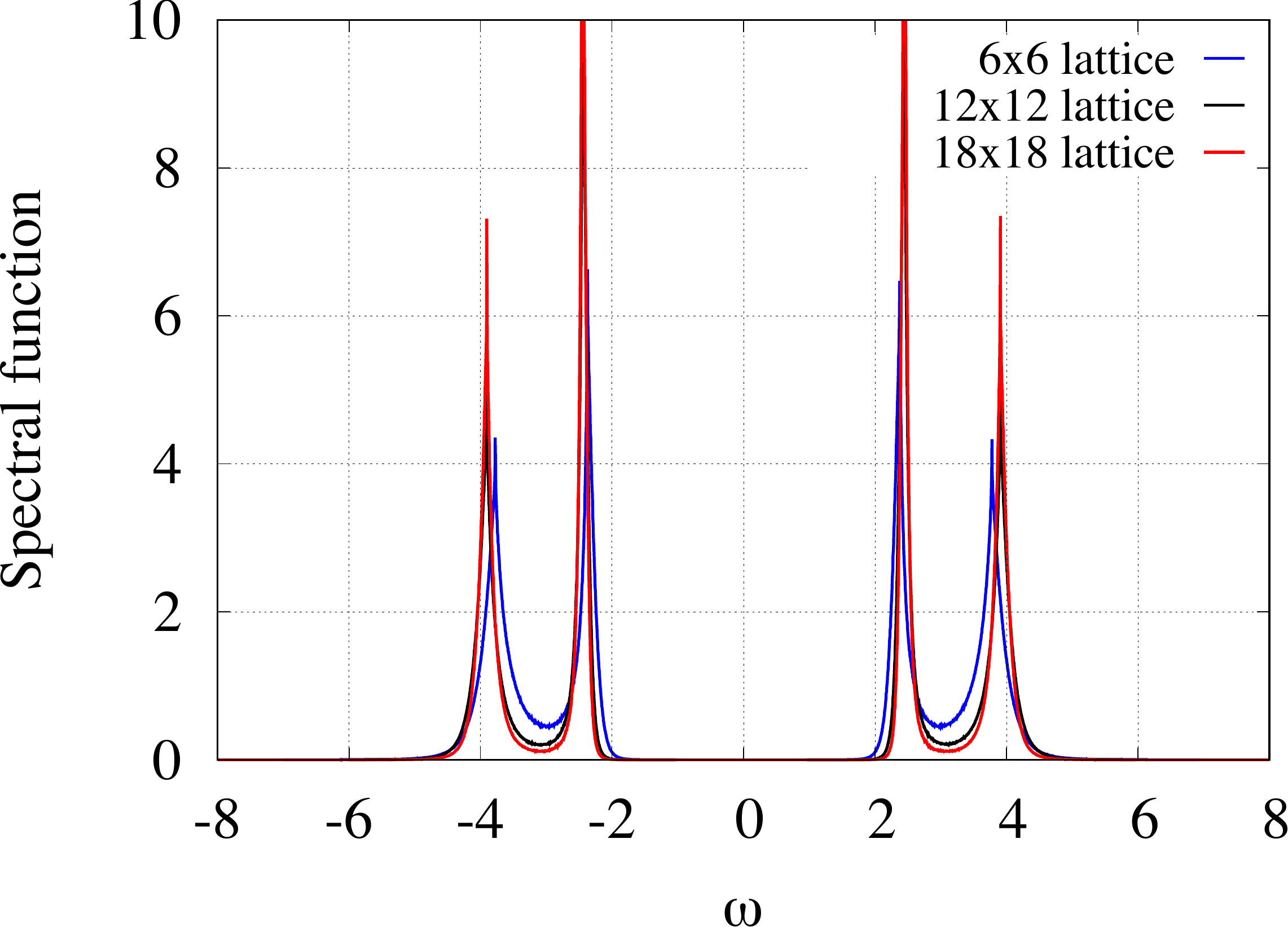}
        \caption{Comparison of the spectral functions at the $\Gamma$-point for different spatial volumes in the instanton gas model with hardcore repulsion. The interaction strength is equal to $U=6.0\kappa$ and inverse temperature $\beta \kappa =20$.}
        \label{fig:SpectralFunctionsGModelCheck}
\end{figure}

As    shown in Fig.~\ref{fig:spectralFunctionsU6Saddles},  at $U =  6\kappa$,  the   instanton  gas spectral   function   compares   remarkably well  with the auxiliary  field  QMC  result of  Fig,~\ref{fig:spectralFunctionsU6QMC}  provided that we  just  shift   it  by  the mass  gap.    Thus, our instanton approach    reproduces  the  salient  features  of  the high energy  spectral  function  that  can  naturally  be  accounted for  within a  parton picture  of the  the  single-particle  spectral  function.  The upper  Hubbard \textit{ band } is most pronounced in the vicinity of the $\Gamma$-point.  In order to quantify this the  relative weight  of the upper  Hubbard band, we plot the share of the lower peak in the whole spectral function (Fig. ~\ref{fig:spectralFunctionsU6Z}). We use the local minimum between peaks as the delimiter. Again, the QMC and instanton gas curves show remarkably similar behaviour, and thus we can conclude that the instanton gas model provides an accurate picture of the single-electron spectral functions both in   the weak- and strong-coupling limits, away from the Dirac point. 

 We now provide a   detailed study of the spectral function  at the  $\Gamma$-point,  where   high -energy spectral  weight is  most pronounced.  
 A comparison of the QMC results with those obtained from   auxiliary field QMC  is shown in Fig.~\ref{fig:SpectralFunctionsG}.      By  construction,  the  instanton approach   satisfies  the  sum  rule,    
 \begin{equation}
 	   \int  d \omega  A(\ve{k},\omega)   =  \pi. 
 \end{equation}
  At  $U=0$,  all the  spectral   weight is located at  $\omega  =  \pm 3  \kappa$.  As  one increases the Hubbard  $U$,   one observes  a  clear    transfer of  spectral  weight  from this  peak  to  higher   frequencies.  This phenomena  already  occurs  priror to  the magnetic  transition  and is pronounced both in the  QMC and in the  instanton   gas  approach.  We have  equally  checked that the observed  transfer of  spectral  weight at large  $U$ is not a finite-size effect. The corresponding data is shown in Fig.~\ref{fig:SpectralFunctionsGModelCheck}, where we compared the results for the instanton gas model on spatial  lattice volumes of $6\times6$, $12\times12$, and $18\times18$. The observed spectral functions are almost identical which is a nontrivial check of the instanton gas model.     On the whole, the comparison between  the QMC and instanton gas  approach is  very  good, especially  if    a rigid  shift is taken  into  account  to  accommodate  for the mass gap that  develops at  $U > U_c \simeq  3.8\kappa $ 
  Hence,  the  instanton gas  model  does  capture the high-energy   spectral   weight.  Within  the parton approach  this  finds  a  natural  interpretation since  the  electron corresponds  to a  composite  object  that  breaks down at  high  energies,  giving  rise to incoherent  spectral  weight.

\begin{center}
  \begin{figure}[]
   \subfigure[]{\includegraphics[width=0.35\textwidth,clip]{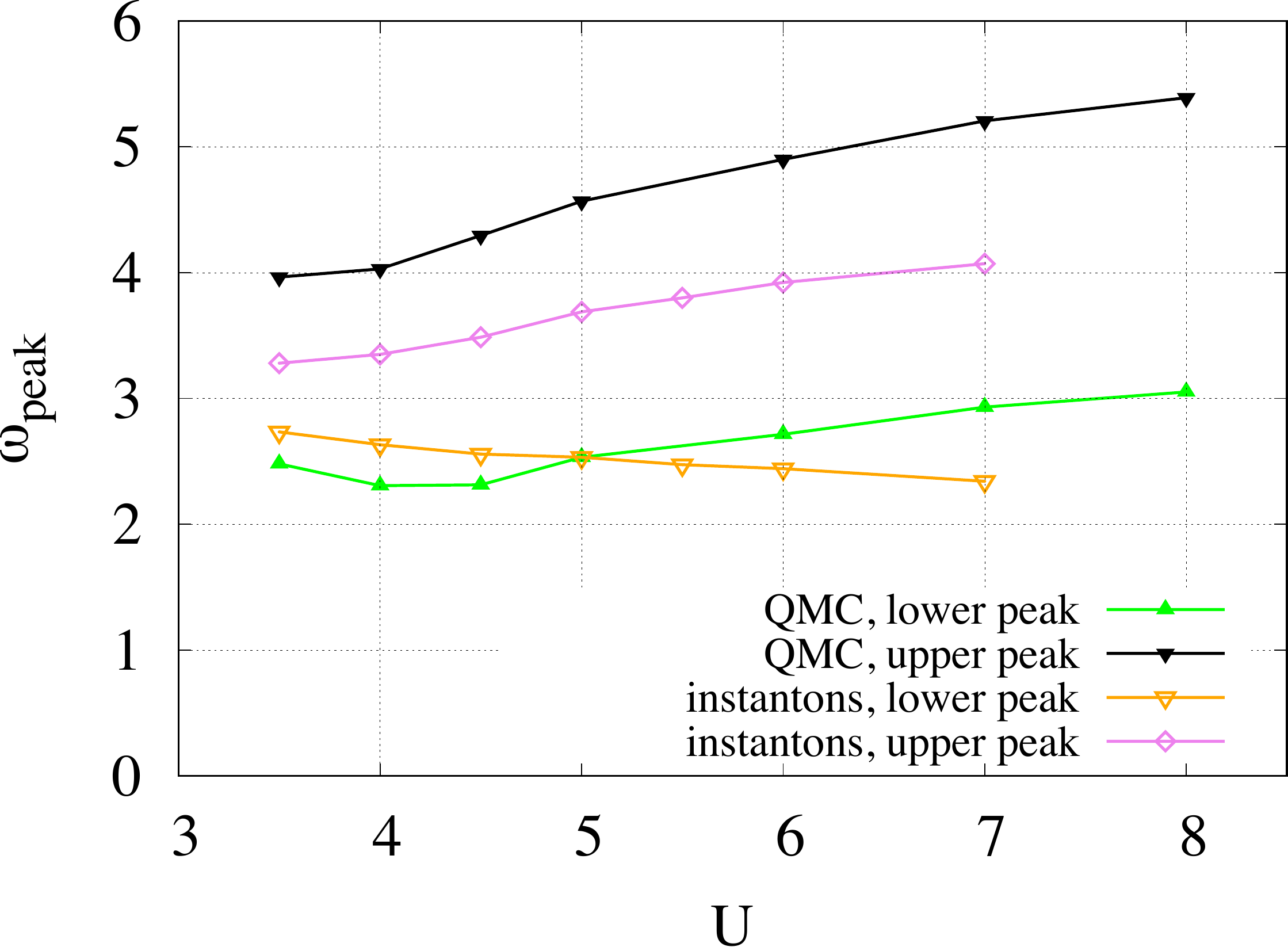}}
       \caption{The energies corresponding to the upper  and  lower peaks in the single  electron spectral function at the $\Gamma$ point in the strong coupling limit $U>3.5$,   where   the two peak  structure  becomes apparent  (see Fig.  \ref{fig:SpectralFunctionsG} for the full profiles).  All data were produced on a   $12\times12$  lattice  at  $N _{\tau}=256$ and $\beta \kappa=20$. The hardcore repulsion model was used for the instanton gas approach. }
   \label{fig:EUdependence}
\end{figure}
\end{center}

For a more detailed analysis of the spectral functions at the $\Gamma$-point, we plot the location of  the upper  and  lower   edges -- as   determined  by  the peak positions --  of  the spectral  function  in Fig.~ \ref{fig:EUdependence}.        Within an  exact  solution to the Hubbard model,  we  expect    the  charge  gap to scale  as  $U$ in the strong coupling limit,  and  the   width  of  the spectral  function, as  defined  by  the difference in energy  between the upper and  lower   edges  of  the single  electron spectral  function,  to scale  as  $\kappa$.     
This  statement is supported  by   numerical    exact  diagonalization of  the 
t-J  model  in which  the    charge  gap   is  infinite  but  the   width of  support of  the  single  particle spectral  function   is set by  the  kinetic energy  \cite{Beran96}. 
As  apparent  from the data in Fig.~\ref{fig:EUdependence}, we  see   that the QMC   supports   this  statement since both  the lower and  upper  edges of  the spectrum grow as  a function of U,  but the  difference  remains,  to  a  first  approximation,  constant.   
As  argued previously,  the  instanton approach  captures  the high energy  physics  but  fails  in the low energy sector, in the  sense  that   a  charge  gap is  not produced.  In fact,  for   Dirac system,  and excluding   exotic  physics  such  as  the  formation of a   quantum  spin liquid  ground state,   a  mass  gap can only  occur    provided  that   symmetry  breaking  occurs.    This   shortcoming of  the instanton approach  shows  up in   Fig.~\ref{fig:EUdependence}:   while  the upper  edge  grows as a function of  $U$,  the lower  edge  actually  decreases.

 \section{\label{sec:Square}Square lattice Hubbard model}

We now briefly apply the same techniques to the Hubbard model on the square lattice in order to investigate the differences and similarities with the hexagonal lattice. The physics of the Hubbard model on the square lattice  at  the  particle-hole  symmetric  point is different,  since    a   Stoner instability 
suggests  that the  AFM insulating phase  is present  for  any  infinitesimal  U   in the  ground  state \cite{PhysRevB.91.125109}. There are, however, some indications \cite{PhysRevB.91.125109, Gull_2008} that there is a crossover from Slater-like to Heisenberg-like fluctuations at around $U=5 \kappa$.   Since  a partial  particle-hole  transformation  maps the repulsive Hubbard model onto the  attractive one,    this parallels the  BCS to  BEC crossover  \cite{Randeria14}. 

\begin{figure}[]
   \centering
   \subfigure[]     {\label{fig:SqLatShares}\includegraphics[width=0.3\textwidth,angle=270]{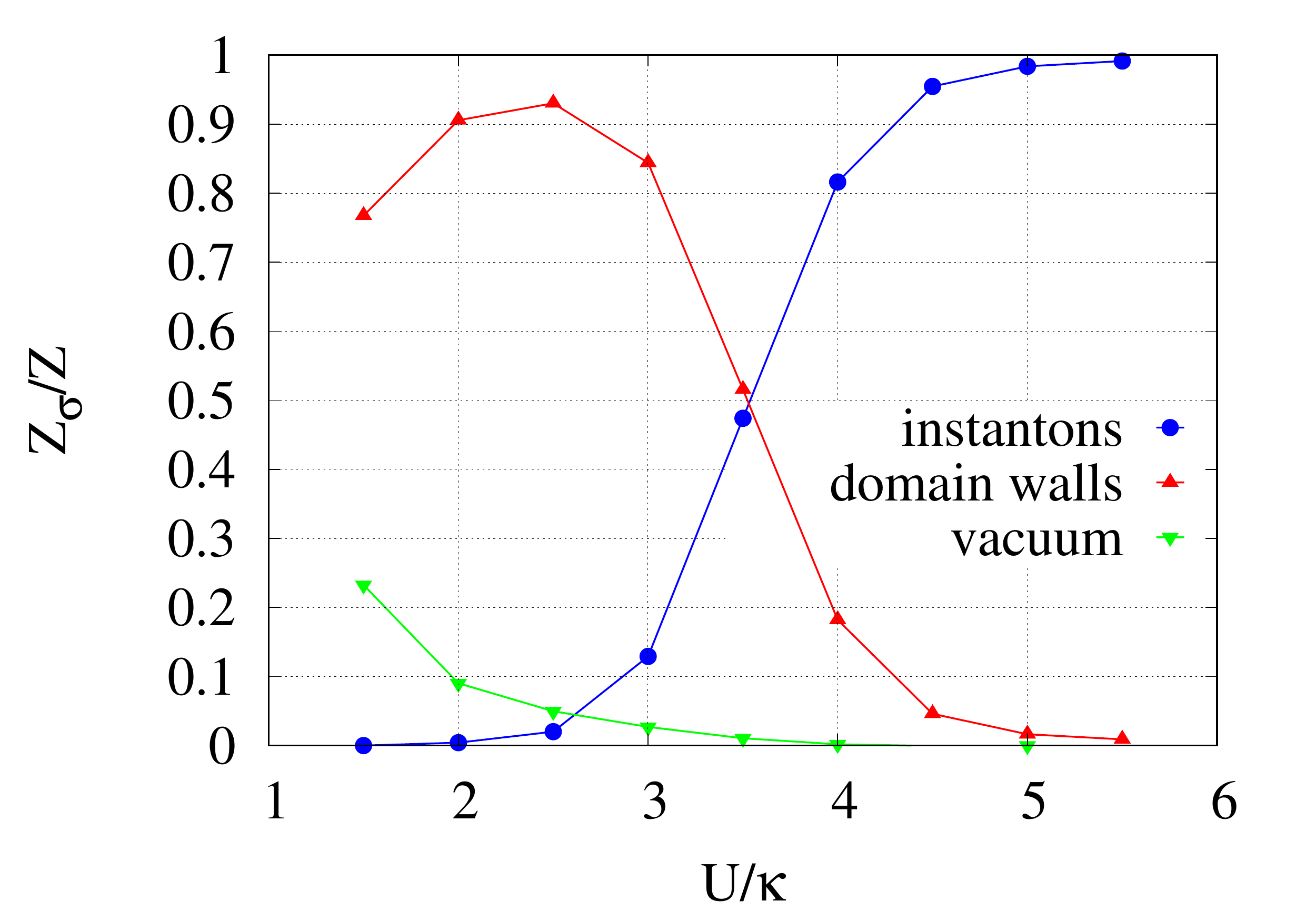}}
   \subfigure[]    {\label{fig:DomainWall}\includegraphics[width=0.3\textwidth,angle=270]{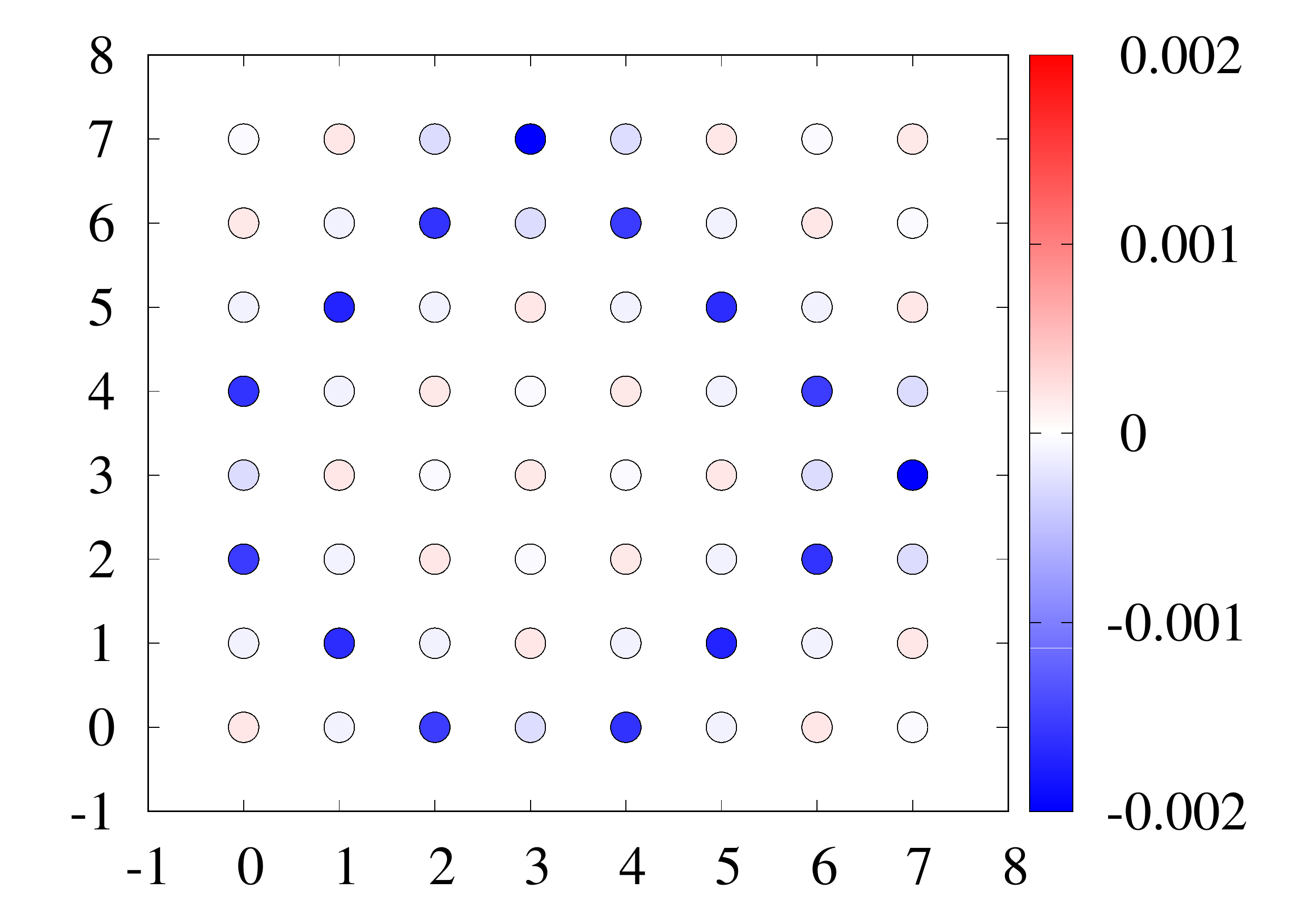}}
          \caption{(a) The relative weight of instantons, domain walls and vacuum saddle in the full partition function for the square lattice Hubbard model. These were obtained  from the configurations generated in QMC on the square lattice, followed by the application of GF. The data was generated on a $8\times8$ square lattice with $\beta \kappa =20$, $N_\tau=512$.  (b) An example of a domain wall configuration at $U=1.5 \kappa$. The color scale shows the value of the auxiliary field $\phi$.}
   \label{fig:SqLat}
 \end{figure}

Now we turn to the picture of the saddle points, obtained with exactly the same procedure as described in Appendix~\ref{sec:AppendixA}.  At large $U$, we mainly find  the same highly localized instantons, with their density increasing with increasing $U$. However, the picture is quite different at $U<5 \kappa$: in this case we observe not only instantons, but also \textit{domain wall} solutions, that are constant in Euclidean time and form barriers  that  divide the lattice in space. An example of such a solution is shown in Fig.~\ref{fig:DomainWall}. It is a spatial map of the charge-coupled auxiliary field $\phi_{\ve{x},\tau}$ (we do not show the Euclidean time dynamics, since the field is independent of time). In the configuration depicted in this plot, the saddle point consists of two domain walls which intersect at a right angle. 

The relative weight of the domain walls, instantons and the vacuum saddle ($\phi_{\ve{x},\tau}=0$) in the partition function as a function of interaction strength $U$ is shown in Fig.~\ref{fig:SqLatShares}. At small $U$, the partition function can be fully described by the integrals attached to the vacuum and \textit{domain wall} saddles. At larger $U$, there is a relatively smooth transition to the instanton-dominated region. Interestingly,   the  crossover  between  these  two regimes coincides with the above-mentioned crossover from a Slater to a Heisenberg antiferromagnet.  From the  present data,  it is  remarkable  to see  that  the saddle point approximation  captures  this crossover.  We expect that the connection between the ground state properties and the saddles can be established using the formalism described in Appendix~\ref{sec:AppendixE}.  However, the detailed study of this subject is beyond the scope of the current paper. 
 
\section{\label{sec:Conclusion}Conclusion}

We have considered the exact saddle points for the path integral formulation of the Hubbard model, where the continuous auxiliary field is coupled to the charge density. The saddle points have been obtained both numerically and analytically without  restriction to constant fields in space and time. Remarkably, the  general saddle point field configuration can be  decomposed into a collection of instantons. An individual instanton is a solution of the classical Euclidean equations of motion for the auxiliary scalar field, which is localized both in space and Euclidean time and determined by taking into account the back reaction of the fermionic determinant.

As a  result of the above, we can define a Gaussian approximation to the partition function, where the Gaussian integral is taken around the N-instanton saddle point. This integral has a well-defined continuum limit with respect to the Euclidean time discretization.  The study  of the two-instanton saddle reveals that the characteristics of this saddle (e.g. action, Hessian, etc) are almost independent of their relative position such that we can treat instantons as weakly interacting classical point-like objects in 3D space-time.  

Using this knowledge, we have constructed the instanton gas model, that can be solved using both analytical approaches and classical grand canonical Monte Carlo simulations. A comparison with unbiased QMC simulations shows that this model gives correct predictions for the structure of the dominant saddle point.  While this feature is probably not so useful at half-filling, it would be interesting to extend this approach away 
from  half-filling. As was shown in \cite{PhysRevD.101.014508}, the instantons for the Hubbard model retain their structure even at non-zero chemical potential. The only difference is that, according to the general logic of the Lefschetz decomposition, these instantons are now shifted in complex space, which means that the auxiliary fields acquire 
complex values. However, the profiles for both the real and imaginary parts of the auxiliary fields remain localized in space and time. 
Thus, it would be interesting to construct a similar instanton gas model for these complex field profiles. If successful, we might obtain an accurate prediction for the dominant saddle point even away from half-filling. Combined with the demonstrated sharpness of the distribution of the instanton density in the thermodynamic limit, this gives us an opportunity to replace the sum over Lefschetz thimbles by one integral over the thimble attached to the  \textit{a priori}  known dominant saddle point.

The  structure of the saddle  points provides a  very interesting  approximation to the  Hubbard model.   Given  the  partition  function of  the instanton  gas, we  can  use  classical  Monte  Carlo  methods  
to  sample   it.    The instanton  configurations  can then be translated into  auxiliary  field  configurations,   for  which the  fermion  determinant 
and    various  equal-time  and  time-displaced correlation  functions can  be  computed.    Using this  scheme,  we  can  elucidate  the  physical content of  a single  instanton  located  at  a space-time point $(X,T)$  by  computing  the spin-spin and  charge-charge  correlation functions.   The individual instanton  generates a local moment  and  concomitant short-ranged  anti-ferromagnetic  fluctuations  in space.    The  instanton  approximation fails  to  capture  long-ranged   AFM order  but  certainly  describes  metallic  states  in the presence  of  short-ranged magnetic  fluctuations.  
By  computing  the   single-particle  spectral  function,  we  have observed  remarkable agreement with   unbiased  quantum Monte  Carlo  results  provided that,  in the magnetically  ordered  state,  we  account for the mass  gap   by  a rigid  shift  in frequency.

As  mentioned  above, the saddle-point approximation fails to capture the formation of long-range AFM order. However, this failure might not be  so important away from half-filling, since the long-range order rapidly breaks down with increased doping. Thus, the proposed instanton gas approach, despite its deficiencies at half-filling, might be even more suitable for the approximate calculations at finite chemical potential, where a severe sign problem hinders our ability to get unbiased results using QMC simulations.

The  instanton  gas  approach is a  thermodynamically  well-defined    approximation  such  that low-temperature  properties  of  metals   subject  to 
anti-ferromagnetic  fluctuations  can  be investigated.  In this  context,   it is very  desirable and  feasible to consider  the  Hubbard model  on the  square lattice  at  half-filling  and in the  strong-coupling limit,  where  we observe  the instanton structure of the path integral.  The  lack of   long-range  magnetic order actually plays  to our advantage  since  a   \textit{ trivial }    gap will not  open   even at the  
particle-hole  symmetric  point  where the   negative  sign problem  is  absent.    
The  properties  of  this  metallic state    and  its relation  to  theories of  nearly  anti-ferromagnetic  Fermi  liquids   reviewed in  \cite{Pines96}    certainly  deserves  further       attention.

\begin{acknowledgments}
Computational resources were provided by the Gauss Centre for Supercomputing e.V. (www.gauss-centre.eu) through the 
John von Neumann Institute for Computing (NIC)
on the GCS Supercomputer JUWELS~\cite{JUWELS} at J\"ulich Supercomputing Centre (JSC).  
This work also benefited from access to the Ir\`ene Joliot-Curie supercomputer of the Tr\`es grand centre de calcul (TGCC) of the Commissariat \`a l’\'energie atomique et aux \'energies alternatives (CEA) in France as part of the project gen2271 awarded by GENCI (Grand Equipement National de Calcul Intensif).
MU  thanks  the  DFG   for financial support  under the  projects  AS120/14-1 and UL444/2-1.  FFA   acknowledges financial support from the DFG through the W\"urzburg-Dresden Cluster of Excellence on Complexity and Topology in Quantum Matter - ct.qmat (EXC 2147, project-id 39085490)  and  the 
SFB1170 on Topological and Correlated Electronics at Surfaces and Interfaces. CW acknowledges support by the Deutsche Forschungsgemeinschaft (DFG, German Research Foundation) through the CRC-TR 211 ``Strong-interaction matter under extreme conditions''~--~project number 315477589~--~TRR 211.
\end{acknowledgments}

\pagebreak
\clearpage
\newpage

\appendix
\counterwithin{figure}{section}

\section{\label{sec:AppendixA}  Observation of instantons  in the QMC data}

  \begin{figure}[]
   \centering
   \subfigure[]  {\label{fig:FlowHistory}\includegraphics[width=0.35\textwidth,clip]{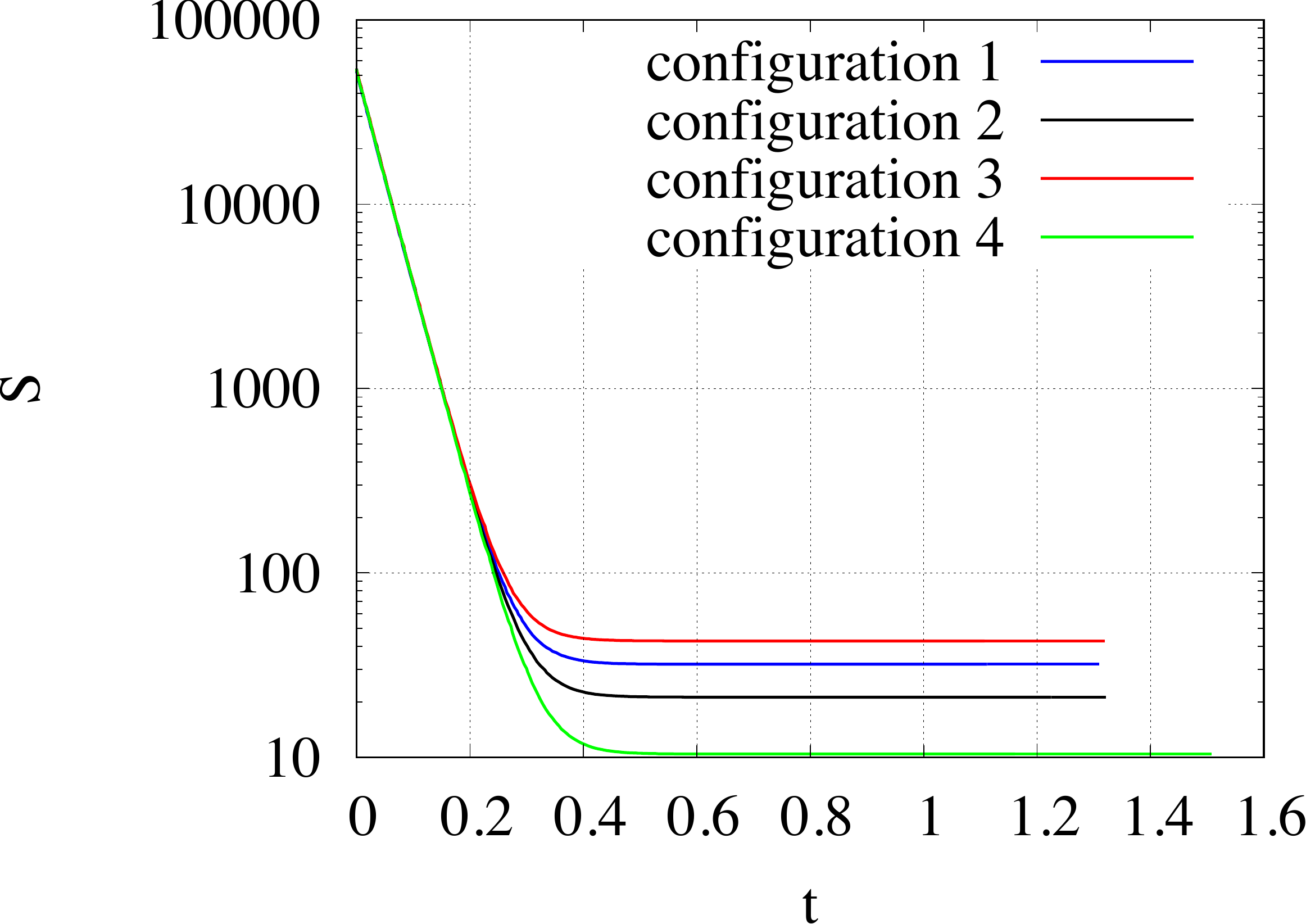}}
   \subfigure[]  {\label{fig:ExampleHistogram}\includegraphics[width=0.35\textwidth,clip]{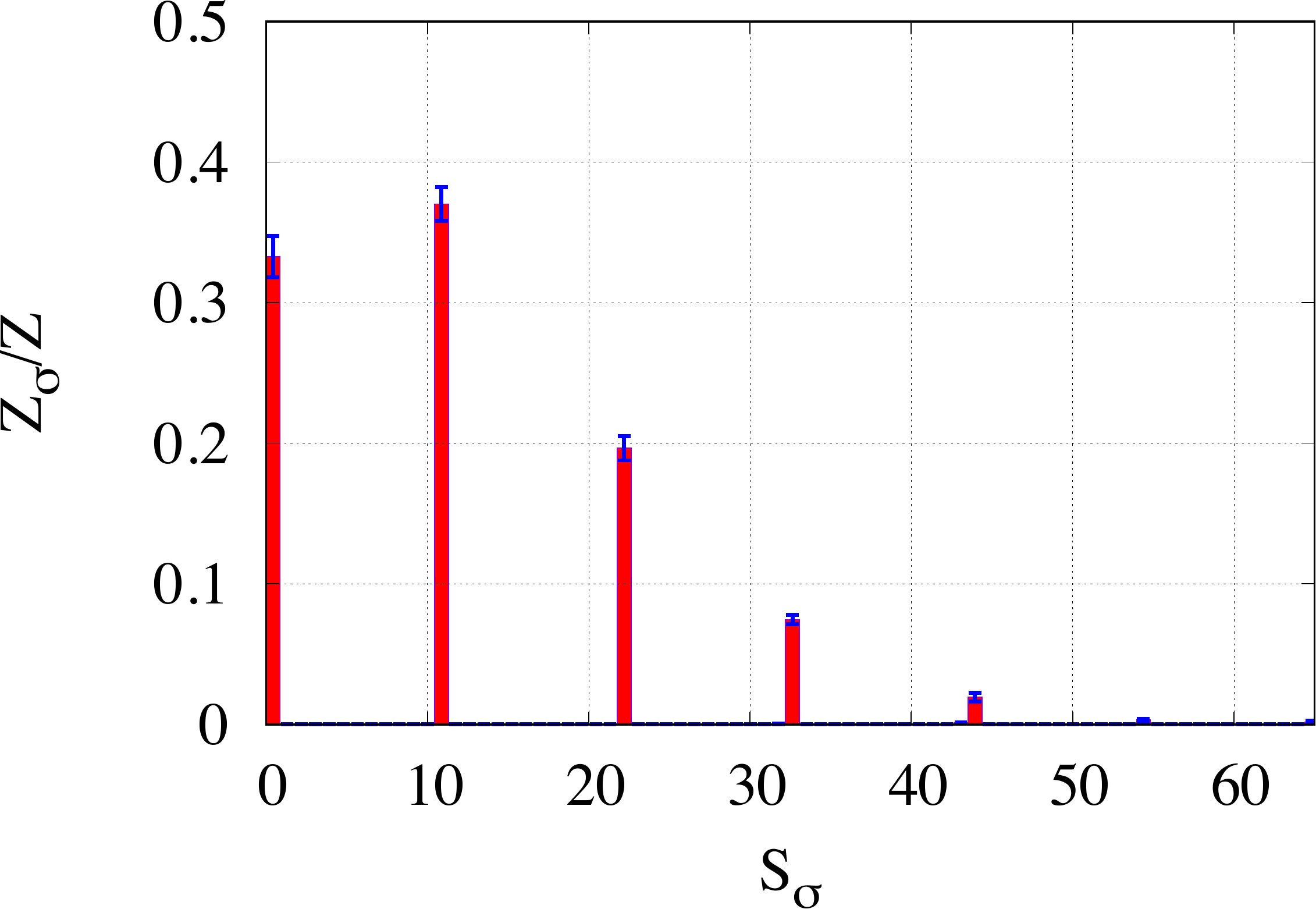}}
       \caption{(a) The profile of the action during the  \textit{ downwards } flow to the saddle points. This procedure essentially amounts to the solution of the gradient flow equation in (\ref{eq:inverse_flow}). Four different configurations are shown, each ending at a different saddle. (b)  The distribution for the final actions obtained after the application of the downward gradient flow to the set of configurations generated with HMC. These calculations were performed on a $12\times12$ lattice with $\beta \kappa=20$ and $N_{\tau}=512$, with interaction strength $U=2.0 \kappa$, and $\alpha=0.99$.}
   \label{fig:HistogramsGeneration}
\end{figure}

This appendix describes technical aspects regarding the determination of saddle points from the QMC data. Since we can only use continuous auxiliary fields in order to construct the saddle point approximation, we employ the standard technique of updating the continuous fields in QMC simulations which goes under the name of hybrid Monte Carlo (HMC). This updating procedure is based on Hamiltonian dynamics for the auxiliary field. The main issue is the absence of ergodicity in such simulations for the Hubbard model when only one auxiliary field is used \cite{Assaad_complex}. Following \cite{Assaad_complex}, we employ the two-field approach to overcome this issue. In this approach, the interaction term is split into two terms using the Fierz identities
\beq \label{eq:Fierz_identity}
\frac{U}{2}\hat q_{\ve{x}}^2 = \frac {\alpha U}{2}\hat q_{\ve{x}}^2 - \frac{(1-\alpha) U}{2} \hat s_{\ve{x}}^2 + (1-\alpha) U \hat s_{\ve{x}},
\eeq
where $\hat s_x = \hat n_{{\ve{x}}, \text{el.}} + \hat n_{{\ve{x}}, \text{h.}}$ is the spin operator, and $\alpha \in [0,1]$ is an extra, non-physical parameter. Thus, in addition to (\ref{eq:continuous_HS_imag}), for $\alpha \neq 1$, one must also introduce a second auxiliary field coupled to the spin
\beq \label{continuous_HS_real}
   e^{\frac{\Delta \tau U (1 - \alpha)}{2}\hat s^2_{\ve{x}}} \! \cong \! \int \! d \chi_{\ve{x}}\, e^{- \frac{\chi^2_{\ve{x}}}{2\Delta \tau U (1-\alpha)} } e^{ \chi_{\ve{x}} \hat s_{\ve{x}}}.
\eeq 
This generalized HS transformation serves several purposes. It solves the ergodicity problems associated with the HMC as infinite energy barriers appear which separate regions where the electron(hole) determinant has different signs. This was first noted in \cite{complex1} and further applied in \cite{Assaad_complex, Ulybyshev:2017}.
This particular representation is also advantageous as it works for non-local interactions, unlike methods that employ discrete auxiliary fields. The form of the functional integral is slightly modified when the spin-coupled field is introduced following \cite{Ulybyshev:2013swa,SmithVonSmekal,Assaad_complex}, taking the form
\begin{align}
  &\mathcal{Z}\!=\! \int\!\mathcal{D} \phi_{\ve{x},\tau}\, \mathcal{D} \chi_{\ve{x},\tau}\, e^ {-S_{\alpha}}  \det M_{\text{el.}} \det M_{\text{h.}},&  \\
   &S_\alpha[\phi_{\ve{x},\tau},\chi_{\ve{x},\tau}] \!=\! \sum_{\ve{x},\tau}  \left[\frac  {\phi_{\ve{x},\tau}^2} {2 \alpha \Delta \tau U}  \!+\!  \frac {(\chi_{\ve{x},\tau}\!-\! (1\!-\!\alpha) \Delta \tau  U)^2} {2 (1\!-\!\alpha) \Delta \tau U}\right],&\nonumber
  \label{eq:Z_continuous}
\end{align}
where the determinants of the fermionic operators are given by
\begin{eqnarray}
 \det M_{\text{el.,h.}} = \det \left[ I +\prod^{N_{\tau}}_{\tau=1}  e^{-\Delta \tau h} \tilde{D}_{\tau} \right]. 
  \label{eq:M_continuous_appendix}
\end{eqnarray}
Here we have introduced the matrices $\tilde D_{\tau} \equiv \diag{ e^{\pm i \phi_{\ve{x},\tau}+\chi_{\ve{x},\tau}} }$, in analogy with the case which only involved the charge-coupled field. 

\begin{figure}[]
   \centering
   \subfigure[]   {\label{fig:histogramAlpha}\includegraphics[width=0.35\textwidth,clip]{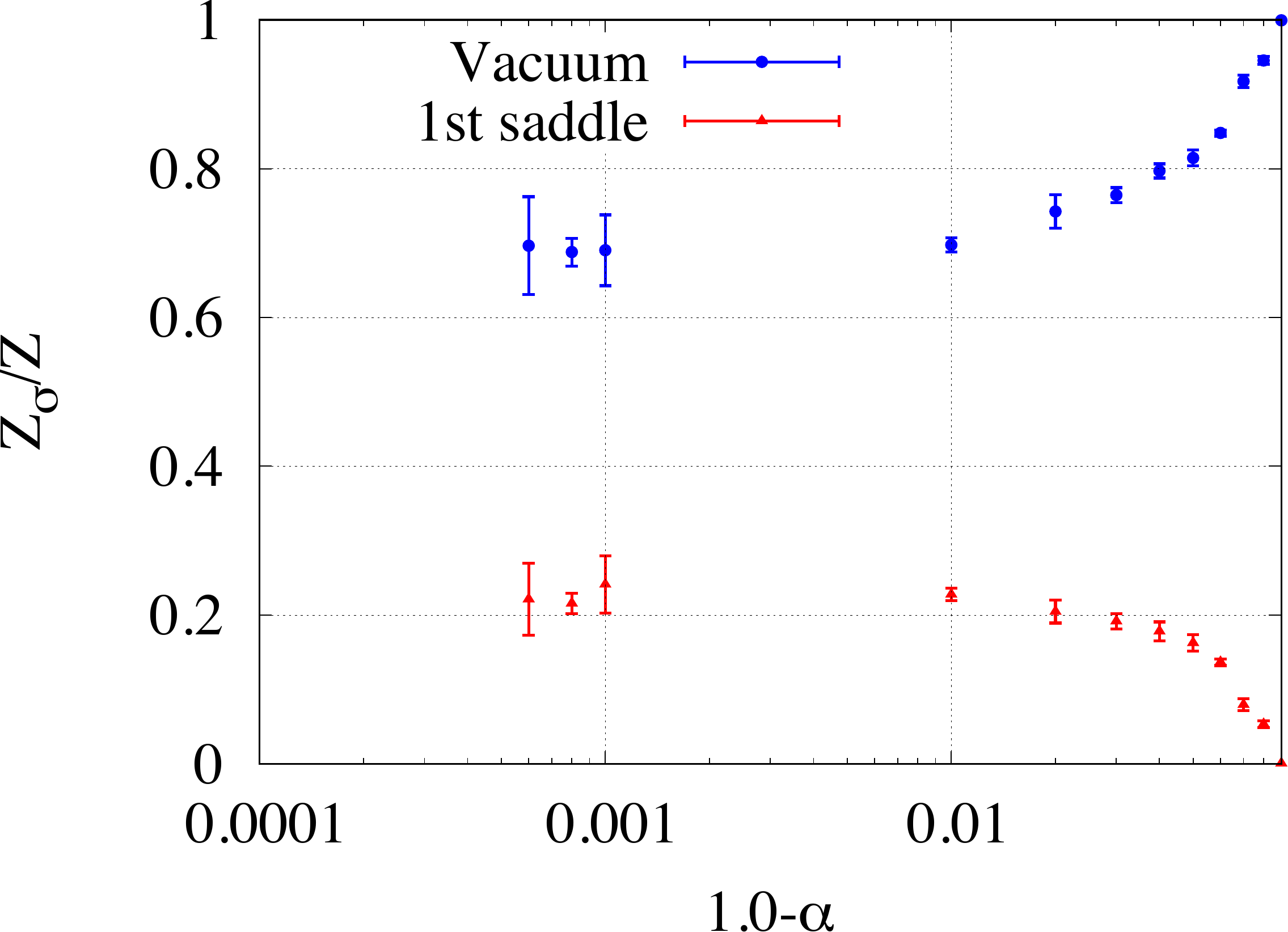}}
   \subfigure[]  {\label{fig:histogramDt}\includegraphics[width=0.35\textwidth,clip]{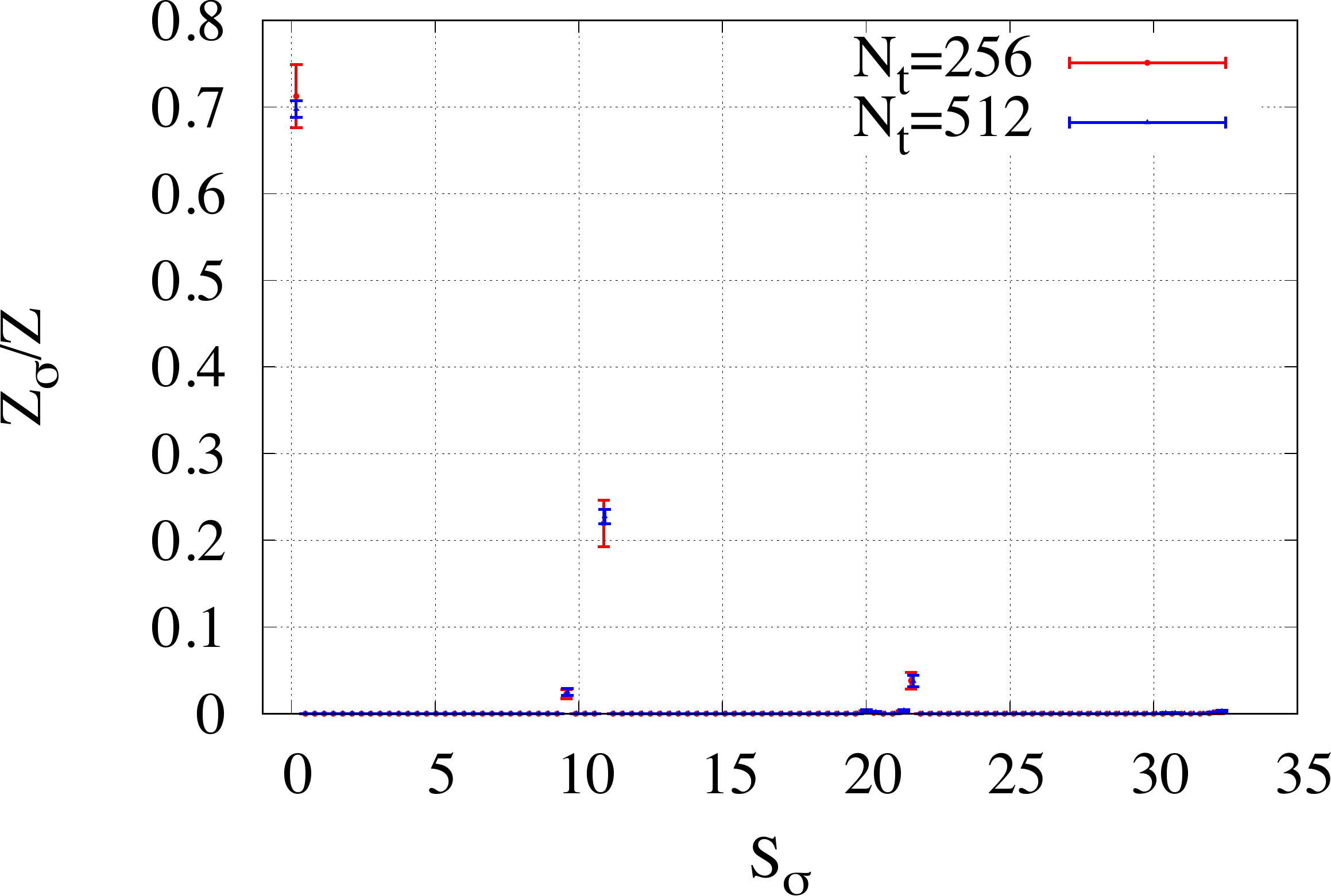}}
      \caption{(a)  Relative weight of the thimble attached to the first non-trivial saddle with respect to the full partition function, $Z_1/Z$, as a function of the $\alpha$ parameter. These calculations were performed on a $6\times6$ lattice with $N _{\tau}=512$, $\beta \kappa=20$. (b) A comparison of the histograms obtained for $N _{\tau}=256$ and $N _{\tau}=512$ with other parameters fixed ($6\times6$ lattice, $\beta \kappa=20$, $\alpha=0.99$). The interaction strength is fixed at $U=2.0 \kappa$ for all plots.}
   \label{fig:histogramsTech}
\end{figure}

The two-field formalism solves the ergodicity issues in the HMC simulations. Therefore, we are free to use the action in (\ref{eq:Z_continuous}) in HMC both for the generation of configurations and for the subsequent solution of the GF equations. Several examples depicting the flow-time history of the action during GF are shown in the Figure \ref{fig:HistogramsGeneration}\textcolor{red}{(a)}.
However, we are interested in the case of $\alpha=1$, whereby only the charge-coupled field $\phi$ is present. In practice, we recover this limit by setting $\alpha\approx 1$ and taking the limit $\alpha\to 1$. If $\alpha$ is reasonably close to 1, we already recover the equidistant discrete peaks in the final histogram for the action after GF (see Fig.~\ref{fig:HistogramsGeneration}\textcolor{red}{(b)} and more detailed study \cite{PhysRevD.101.014508}). The weights of the peaks are stable in the limit $\alpha \to 1$, as demonstrated in Fig.~\ref{fig:histogramsTech}\textcolor{red}{(a)}. Thus, our strategy can indeed be used to   determine the structure of saddle points in the limit $\alpha=1$. We always use $\alpha=0.99$ in the QMC calculations presented in this paper.

As a final check, we repeat the simulations for the same physical parameters but  at different values of  $N_\tau$. This is to ensure that systematic effects due to discretization in Euclidean time are under control. As one can see in Fig.~\ref{fig:histogramsTech}\textcolor{red}{(b)}, the histograms do not change. We can therefore conclude that the continuum limit $\Delta \tau \rightarrow 0$ indeed exists and with our typical setup of $\beta \kappa=20$ and $N_\tau=256$, we are reasonably close to it.

\section{\label{sec:AppendixB}Analytical solution for individual instanton}

In developing the functional integral approach to the Hubbard model, we have integrated out the fermionic degrees of freedom leaving us with a theory that only involves bosonic degrees of freedom. The tradeoff is that the theory involves a non-local quantity, namely the fermionic determinant. In order to obtain the saddle point equations for the auxiliary field, $\phi$, we must first re-exponentiate the fermion determinant which gives the following form for the effective Euclidean action
\begin{eqnarray}
S=S_{B} - \ln (\det M_{el.} \det M_{h.}).
 \label{eq:action_continuous}
\end{eqnarray}
In order to make the process of taking the continuum limit in Euclidean time clearer, we make the trivial rescaling $\phi_{\ve{x}, \tau} \rightarrow \Delta \tau \phi_{\ve{x}, \tau}$. As a result, the bosonic action reads as 
\begin{eqnarray}
 S_B[\phi] = \sum_{{\ve{x}},\tau}  \frac  {\phi_{{\ve{x}},\tau}^2  \Delta \tau } {2 U},
 \label{eq:SB_rescaled}
\end{eqnarray}
and the exponents in the ``even" blocks of the fermionic determinants are written as  $D_{2\tau} \equiv \diag{e^{i \Delta \tau \phi_{{ \ve{x}},\tau}} }$. 

From (\ref{eq:SB_rescaled}), one sees that the bosonic part of the action, $S_B$, is Gaussian, and thus any potentially non-trivial solution to the saddle-point equations for the auxiliary fields can only be achieved by taking into account the contribution from the fermion determinant in the background of a given bosonic field configuration. 

In order to obtain the saddle point equation for the bosonic field $\phi_{\ve{x},\tau}$, it is convenient to express the electron(hole) operator in the following basis
\begin{eqnarray}
M_{el.}=\nonumber \\ \left(\begin{matrix}
I   & D_1  & 0 & 0 &...&...& 0 \\
0   & I  & D_2 & 0 &...&...& 0  \\
... & ... & \ddots &  \ddots  &...&...&... \\
...&...&...&  I &   D_{2\tau} &...&... \\
... & ... &...&...& I & D_{2\tau+1} & ... \\
... & ... & ... &  ... & ...& \ddots& \ddots \\
 - D_{2N_{\tau}} & ... & ... &  ... & ...& ...& I \\
                  \end{matrix}\right),
   \label{eq:ferm_operator_extended}  
\end{eqnarray}
where each entry represents a block of size $N_S \times N_S$. 
In a similar way, the fermion propagator can be written in terms of $N_S \times N_S$  blocks
\begin{eqnarray}
M^{-1}_{el.}=\left(\begin{matrix}
g^1  & ...  & ...&...&...&...& \bar g^{2 N_\tau} \\
\bar g^1 & g^2  & ... & ... &...&...&...  \\
... & \bar g^2 & g^3 & ... &...&...&... \\
...&...&\ddots &  \ddots &  ... &...&... \\
... & ... &...&\bar g^\tau & g^\tau  & ... & ... \\
... & ... & ... &  ... & \ddots & \ddots& ... \\
... & ... & ... &  ... & ...& ...& g^{2 N_\tau} \\
                  \end{matrix}\right).
   \label{eq:ferm_propagator_extended} 
\end{eqnarray}
The off-diagonal blocks $\bar g_i$ satisfy the following relation 
\beq \label{eq:BSS_QMC_forward_prop}
\bar{g}^{\tau+1} = D^{-1}_{\tau+1} \bar{g}^{\tau} D_{\tau},
\eeq 
which is reminiscent of the   \textit{ forward } propagation relation for the equal-time Green's function in the BSS-QMC algorithm  \cite{Assaad08_rev}. One can relate the blocks in (\ref{eq:ferm_propagator_extended}) on the diagonal to the blocks below the diagonal using 
\beq \label{eq:unbarred_relation_barred_BSS}
g^{\tau} = I - D_{\tau} \bar{g}^{\tau},
\eeq
which simply follows from $M_{el.}M^{-1}_{el.} = I$.
We note that the Euclidean time index for the blocks $g^{\tau}$ and $\bar g^{\tau}$ takes values from $1$ to $2 N_\tau$, in accordance with the employed scheme for the decomposition of the Boltzmann weight at each time slice. Using the well-known relation for the derivative of the logarithm of the fermion determinant
\begin{eqnarray}
\label{eq:det_der}
 \frac{\partial \ln \det M} {\partial \phi_{\ve{x},\tau}} =  \Tr\left( { M^{-1} \frac {\partial M}{\partial \phi_{\ve{x},\tau}} }\right),
 \end{eqnarray}
one can obtain the following expression for the derivative of the action (\ref{eq:action_continuous}) with respect to the bosonic auxiliary fields
\begin{eqnarray}
\label{eq:action_der}
 \frac {\partial S}{\partial \phi_{\ve{x}, \tau}} = \Delta \tau  \frac{\phi_{\ve{x}, \tau}}{U} -   \\ \Delta \tau  \left(\bar g^{2\tau}_{\ve{x}\ve{x}} i e^{i \Delta \tau \phi_{\ve{x},\tau}} - (\bar g^{2\tau}_{\ve{x}\ve{x}})^* i e^{-i \Delta \tau \phi_{\ve{x},\tau}}\right). \nonumber
 \end{eqnarray}
 This relation is used not only in numerically determining the fermionic  \textit{ force }  in HMC calculations, but can also be used to determine the saddle points of Eq.~(\ref{eq:action_continuous}).
 These are obtained from the relation
 \begin{eqnarray}
\label{eq:action_der_zero}
 \frac {\partial S}{\partial \phi_{\ve{x},\tau}} = 0.
 \end{eqnarray}
 One then obtains the following form for the saddle point equation
 \beq \label{eq:saddle_point}
 \phi_{\ve{x},\tau} = -U \operatorname{Im} \{ \bar{g}^{2\tau}_{xx} e^{i \Delta \tau \phi_{\ve{x},\tau}}\},
 \eeq
 which relates the bosonic field to the Green's function. In the continuum limit, $\Delta \tau \rightarrow 0$, the saddle point equation (\ref{eq:saddle_point}) becomes
 \beq  \label{eq:saddle_point_cont}
 \phi_{\ve{x},\tau} =- U \operatorname{Im} \bar g^{2\tau}_{\ve{x}\ve{x}},
 \eeq 
 where we have taken into account that $\operatorname{Re} \bar g^\tau_{xx}=\frac{1}{2}$, for all $\tau$, which follows from particle-hole symmetry. In order to close the system of equations, we add the equations for the fermionic propagator at a given background of the auxiliary field.  Applying the BSS-QMC forward propagation relation in (\ref{eq:BSS_QMC_forward_prop}) twice, one obtains the following equations:
 \begin{eqnarray}
\label{eq:bar_g_evolution} 
&&\bar g^{2\tau+2} = D^{-1}_{2\tau+2} D^{-1}_{2\tau+1}  \bar g^{2\tau} D_{2\tau} D_{2\tau+1},  \\
&& = \diag{e^{-i \Delta \tau \phi_{\ve{x},\tau+1}}} e^{\Delta \tau h} \bar g^{2\tau} \diag{e^{i \Delta \tau \phi_{\ve{x},\tau}}} e^{-\Delta \tau h}. \nonumber 
 \end{eqnarray}
The various terms on the right-hand side of (\ref{eq:bar_g_evolution}) can be expanded to linear order in $\Delta \tau$. Then, remembering the saddle-point relation (\ref{eq:saddle_point_cont}), $\phi$ can be eliminated in favor of $\bar g$. Finally, after taking the continuum limit, $\Delta \tau \rightarrow 0$, one obtains 
\begin{widetext}
\begin{eqnarray}
\label{eq:full_g_system}
\left\{ { \frac{d g_{\ve{x}\ve{x}}(\tau)}{d\tau} = - \kappa \sum_{\langle{\ve{x}},{\ve{y}}\rangle}(g_{\ve{x}\ve{y}}(\tau) - g_{\ve{y}\ve{x}}(\tau)) }\atop{ \frac{d g_{\ve{x}\ve{y}}(\tau)}{d\tau}= i U g_{\ve{x}\ve{y}} (\operatorname{Im} g_{\ve{x}\ve{x}}(\tau) -  \operatorname{Im} g_{\ve{y}\ve{y}}(\tau)) -  \kappa \left( \sum_{\langle{\ve{z}},{\ve{x}}\rangle}  g_{\ve{x}\ve{z}}(\tau)- \sum_{\langle{\ve{z}},{\ve{y}}\rangle} g_{\ve{z}\ve{y}}(\tau)\right) } \right. .
 \end{eqnarray}
\end{widetext}
This set of equations completely determines the semiclassical description of fermions propagating in the background of an instanton. In particular, the correlations between winding and the number of instantons and anti-instantons can be understood from this set of equations (see below).

Alternatively, this set of equations can be also derived from the Euclidean-time Heisenberg equations of motion
\begin{eqnarray}
\label{eq:Heisenberg}
\frac{d\hat A}{d\tau}=[\hat H, \hat A],
 \end{eqnarray}
 where we consider the following fermion bilinear operators  
 \begin{eqnarray}
\label{eq:A_op_examples}
\hat A=\hat a^\dag_{\ve{x}} \hat a_{\ve{y}}, \, \hat b^\dag_{\ve{x}} \hat b_{\ve{y}}, \, \hat a^\dag_{\ve{x}} \hat a_{\ve{x}}, \, \hat b^\dag_{\ve{x}} \hat b_{\ve{x}}
 \end{eqnarray}
 and the Hamiltonian $\hat H$ is taken from Eq. (\ref{eq:Hamiltonian}).
 
After substitution of (\ref{eq:A_op_examples}) into (\ref{eq:Heisenberg}),  the operator analogue of the equations (\ref{eq:full_g_system}) can be written as 
\begin{widetext}
\begin{eqnarray}
\label{eq:full_g_system_mean_field}
\left\{   {\frac{d \hat a^\dag_{\ve{x}}\hat a_{\ve{x}}}{d\tau} = -\kappa \sum_{\langle{\ve{x}},{\ve{y}}\rangle}(\hat a^\dag_{\ve{y}} \hat a_{\ve{x}} - \hat a^\dag_{\ve{x}} \hat a_{\ve{y}}) } \atop {\frac{d \hat a^\dag_{\ve{x}}\hat a_{\ve{y}}}{d\tau} = U \hat a^\dag_{\ve{y}} \hat a_{\ve{x}} (\hat b^\dag_{\ve{x}} \hat b_{\ve{x}} - \hat b^\dag_{\ve{y}} \hat b_{\ve{y}} )-\kappa \left(\sum_{\langle{\ve{z}},{\ve{y}}\rangle} \hat a^\dag_{\ve{z}} \hat a_{\ve{x}} -  \sum_{\langle{\ve{z}},{\ve{x}}\rangle} \hat a^\dag_{\ve{y}} \hat a_{\ve{z}} \right) }      \right.
 \end{eqnarray}
\end{widetext}
These operator relations, when applied to expectation values, take the exact same form as (\ref{eq:full_g_system}) in the mean-field limit. As usual, when working in the mean-field approximation, one assumes that the expectation of four-fermion terms factorize
\begin{eqnarray}
\label{eq:mean_field}
\langle \hat a^\dag_{\ve{x}} \hat a_{\ve{y}} \hat b^\dag_{\ve{x}} \hat b_{\ve{x}}  \rangle \approx \langle \hat a^\dag_{\ve{x}} \hat a_{\ve{y}} \rangle \langle \hat b^\dag_{\ve{x}} \hat b_{\ve{x}}  \rangle
 \end{eqnarray}
 After this factorization, we arrive at exactly the same equations as (\ref{eq:full_g_system}) taking into account that $g_{xy}=\langle  \hat a_{\ve{x}} \hat a^\dag_{\ve{y}}  \rangle$

The system of equations (\ref{eq:full_g_system}) can be further simplified if we take into account the fact that the exact instanton solutions observed in the QMC results of the previous section are ultra-local in space. Namely, the bosonic field $\phi_{{\ve{x}},\tau}$ is sharply concentrated (measured by its magnitude $|\phi|$) at one lattice site. A further simplification occurs if one takes into account the $C_3$-symmetry of the hexagonal lattice, and the rapid decay of the equal-time fermionic propagator with increased spatial separation between source and sink. It turns out that this rapid decay is also a consequence of the observed locality of the instanton field configurations. In fact, the 
propagator is identical to that of freely propagating fermions everywhere except in the close vicinity of the instanton  \textit{core}. Our assumptions about the equal-time fermion Green's function, computed in the background of an instanton centered at spatial lattice site $x$, can be summarized as follows: we take into account only $\operatorname{Im} g_{\ve{x}\ve{x}}(\tau)$ and $g_{\langle xy \rangle}(\tau)$, and the latter components of the Green function are equal for all three nearest neighbours.

Under these assumptions, (\ref{eq:full_g_system}) simplifies greatly and takes the form 
\begin{eqnarray}
\label{eq:g_evolution_simplified}
\left\{ { \frac{d}{d\tau} \operatorname{Im} g_{\ve{x}\ve{x}} (\tau) = 6 \kappa \operatorname{Im} g_{\ve{x}\ve{y}} (\tau) } \atop {\frac{d}{d\tau} \operatorname{Im} g_{\ve{x}\ve{y}} (\tau) = i U g_{\ve{x}\ve{y}}(\tau) \operatorname{Im} g_{\ve{x}\ve{x}}(\tau) + i \kappa  \operatorname{Im} g_{\ve{x}\ve{x}}(\tau) } \right.  . 
 \end{eqnarray}
Separating the real and imaginary parts of the above equations gives the following set of coupled, first-order differential equations
\beq \label{eq:d_evolution}
\dot{d}(\tau) = 6 \kappa b(\tau), \\ \label{eq:b_evolution}
\dot{b}(\tau) = U d(\tau) \left( a(\tau) + G^{-1} \right), \\ \label{eq:a_evolution}
\dot{a}(\tau) = -U \dot{b}(\tau) d(\tau), 
\eeq 
where $g_{\ve{x}\ve{y}} (\tau) = a(\tau) + ib(\tau)$, $\operatorname{Im} g_{\ve{x}\ve{x}} = d(\tau)$, and we have defined the dimensionless ratio $G \equiv U/\kappa$. From (\ref{eq:b_evolution}) and (\ref{eq:a_evolution}), it is straightforward to see that the solutions can be written in the form
\beq \label{eq:a_solution}
a(\tau) = -G^{-1} + R \cos \theta(\tau), \\ \label{eq:b_solution}
b(\tau) = R \sin \theta(\tau), \\ \label{eq:d_solution}
d(\tau) = \frac{\dot{\theta}(\tau)}{U},
\eeq 
where $R$ is a dimensionless constant determined by the initial conditions far away from the center of instanton, where the Green's function $g_{\ve{x}\ve{y}} (\tau)$ tends to its vacuum value. For the imaginary part, this means that $\operatorname{Im} g_{\ve{x}\ve{y}} = b \rightarrow 0$, thus $\theta \rightarrow 0$. For the real part, this means that $\operatorname{Re} g_{\ve{x}\ve{y}}|_{vac.} = -G^{-1} + R$.   Finally, inserting (\ref{eq:d_solution}) into (\ref{eq:d_evolution}) one obtains a second-order differential equation for the angle
\beq \label{eq:theta_solution}
\ddot{\theta}(s) = \sin \theta(s), 
\eeq 
where we have introduced the rescaled Euclidean time $s \equiv \tau \sqrt{6\kappa U R}$. One recognizes (\ref{eq:theta_solution}) as the equation of motion satisfied by a physical pendulum where the angle between the vertical and the pendulum has been shifted by $\pi$. Thus, the vacuum corresponds to the upper position of the pendulum, and the instanton solution corresponds to the trajectory $\theta(\tau)$, which starts near the upper position of the pendulum, spends a large time in its vicinity, then quickly performs a rotation through the bottom position.  If the initial velocity $\dot \theta$ is large enough to make one or more full rotations during the period $s_{full} = \beta \sqrt{6\kappa U R}$, we have a solution with $N_{inst.}$ instantons. If the initial velocity is not large enough in order to pass over the highest point, the pendulum goes in the opposite direction during the second half of the period and we have  an  instanton-anti-instanton solution. 

The number of instantons can be connected to the initial conditions of the pendulum using the analogy with classical mechanics. Energy conservation in this case takes the form
\beq \label{eq:energy_conservation}
\frac{\dot \theta^2}{2} + \cos \theta=E_0. 
\eeq 
Then, the initial conditions for the $N_{inst.}$ solution can be written as $\theta|_{\tau=0}=0$, $\dot \theta|_{\tau=0}=\sqrt{2 (E_0-1)}$, and $E_0$ is defined by the number of instantons:
\beq \label{eq:energy_N_inst}
\frac{s_{full}}{N_{inst.}}=2\int^\pi_0 \frac{d\theta}{\sqrt{2(E_0-\cos\theta)}}
\eeq

\begin{figure}[]
   \centering
   \subfigure[]   {\label{fig:1instantonV}\includegraphics[width=0.15\textwidth,angle=270]{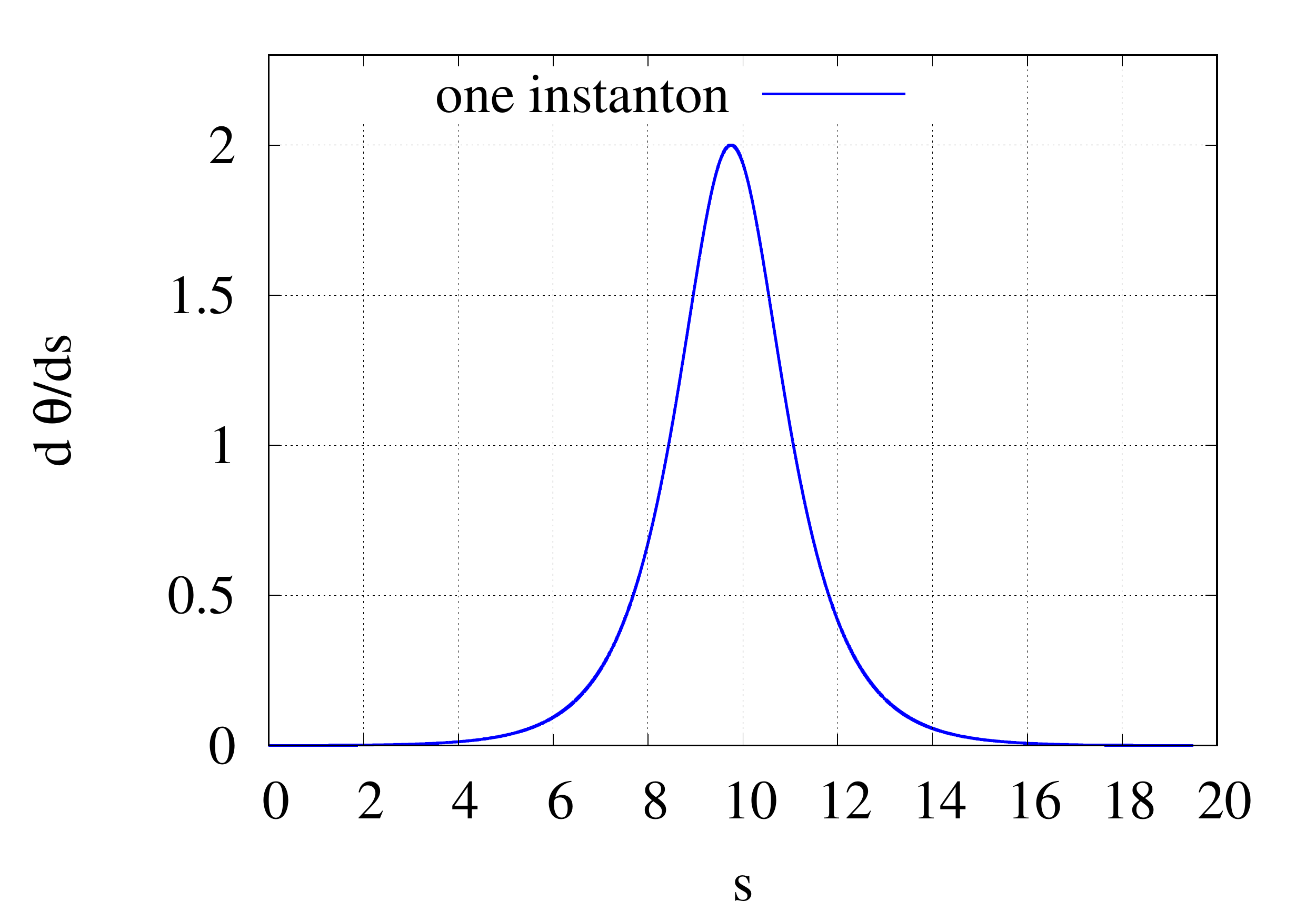}}
   \subfigure[] 
    {\label{fig:1instantonX}\includegraphics[width=0.15\textwidth,angle=270]{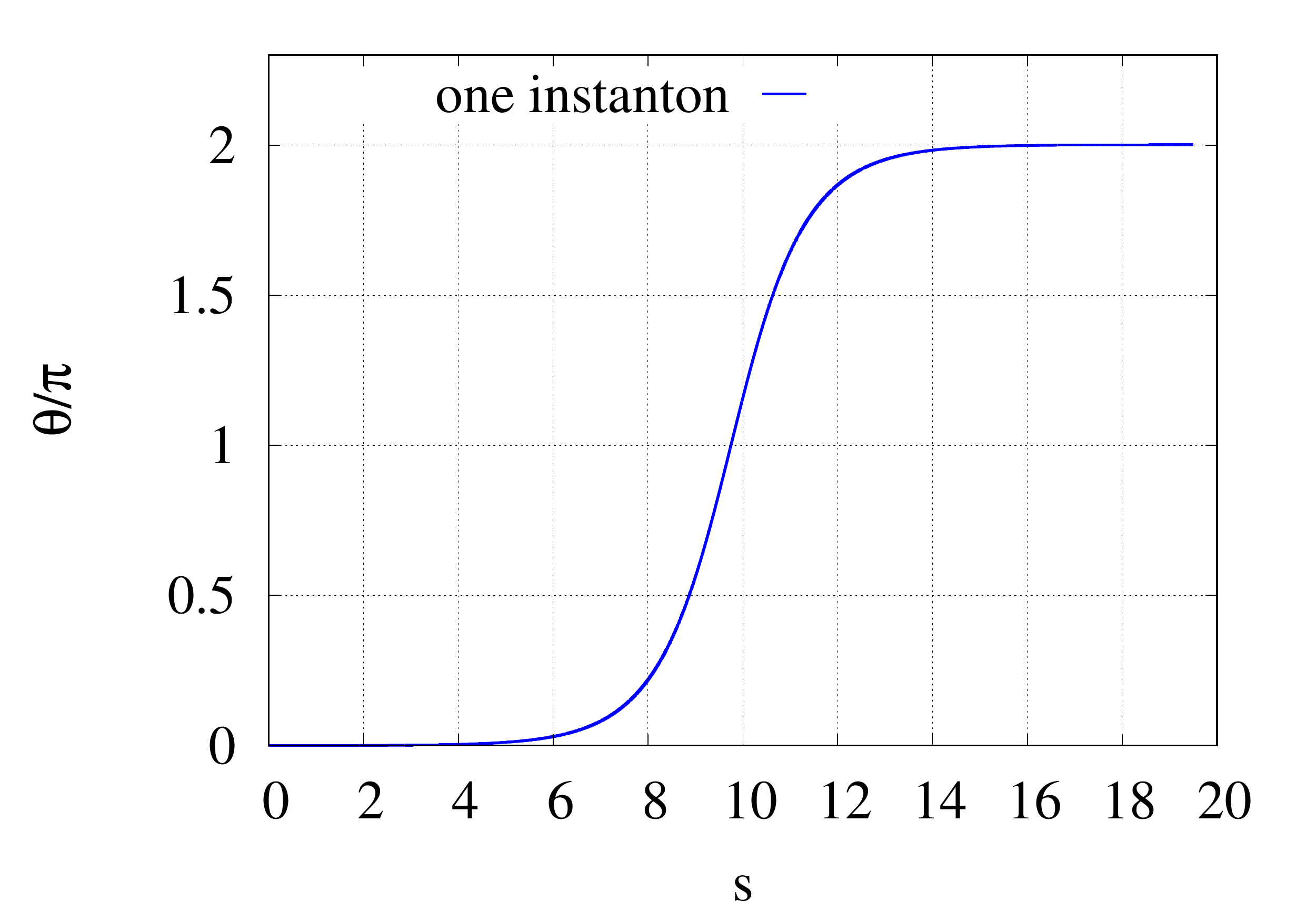}}
    \subfigure[] 
     {\label{fig:AinstantonV}\includegraphics[width=0.15\textwidth,angle=270]{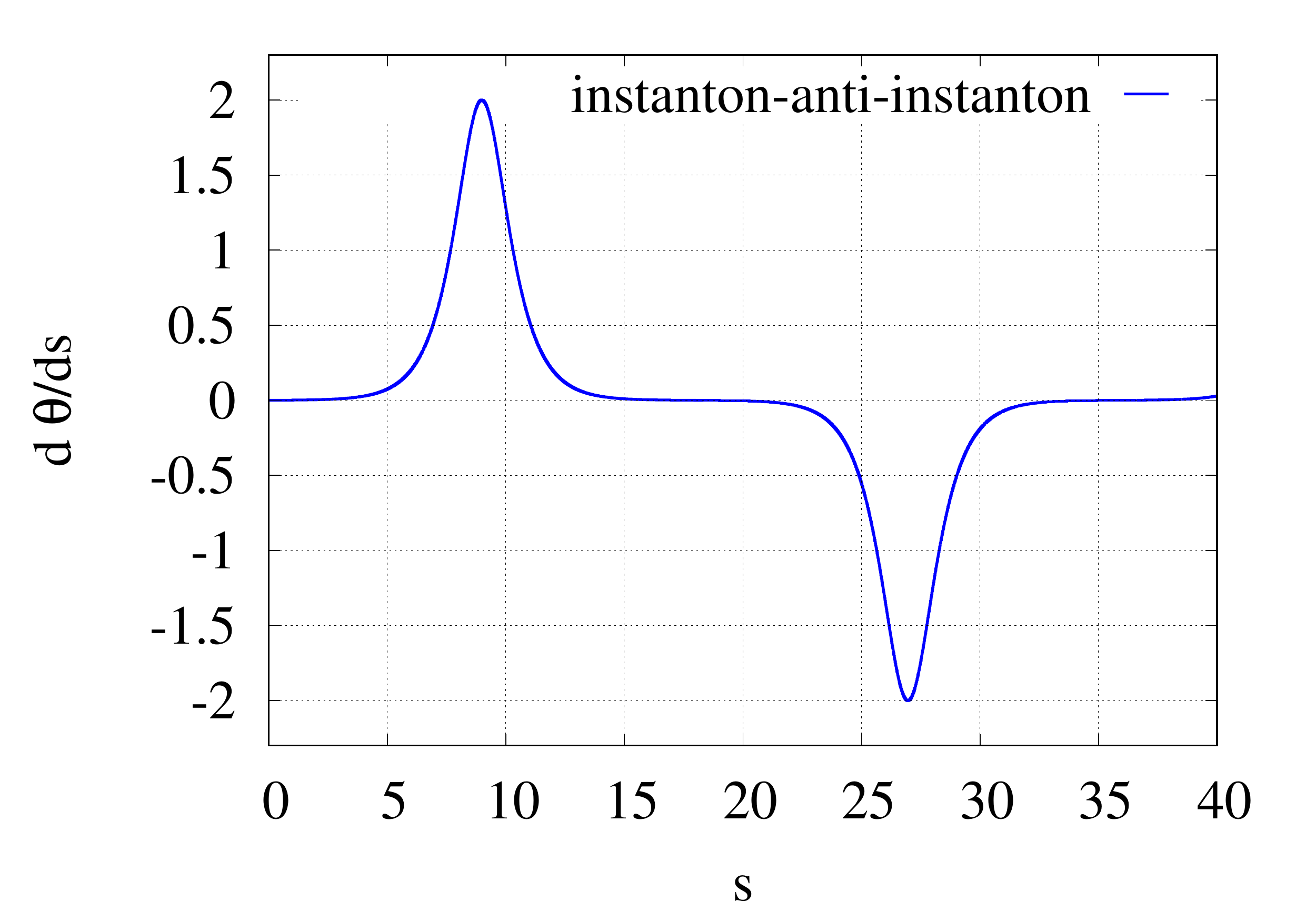}}
     \subfigure[] 
    {\label{fig:AinstantonX}\includegraphics[width=0.15\textwidth,angle=270]{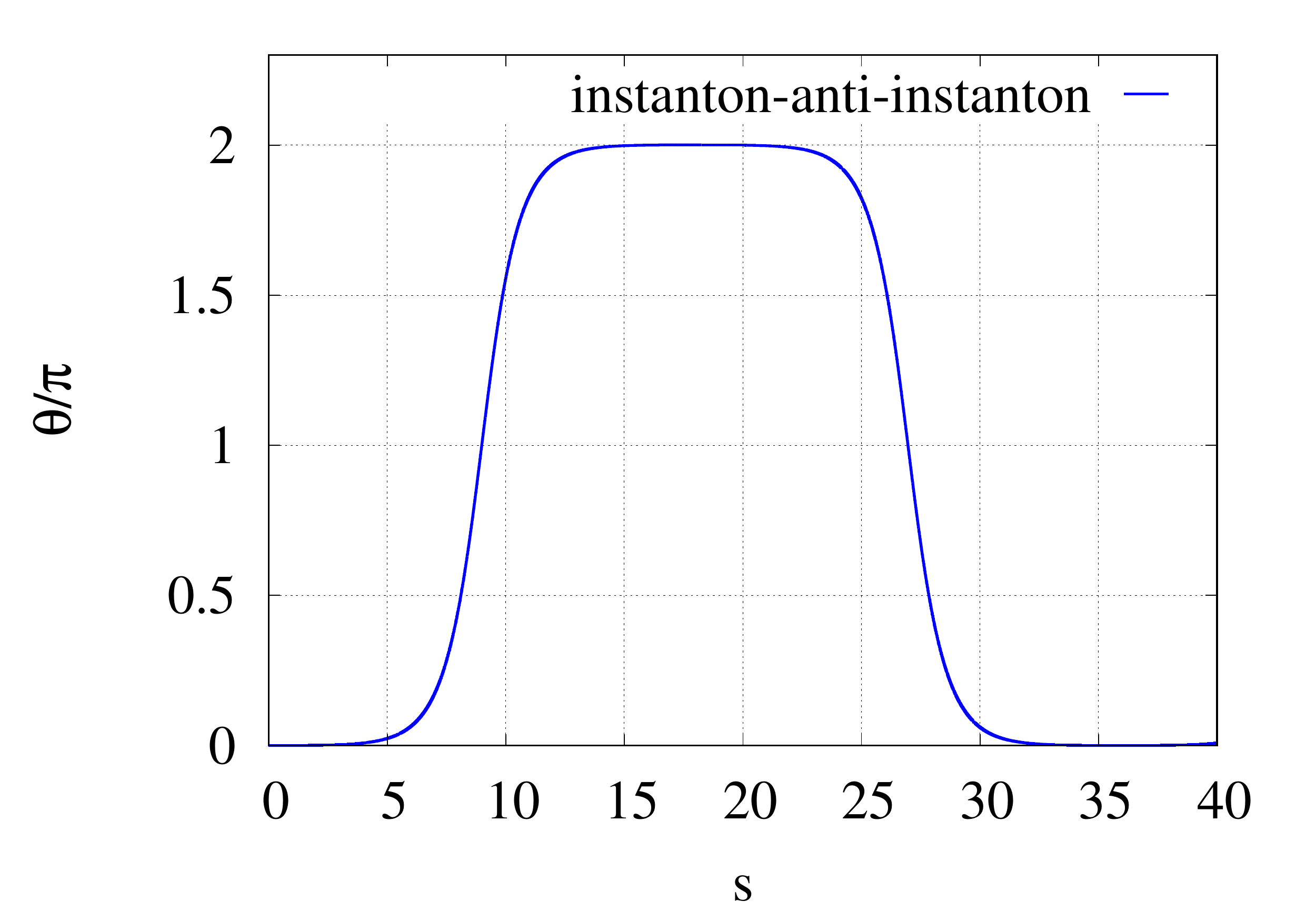}}
     \caption{Analytical profiles for instantons obtained from Eq. (\ref{eq:theta_solution}) for the case of single-instanton (a and b) and instanton-anti-instanton (c and d) solutions. Figures (a) and (c) show the derivative $\dot{\theta}$, while the plots (b) and (d) show the $\theta$ angle itself.}
   \label{fig:analytical_instantons}
\end{figure}

Example solutions of the equation (\ref{eq:theta_solution}), with initial conditions corresponding to a single instanton and instanton-anti-instanton pair are shown in Figure~\ref{fig:analytical_instantons}. One can see how the single instanton solution corresponds to the transition of $\theta$ angle between two equivalent values $0$ and $2 \pi$, while $\theta$ returns to $0$ in the case of the instanton-anti-instanton saddle. This observation allows us to introduce the winding number
\beq \label{eq:winding}
W=\frac{1}{2 \pi} \int_0^\beta d \tau \theta(\tau),
\eeq 
which is equal to the difference between the number of instantons and anti-instantons at a  given  site.

\section{\label{sec:AppendixC}Hessians for N-instanton saddle points}

  \begin{figure}[]
   \centering
   \subfigure[]   {\label{fig:Hessian1}\includegraphics[width=0.35\textwidth,clip]{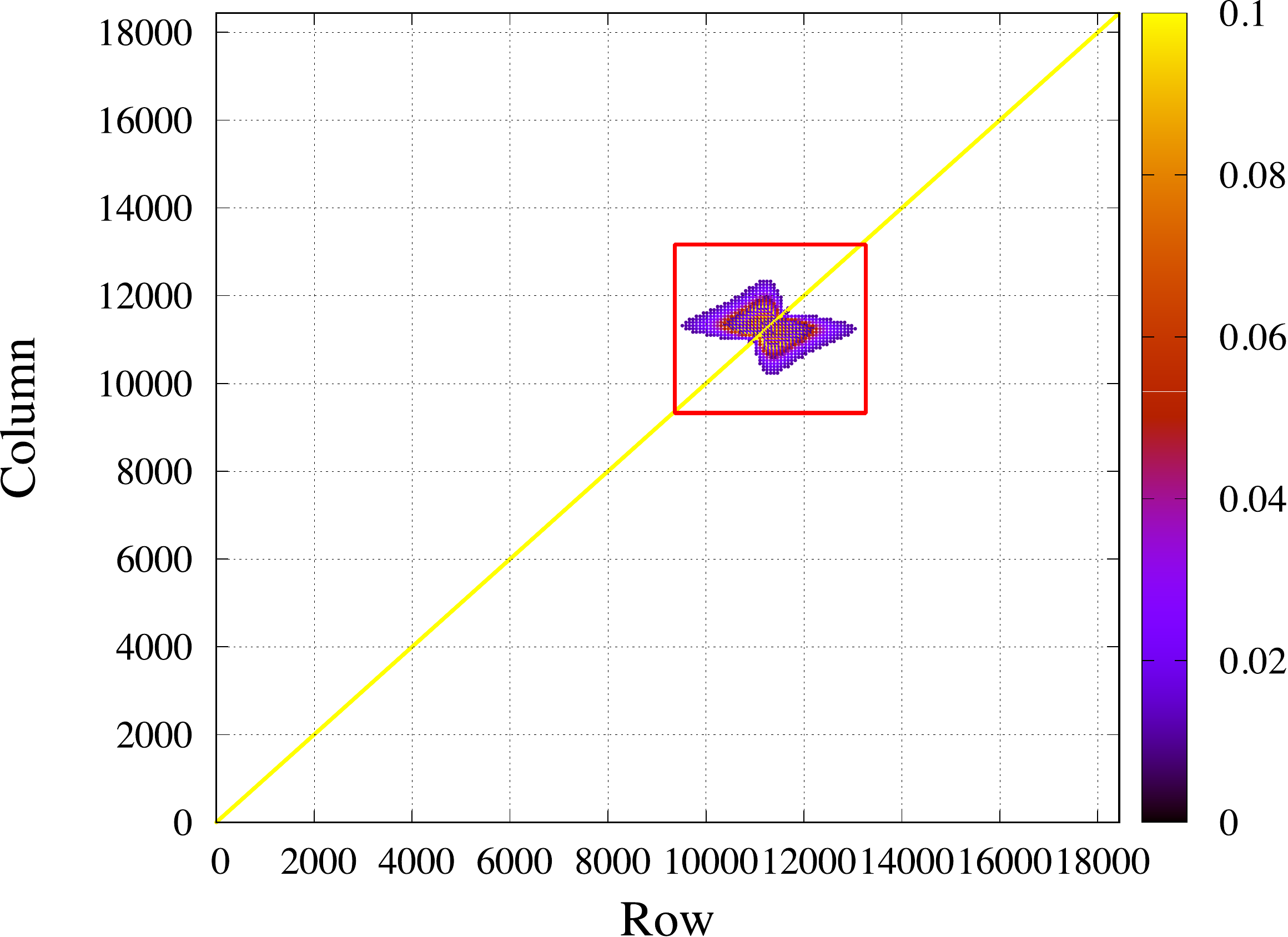}}
   \subfigure[]  {\label{fig:Hessian2}\includegraphics[width=0.35\textwidth,clip]{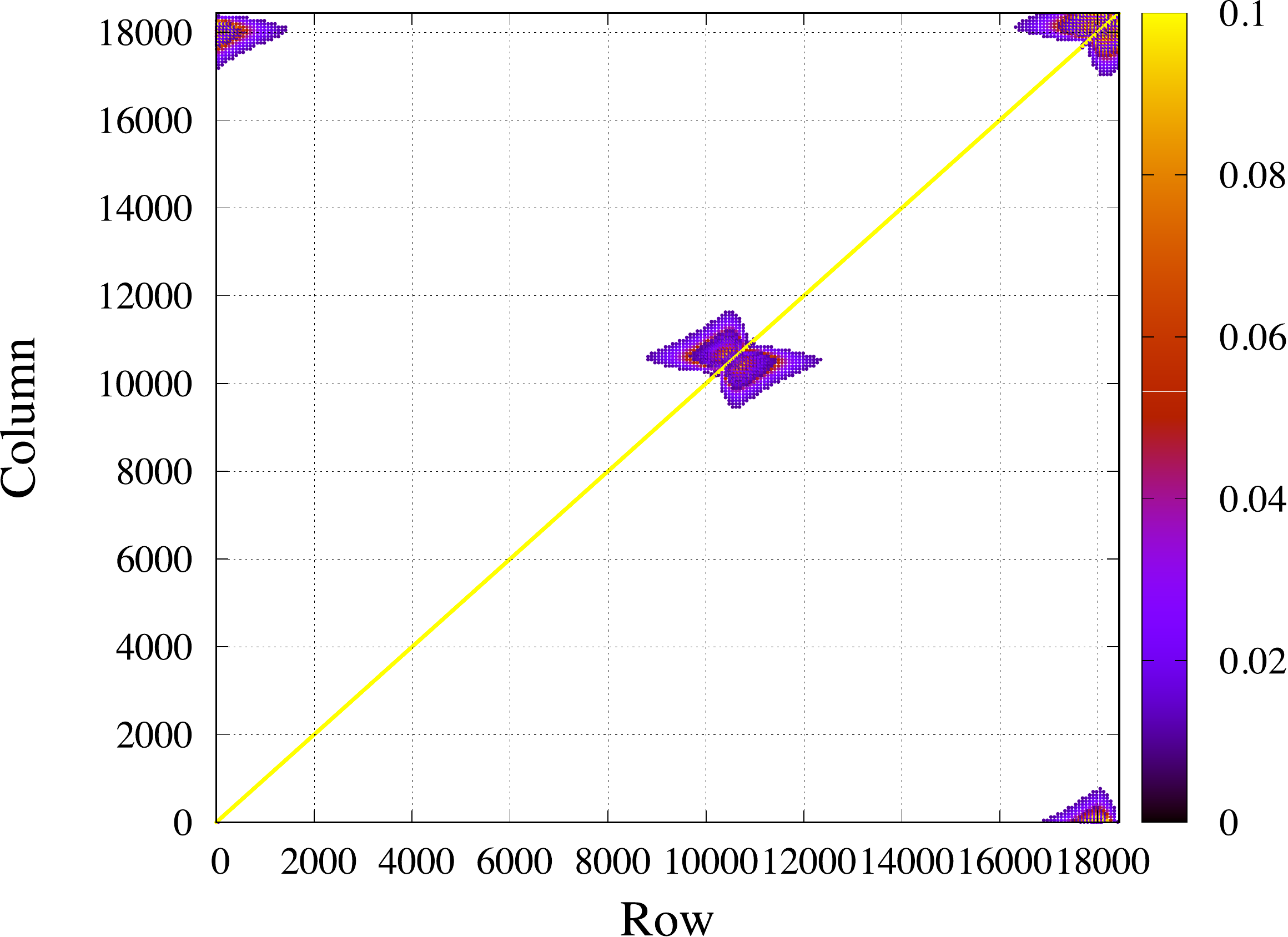}}
     \caption{(a) Absolute values of the elements of the reduced Hessian for the one-instanton saddle. The red rectangle denotes the region of the Hessian matrix which is sufficient to calculate the full determinant with high precision. (b) The elements of the reduced Hessian for the two-instanton saddle. To highlight the most important part of the matrix, we only plot the points where the element of the matrix is larger than 0.01. The conversion of the 2+1D coordinates to a linear index is done according to the rule (\ref{eq:conversion_index}). 
      These calculations were performed on a $6\times6$ lattice with $N _{\tau}=256$, $\beta  \kappa=20$, $U=5.0 \kappa$.}
   \label{fig:Hessians}
\end{figure}

In this Appendix the properties of the Hessians around saddle points containing one or more instantons are discussed in further detail. This is necessary, as the treatment of the Hessian is a crucial ingredient of the instanton gas model. 

For the construction of the analytical saddle point approximation, we need the ratio of the determinants for the one-instanton saddle and the vacuum saddle. This follows from the fact that all the weights of the non-trivial saddles can be computed in relation to the weight of the trivial vacuum saddle point. Thus, after excluding the zero mode corresponding to the translations in Euclidean time direction (\ref{eq:det_H_perp}), we arrive at the expression
\begin{eqnarray}
\frac {\det \mathcal{H}^{(1)}_\perp}{\det \mathcal{H}^{(0)} } &=&  \det \left( \mathcal{\tilde H}^{(1)}_\perp \right) \nonumber \\ &=& \det \left( \left( \mathcal {H}^{(1)} + \mathcal {P}^{(1)} \right) \left( \mathcal{H}^{(0)} \right)^{-1} \right).
  \label{eq:det_ratio}
\end{eqnarray}
With high accuracy, the projection operator $\mathcal {P}^{(1)} $ to the zero mode direction can be computed through a finite-difference derivative of the instanton field configuration in the Euclidean time direction. This is due to the fact that this derivative approximates the direction which is tangent to the valley in configuration space which is formed by the degenerate saddle
\begin{eqnarray}
\mathcal{P}^{(1)}_{i j} = \frac{ \mathcal{V}_i \mathcal{V}_j}{ |\mathcal{V}|^2} \\
 \mathcal{V}_{i(\ve{x}, \tau)}=  \phi^{(X, T)}_{\ve{x}, \tau}-\phi^{(X, T+\Delta \tau)}_{\ve{x}, \tau}. 
  \label{eq:projection1}
\end{eqnarray}
Here $(X, T)$ refers to the location of the instanton center, and the one-dimensional indices are related to the 2+1D coordinate $(\ve{x}, \tau)$ via the expression:
\begin{eqnarray}
i=2 N_1 N_2 \tau + x_0  N_1 N_2 + x_1 N_2 + x_2
  \label{eq:conversion_index}
\end{eqnarray}
with
\begin{eqnarray}
x_1=0...N_1-1, \nonumber \\
x_2=0...N_2-1, \\
x_0=0,1 \nonumber
  \label{eq:indexes}
\end{eqnarray}
being the two 2D coordinates and sublattice indices defining the spatial position of the lattice site. Here $N_1$ and $N_2$ are the lattice sizes in the two spatial directions. 

Due to the fact that the instanton configurations are local, the reduced Hessian $\mathcal{\tilde H}^{(1)}_\perp$ matrix is quite sparse: if the second derivative involves the fields far away of the center of the instanton, the corresponding elements of $\mathcal {H}^{(1)}$ are indistinguishable from the ones in $\mathcal {H}^{(1)}$  and they compensate each other in (\ref{eq:det_ratio}). 

In order to demonstrate the sparsity of the $\mathcal{\tilde H}^{(1)}_\perp$ matrix, we display a visualization of its elements $\left( \mathcal{\tilde H}^{(1)}_\perp \right)_{i j}$ in Figure \ref{fig:Hessians}\textcolor{red}{(a)}. The indices $i, j$ are connected to the corresponding 2+1D coordinates of the fields via the same expressions (\ref{eq:conversion_index}). Figure  \ref{fig:Hessians}\textcolor{red}{(a)} shows that the matrix is indeed quite sparse: elements, which substantially deviate from zero are located along the main diagonal and in the vicinity of the index which maps back to the coordinates of the center of the instanton.
In fact, we have checked that it is sufficient to compute the determinant of the small block encompassing the center of the instanton, illustrated by the red rectangle in Fig.~\ref{fig:Hessians}\textcolor{red}{(a)}.

Now let us now consider the two-instanton saddle point. Unlike the previous consideration in Section \ref{subsec:Interaction}, we neglect the interaction effects. This means that we assume that the instantons are far away from each other (which is generally true for the saddles with a low density of instantons), and we neglect the change in the action caused by the shift of one instanton with respect to another. Thus, there are two zero modes and the reduced Hessian $\mathcal{\tilde H}^{(2)}_\perp$ for the two-instanton saddle point is defined as 
\begin{eqnarray}
\frac {\det \mathcal{H}^{(2)}_\perp}{\det \mathcal{H}^{(0)} }=  \det \left( \mathcal{\tilde H}^{(2)}_\perp \right) = \nonumber \\ \det \left( \left( \mathcal {H}^{(2)} + \mathcal{P}^{(1)}+\mathcal {P}^{(2)} \right) \left( \mathcal{H}^{(0)} \right)^{-1} \right),
  \label{eq:det_ratio2}
\end{eqnarray}
where the projectors to the zero modes are computed via finite differences corresponding to the shifts of only one of the two instantons:
\begin{eqnarray}
\mathcal{P}^{(I)}_{i j} = \frac{ \mathcal{V}^{(I)}_i \mathcal{V}^{(I)}_j}{ |\mathcal{V}^{(I)}|^2} \nonumber \\
 \mathcal{V}^{(1)}_{i(\ve{x}, \tau)}=  \phi^{ ((X^{(1)}, T^{(1)}), (X^{(2)}, T^{(2)}))}_{\ve{x}, \tau}-  \nonumber \\ \phi^{ ((X^{(1)}, T^{(1)}+\Delta \tau),(X^{(2)}, T^{(2)}))}_{\ve{x}, \tau} \\
 \mathcal{V}^{(2)}_{i(\ve{x}, \tau)}=  \phi^{((X^{(1)}, T^{(1)}),(X^{(2)}, T^{(2)}))}_{\ve{x}, \tau}- \nonumber \\ \phi^{((X^{(1)}, T^{(1)}),(X^{(2)}, T^{(2)}+\Delta \tau))}_{\ve{x}, \tau}.  \nonumber
  \label{eq:projection2}
\end{eqnarray}
Here $\phi^{((X^{(1)}, T^{(1)}), (X^{(2)}, T^{(2)}))}_{\ve{x}, \tau}$ refers to the field configuration for two instantons, whose centers are located at the points $(X^{(1)}, T^{(1)})$ and  $(X^{(2)}, T^{(2)})$.

\begin{figure}
        \centering
        \includegraphics[width=0.35\textwidth, angle=0]{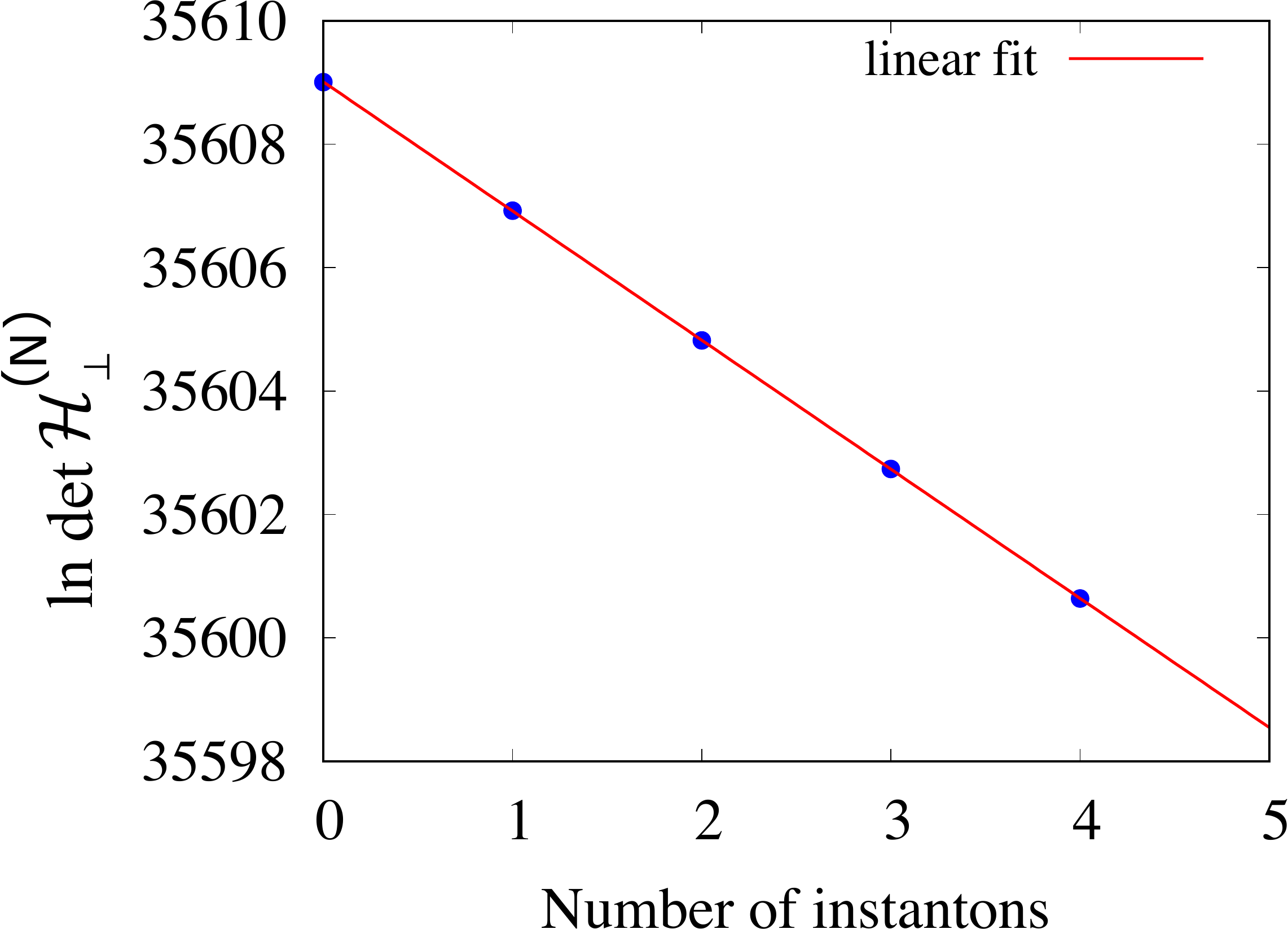}
        \caption{Numerical proof for the Eq.   (\ref{eq:general_hessian}). $\ln \det \mathcal{H}^{(0)}$ is shown for the vacuum, and $\ln \det \mathcal{H}^{(N)}_\perp$ is shown for the $N$-instantons saddle. The calculations were performed on a $6\times6$ lattice with $\beta \kappa=20$, $U=2\kappa$ and $N_\tau=256$.}
        \label{fig:general_hessian}
\end{figure}

The reduced Hessian matrix, $\mathcal{\tilde H}^{(2)}_\perp$, is shown in Figure \ref{fig:Hessians}\textcolor{red}{(b)}. If the instantons are far away from each other, the identical non-zero blocks corresponding to the single-instanton configurations are split along the main diagonal. This means that within the approximation of non-interacting instantons, the determinant of the reduced Hessian $\mathcal{\tilde H}^{(N)}_\perp$ for the $N-$instanton saddle point can be approximately computed as
\begin{eqnarray}
\frac {\det \mathcal{H}^{(N)}_\perp}{\det \mathcal{H}^{(0)} } \approx \left[ \det \left( \left( \mathcal {H}^{(1)} + \mathcal {P}^{(1)} \right) \left( \mathcal{H}^{(0)} \right)^{-1} \right) \right]^N. 
  \label{eq:general_hessian}
\end{eqnarray}
Due to its simplicity, this expression will be used in the construction of the analytical partition function for the instanton gas model. A numerical proof of this expression is presented in Figure \ref{fig:general_hessian}.

\section{\label{sec:AppendixD}Grand canonical Monte Carlo for instanton gas model}

In this Appendix we describe  the algorithm  used for sampling the instanton gas model  with the grand canonical partition function given by  Eq.~(\ref{eq:ClassicalModelInt}).   
The  state of the system is described by the set of $N$ coordinates $\{X_i, T_i\}$, $i=1...N$. $T_i\in(0,\beta)$ is the Euclidean time coordinate of the $i$-th instanton and  
\beq
X_i=(\nu_i,\ve{r}_i),
\label{eq:XDefinition}
\eeq
where $\nu_i=-1,1$ is the instanton-anti-instanton index, and $\ve{r}_i$ is the spatial coordinate of the $i$-th instanton. Thie spatial coordinate contains three components, including the sublattice index (see Eq.  (\ref{eq:conversion_index})).

The grand Canonical Monte Carlo utilizes a Markov chain where each update consists of  two  stages:  in the first stage, we update each coordinate of the instantons one by one, and in the second we change the total number of instantons.

  \begin{figure}[]
   \centering
   \subfigure[]  {\label{fig:instantonsNumberModelFullLat6}\includegraphics[width=0.35\textwidth,clip]{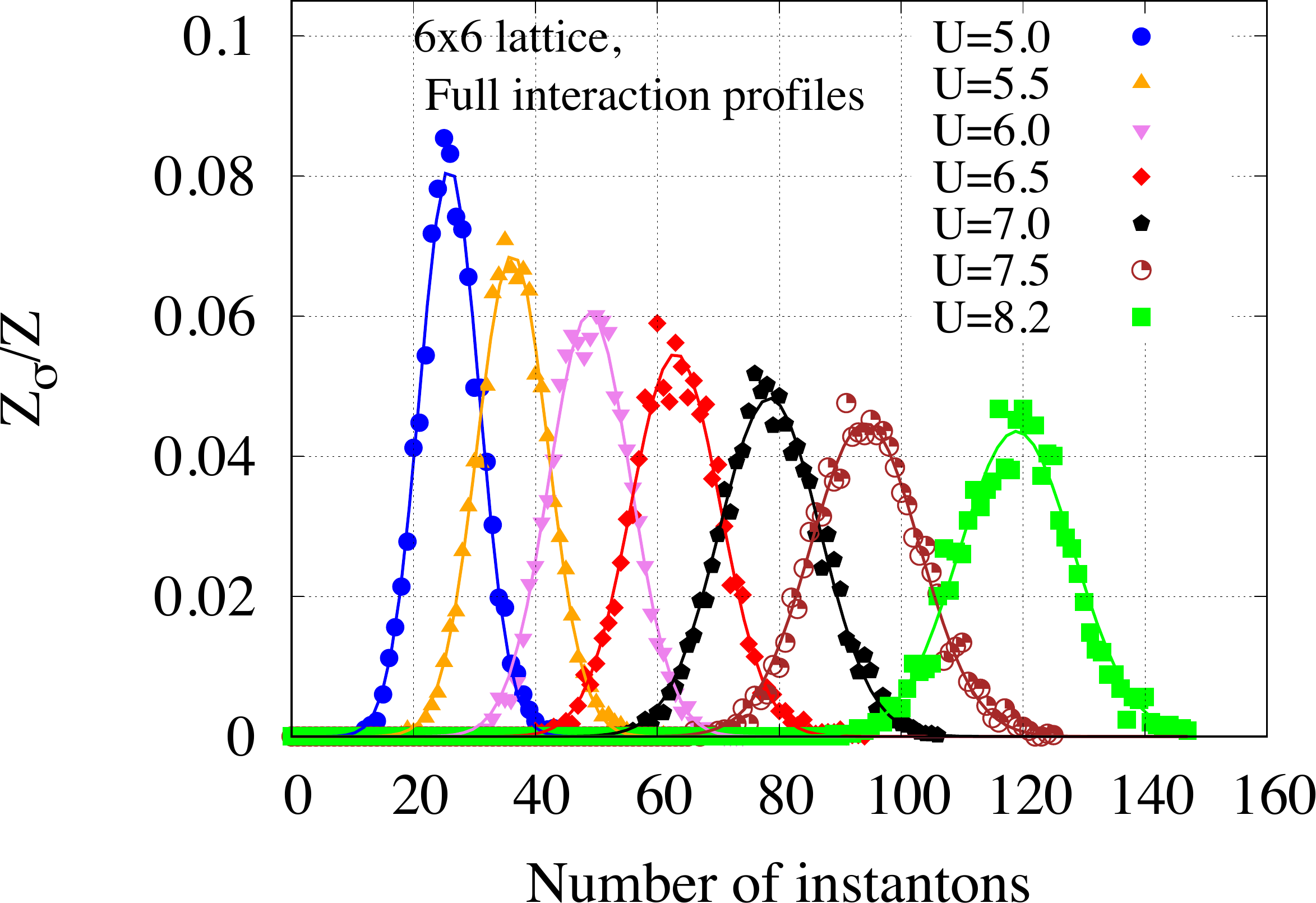}}
   \subfigure[]   {\label{fig:instantonsNumberModelFullLat12}\includegraphics[width=0.35\textwidth,clip]{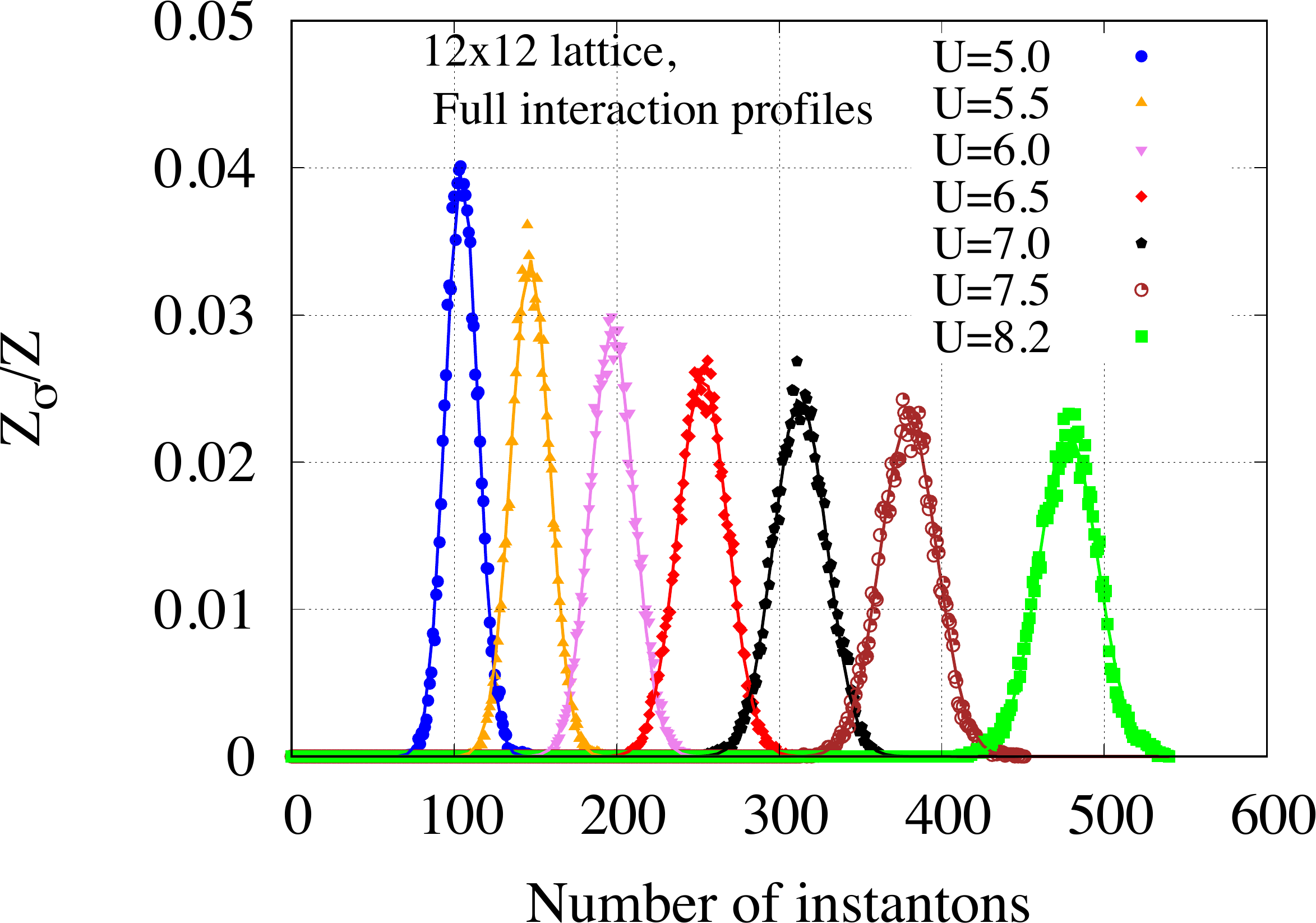}}
     \subfigure[]   {\label{fig:instantonsNumberModelRep}\includegraphics[width=0.35\textwidth,clip]{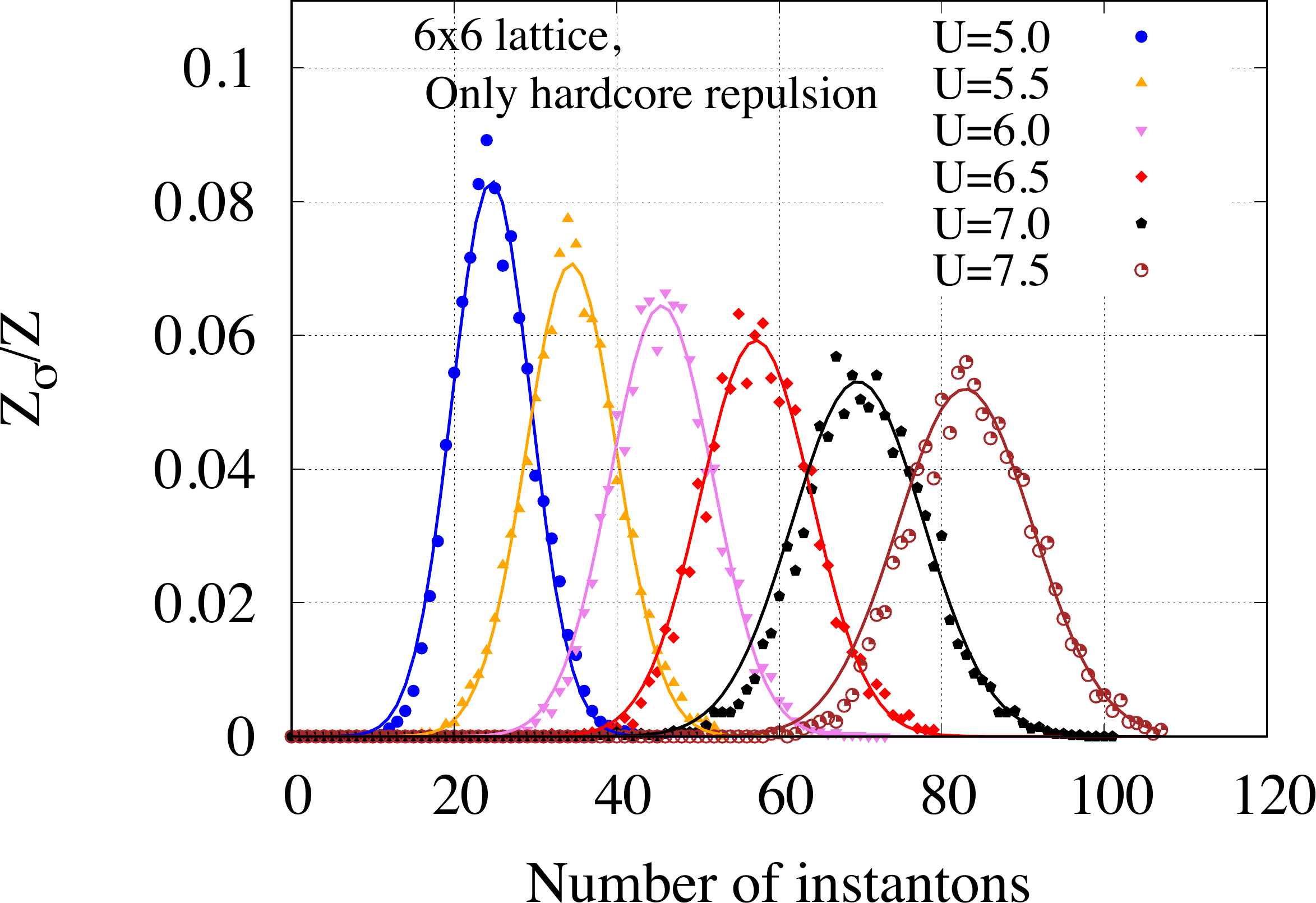}}
      \caption{The distribution of the number of instantons obtained from classical grand canonical Monte Carlo for instantons. The first two plots (a and b) show the results for the model which incorporates the full interaction profile (obtained from the $6\times6$ and $12\times12$ lattices), while the last plot (c) shows the distribution for the case where only a hardcore repulsion between the instantons is taken into account. For all of these calculations, $\beta \kappa=20$. Gaussian fits are also included (shown with lines of the same colors as the corresponding data sets).}
   \label{fig:instantonsNumberModel}
 \end{figure}

The individual updates of the instantons' coordinates $(X_i, T_i)\rightarrow(\tilde X_i, \tilde T_i)$ are made according to the standard Metropolis algorithm. The proposal distribution is defined according to the following rules: 
\begin{itemize}
    \item The new value of the Euclidean time coordinate $\tilde T_i$ is chosen according to the Gaussian distribution, with standard deviation $D_T$, centered about the old coordinate. Here $D_T$ is used as the set up parameter to tune the acceptance rate. 
    \item The new type of the instanton $\tilde \nu_i$ is chosen between instanton (1) and anti-instanton (-1) value with equal probability. 
     \item Proposals for the spatial coordinates and the sublattice index in $\tilde X_i$ are made simultaneously: we chose whether to move the instanton to one of the nearest-neighbours or to leave it at the same site. The probability is equal (25\%) for each variant, since we have three nearest neighbours on the hexagonal lattice. 
\end{itemize}
The Metropolis accept-reject step is made on the basis of the difference between the probability density for the old and the new configurations after the update of the coordinates for the $i$-th instanton. According to    Eq.~(\ref{eq:ClassicalModelInt}), 
the probability to accept   the new coordinates $(\tilde X_i, \tilde T_i)$ reads:
\begin{eqnarray}
P^{(1)}_i=\mbox{min}\left( e^{-\Delta \mathcal{E}^{(1)}_i}; 1 \right),
\label{eq:metropolis_coordinate}
\end{eqnarray}
where 
\begin{eqnarray}
\Delta \mathcal{E}^{(1)}_i=\sum_{j=1}^{N} \left(  U^{(2)} (\tilde X_i, X_j, \tilde T_{i} -T_{j}) \right. \nonumber  \\  \left.  -U^{(2)} (X_i, X_j, T_{i}-T_{j}) \right).
\label{eq:DeltaU1}
\end{eqnarray}

  \begin{figure}[]
   \centering
   \subfigure[]     {\label{fig:DistributionCheckCenter}\includegraphics[width=0.35\textwidth,clip]{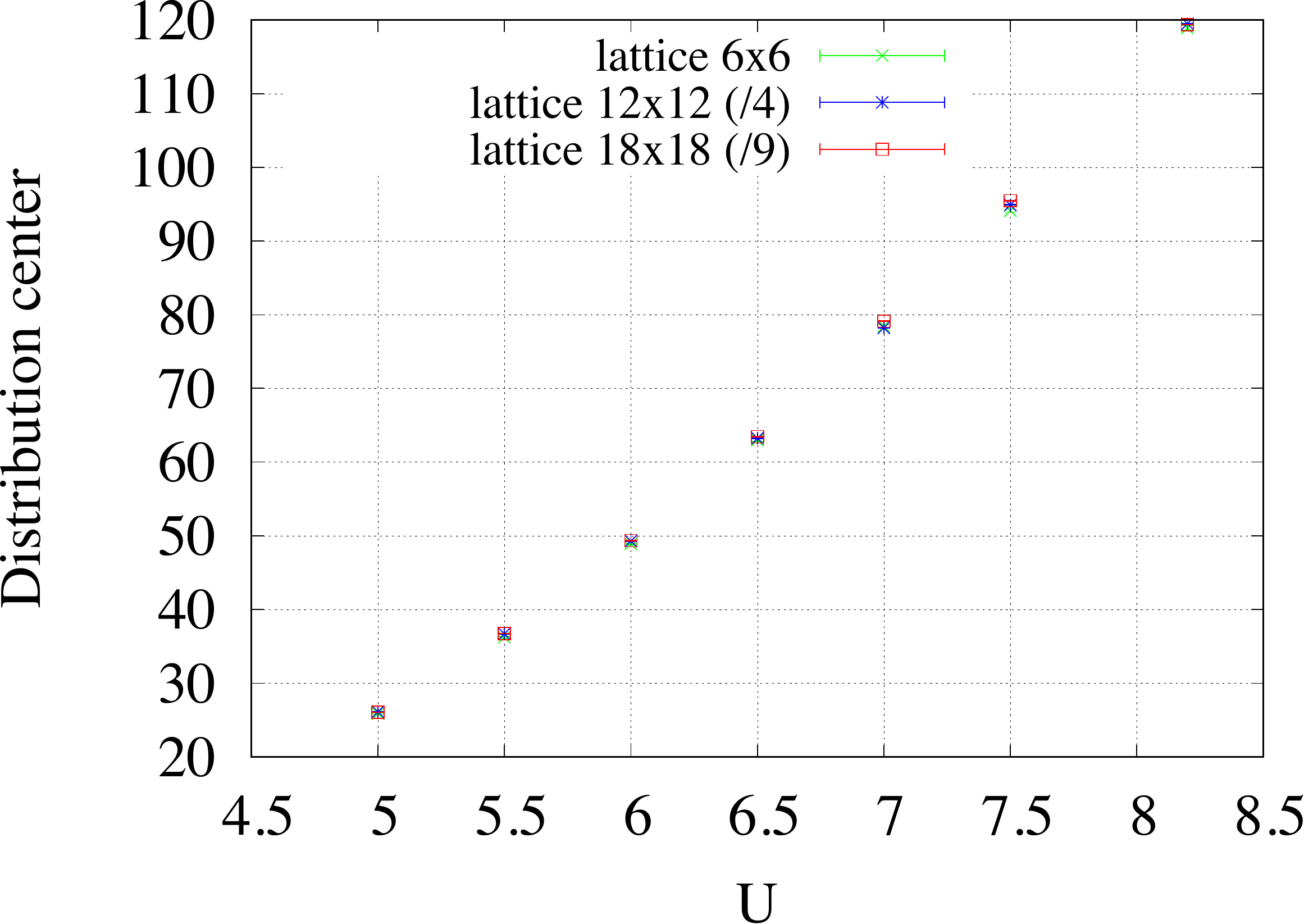}}
   \subfigure[]    {\label{fig:DistributionCheckWidth}\includegraphics[width=0.35\textwidth,clip]{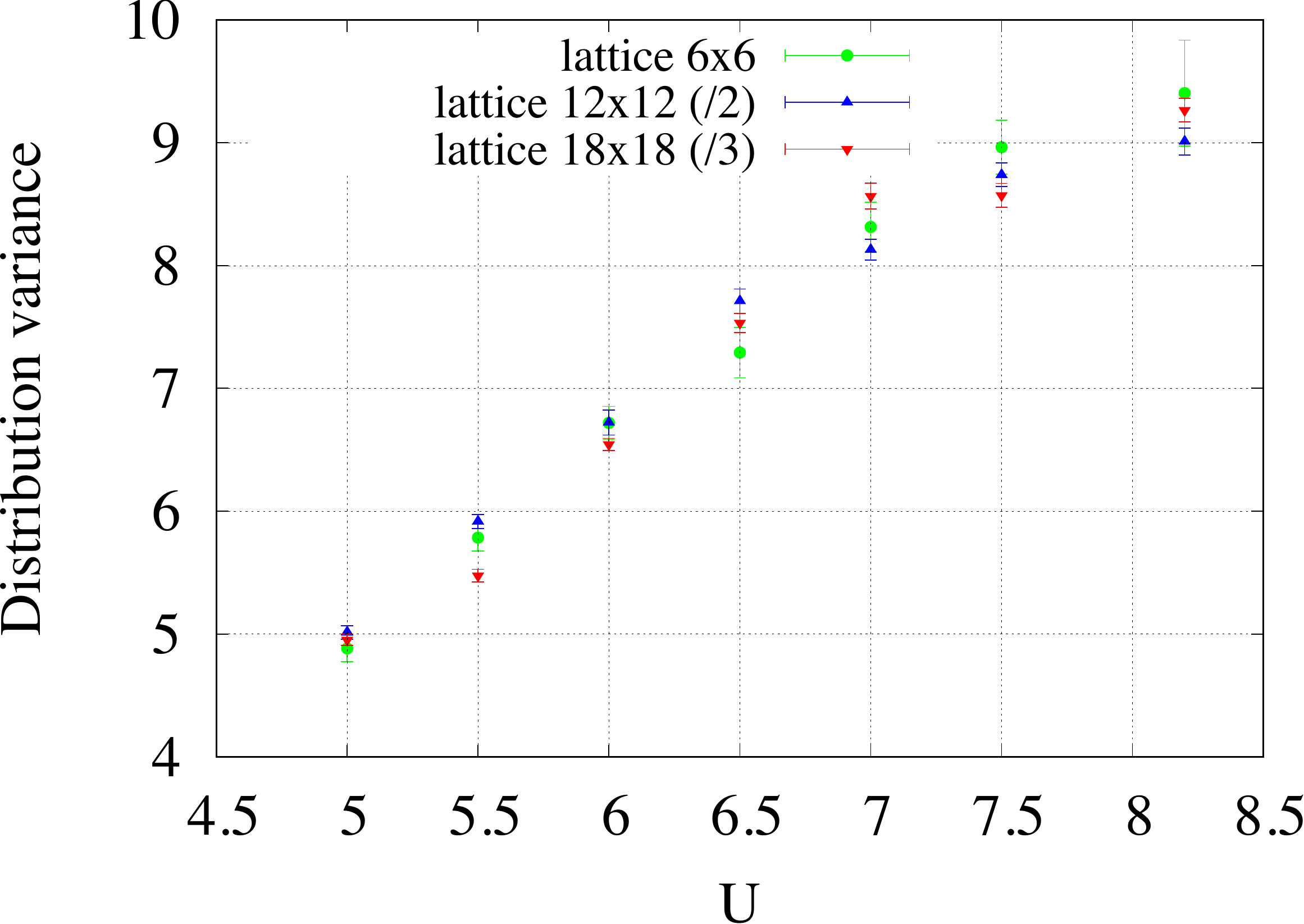}}
          \caption{(a) The average number of instantons, taken as the center of the distribution, from the classical grand canonical Monte Carlo simulations of the instanton gas model, taking into account only hardcore repulsion. (b) The variance of the distribution for the number of instantons from the same simulations. All data are obtained at $\beta \kappa=20$. Note the rescaling of the data points for the $12\times12$ and $18\times18$ lattices.}
   \label{fig:DistributionCheckModel}
 \end{figure}

The update of the configuration size $N$  is made according to the following algorithm:
\begin{itemize}
    \item We choose whether to increase or decrease the configuration size N by one with equal probability. 
   
    \item If we have chosen to increase the number of instantons, $N\rightarrow N+1$, we generate the new coordinates $(\tilde X_{\tilde i}, \tilde T_{\tilde i})$ with uniform distributions and insert them in the configuration at a random index $\tilde i=1... N+1$. The combined total proposal probability $\mathcal{T}_{N\rightarrow N+1}$ can be written as
  \begin{eqnarray}  
    \mathcal{T}_{N\rightarrow N+1}=\frac{1}{N+1} \frac{1}{\beta} \frac{1}{2 N_S}.
    \label{eq:T_N_N+1}
\end{eqnarray}  
    This expression reflects the uniform distribution of the index $\tilde i$ of the new instanton, and also the uniform distributions of the spatial coordinates, Euclidean time coordinate, as well as the instanton-anti-instanton index. The probability of the inverse process corresponds to the simple choice of one instanton for deletion. Thus 
\begin{eqnarray}  
    \mathcal{T}_{N+1\rightarrow N}=\frac{1}{N+1}.
    \label{eq:T_N+1_N}
\end{eqnarray}  
    These expressions are then combined into the Metropolis probability for the acceptance of the new configuration with the additional instanton:
\begin{eqnarray}
P^{(2)}_{\tilde i}=\mbox{min}\left( \frac{\frac{1}{(N+1)!} e^{\tilde \gamma-  \Delta \mathcal{E}^{(2)}_{\tilde i} }  \mathcal{T}_{N+1\rightarrow N} }{ \frac{1}{(N)!} \mathcal{T}_{N\rightarrow N+1}  }   ; 1 \right),
\label{eq:metropolis_N_N+1_initial}
\end{eqnarray}
where
\begin{eqnarray}
\Delta \mathcal{E}^{(2)}_{\tilde i}=\sum_{j=1; j\neq \tilde i}^{N+1} U^{(2)} (X_{\tilde i}, X_j, T_{\tilde i} -T_{j}).
\label{eq:DeltaU3}
\end{eqnarray}
Note, that in this case, unlike the case of Eq. (\ref{eq:metropolis_coordinate}), we should take into account the changing factorials in  (\ref{eq:ClassicalModelInt}). The final expression for the acceptance probability reads as
\begin{eqnarray}
P^{(2)}_{\tilde i}=\mbox{min}\left( \frac{e^{\tilde \gamma - \Delta \mathcal{E}^{(2)}_{\tilde i}}2 N_S \beta}{N+1}; 1 \right),
\label{eq:metropolis_N_N+1_final}
\end{eqnarray}

 \item If we have chosen to decrease the number of instantons, $N\rightarrow N-1$, we select one of the instantons for removal (again with equal probabilities). Thus, the proposal probabilities for the forward and inverse transitions can be written as:
 \begin{eqnarray}  
    \mathcal{T}_{N\rightarrow N-1}=\frac{1}{N}.
    \label{eq:T_N_N-1}
\end{eqnarray}  
and
  \begin{eqnarray}  
    \mathcal{T}_{N-1\rightarrow N}=\frac{1}{N} \frac{1}{\beta} \frac{1}{2 N_S}.
    \label{eq:T_N-1_N}
\end{eqnarray}  
 Subsequently, the Metropolis acceptance probability can be obtained in the same manner as Eq. (\ref{eq:metropolis_N_N+1_final}):
 \begin{eqnarray}
P^{(3)}_{\tilde i}=\mbox{min}\left( \frac{ N e^{-\tilde \gamma + \Delta \mathcal{E}^{(3)}_{\tilde i}}}{2 N_S \beta}; 1 \right),
\label{eq:metropolis_N_N+1_final}
\end{eqnarray}
where 
\begin{eqnarray}
\Delta \mathcal{E}^{(3)}_{\tilde i}= \sum_{j=1, j\neq \tilde i}^{N} U^{(2)} (X_{\tilde i}, X_j, T_{\tilde i} -T_{j}), 
\label{eq:DeltaU2}
\end{eqnarray}  
and $\tilde i$ is the index of the instanton selected for deletion. \end{itemize}

As a test of the classical Monte Carlo, we plot the distributions of the instanton number for different interaction potentials (Fig.~  \ref{fig:instantonsNumberModel}). As one can see, the distributions are perfectly fitted by Gaussian curves, in full agreement with the QMC data displayed in the Fig.~\ref{fig:instantonsNumberQMC}.  We also notice that the distributions are only slightly dependent on the exact form of the interaction profiles: one can compare Fig.~\ref{fig:instantonsNumberModel}\textcolor{red}{(a)}, which corresponds to the full interaction profiles with  Fig.~\ref{fig:instantonsNumberModel}\textcolor{red}{(c)}, where only hardcore repulsion of the two instantons at the same site was taken into account. 

We also check that the center of the distribution scales linearly with the lattice volume $V$ (Fig.~\ref{fig:DistributionCheckModel}\textcolor{red}{(a)}) and the width of the distribution scales as $\sqrt{V}$ (Fig.~\ref{fig:DistributionCheckModel}\textcolor{red}{(b)}), again in agreement with the QMC data.

\section{\label{sec:AppendixE}Instantons and the Gutzwiller projection}

In this Appendix we show how to establish a  connection between the instantons and the ground-state wave function.  In order to characterise the properties of the ground state following from the instanton approximation in a more intuitive way, we choose to work in the basis of occupation numbers. In this basis, the state at each site $\ve{x}$ is labeled by two numbers $n_{\ve{x}, e.}=0,1$ and  $n_{\ve{x}, h.}=0,1$, which characterize the number of electrons and hole. Due to the fact taht the creation-annihilation operators for electrons and holes, used in the Hamiltonian (\ref{eq:Hamiltonian}), are directly connected to those for electrons with spin up and spin down:
\beq
\hat a^\dag_{\ve{x}} = \hat a^\dag_{\ve{x}, \uparrow} \\ \nonumber
\hat b^\dag_{\ve{x}} = \pm \hat a_{\ve{x}, \downarrow},
\label{eq:el_holes_spin}
\eeq
where the sign in the latter equation alternates depending on the sublattice index, the states with fixed number of electrons and holes can be rewritten in terms of electrons with spin up and spin down. Here is an example for a single site:
\beq
|n_{e.}=0;  n_{h.}=0\rangle \rightarrow |n_{\uparrow}=0;  n_{\downarrow}=1\rangle, \nonumber \\ \nonumber
|n_{e.}=0;  n_{h.}=1\rangle \rightarrow |n_{\uparrow}=0;  n_{\downarrow}=0\rangle, \\ \nonumber
|n_{e.}=1;  n_{h.}=0\rangle \rightarrow |n_{\uparrow}=1;  n_{\downarrow}=1\rangle, \\ 
|n_{e.}=1;  n_{h.}=1\rangle \rightarrow |n_{\uparrow}=1;  n_{\downarrow}=0\rangle.
\label{eq:el_holes_state}
\eeq
Thus we can always return to the representation in terms of spin up and spin down electrons, despite the fact we are working in terms of electrons and holes for numerical convenience.

\begin{figure}
        \centering
        \includegraphics[width=0.35\textwidth, angle=270]{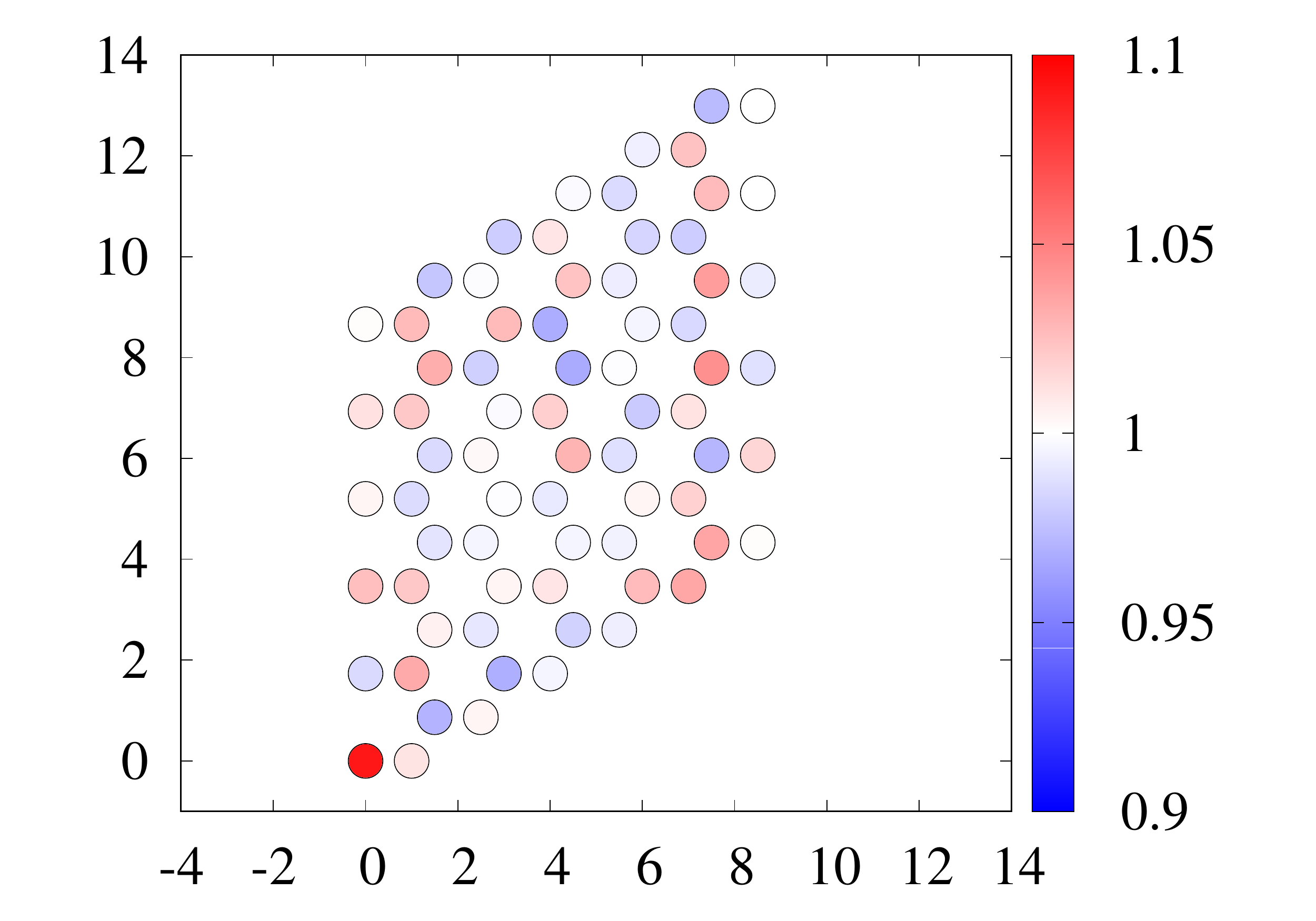}
        \caption{The ratio of frequencies $\mathcal{R}_{\ve{x}}$ from Eq. (\ref{eq:RatioFrequency}). These calculations were performed on the one instanton saddle on a $6\times6$ lattice with $\beta\kappa=20$, $U=6 \kappa$ and $N_\tau=512$. The instanton is located at the origin. }
        \label{fig:GutzProbDistr}
\end{figure}

The general wavefunctions for each configuration $\{n_{\ve{x}, e.}; n_{\ve{x}, h.} \}$ can be obtained as
\beq
|n_{\ve{x}, e.}; n_{\ve{x}, h.} \rangle = \prod_{\ve{x}: n_{\ve{x}, e.}=1}  \hat a^\dag_{\ve{x}}  \prod_{\ve{x}: n_{\ve{x}, h.}=1}   \hat b^\dag_{\ve{x}} |0 \rangle,
\label{eq:basis_occupation_numbers}
\eeq
where $|0 \rangle$ is the quantum state corresponding to  the  empty lattice. Here we consider only the states at half-filling:
\beq
\sum_{\ve{x}} n_{\ve{x}, e.} = \sum_{\ve{x}} n_{\ve{x}, h.} = \frac{N_S}{2}=V.
\label{eq:occupation_numbers_half_filling}
\eeq
Our aim is to look at the decomposition of interacting ground state $|\Omega_{int.}\rangle$ (as it appears in the instanton gas approximation) in terms of the vectors $|n_{\ve{x}, e.}; n_{\ve{x}, h.} \rangle$.   
Thus we need to compute the scalar products $\langle\Omega_{int.}| n_{\ve{x}, e.}; n_{\ve{x}, h.} \rangle$ within the instanton gas approximation. It is convenient to start with the vacuum state for the tight-binding Hamiltonian $|\Omega_{tb.}\rangle$, which is defined as a filled Dirac sea
\beq
|\Omega_{tb.} \rangle = \prod_{\ve{k};\sigma= 1,2}  \hat c^\dag_{\ve{k}, \sigma}|0 \rangle,
\label{eq:tight-binding_vacuum}
\eeq
where 
\beq
\hat c^\dag_{\ve{k}, 1}=\sum_{\ve{x}} V^{(-)}_{\ve{x}} (\ve{k}) \hat a^\dag_{\ve{x}} \\ \nonumber
\hat c^\dag_{\ve{k}, 2}=\sum_{\ve{x}} V^{(-)}_{\ve{x}} (\ve{k}) \hat b^\dag_{\ve{x}}
\label{eq:tight-binding_vacuum}
\eeq
is the creation operator for the state with negative energy and $V^{(-)}_{\ve{x}}(\ve{k})$ is the corresponding eigenvector of single-particle tight-binding Hamiltonian for the momentum $\ve{k}$.  
Thus we have
\beq
|n_{\ve{x}, e.}; n_{\ve{x}, h.} \rangle = \hat A(\{n_{\ve{x}, e.}; n_{\ve{x}, h.} \}) |\Omega_{tb.} \rangle,
\label{eq:A_operator1}
\eeq
where 
\beq
\hat A(\{n_{\ve{x}, e.}; n_{\ve{x}, h.} \}) = \prod_{\ve{x}: n_{\ve{x}, e.}=1}  \hat a^\dag_{\ve{x}}  \prod_{\ve{x}: n_{\ve{x}, h.}=1}   \hat b^\dag_{\ve{x}} \prod_{\ve{k};\sigma}  \hat c_{\ve{k}, \sigma}
\label{eq:A_operator2}
\eeq
Within the  formalism of  projective QMC (PQMC) \cite{Sorella89,Assaad08_rev}, we consider the following combination of traces: 
\beq
\left.
\mathcal{A}=\frac{\operatorname{Tr}\left( e^{-\hat H\beta} \hat A e^{-(\hat H_{0}-E^{vac}_{tb})\beta_P} \right) }{\operatorname{Tr}\left( e^{-\hat H\beta} \right) }\right|_{{\beta \rightarrow \infty}\atop {\beta_P \rightarrow \infty}} = \nonumber \\ \langle O_{int.}| \hat A| O_{tb}\rangle  \langle  O_{tb}|O_{int.} \rangle, 
\label{eq:projection_traces}
\eeq
where $\beta_P$ serves as the projection parameter, $\hat H$ is the full interacting Hamiltonian (\ref{eq:Hamiltonian}), $\hat H_0$ is its tight-binding part and $E^{vac}_{tb}$ is the energy corresponding to the tight-binding ground state $|\Omega_{tb.} \rangle$.  
After the Trotter decomposition  (\ref{eq:partition_function_trace}) and HS decomposition (\ref{eq:boltzmann_weight_trotter}) are made in both traces and the fermionic fields are integrated out, we arrive at the expression 
\beq
 \mathcal{A} \approx \sum_{\{\phi_{inst.} \}} A^{e.}_{\{\ve{x}: n_{\ve{x}, e.}=1\}} (\{\phi\})  A^{h.}_{\{\ve{x}: n_{\ve{x}, h.}=1\}} (\{\phi\}) \nonumber  \\ \times  \frac{|\det M^{proj.}_{el.}(\{ \phi\})|^2 }{|\det M_{el.}(\{ \phi\})|^2}. 
 \label{eq:projection_traces1}
\eeq
Here $\det M_{el.}(\{ \phi\})$ is defined in  (\ref{eq:M_continuous}) and $\det M^{proj.}_{el.}(\{ \phi\})$ is essentially the same except that it includes the additional exponent with projection:
\begin{eqnarray}
 \det M^{proj.}_{el.} = \det \left[ I + e^{-\beta_P h} \prod^{N_\tau}_{\tau=1} D_{2\tau-1} D_{2\tau} \right].
 \label{eq:M_projection}
\end{eqnarray}
The observables in (\ref{eq:projection_traces1}) are defined as the determinants of $V\times V$ matrices:
\beq
  A^{e.}_{\{\ve{x}\}} =  \begin{vmatrix}
\tilde g_{\ve{x}_1 \ve{k}_1}  & ... & \tilde g_{\ve{x}_1 \ve{k}_V}\\
\vdots & ... & \vdots\\
\tilde g_{\ve{x}_V \ve{k}_1} & ... & \tilde g_{\ve{x}_V \ve{k}_V}
\end{vmatrix} 
 \label{eq:A_el}
\eeq
where 
\beq
 \tilde g_{\ve{x}_i \ve{k}_j} = \sum_{\ve{y}} V^{(-)}_{\ve{y}} (\ve{k}_j)   g_{\ve{x}_i \ve{y}}
 \label{eq:tilde_g}
\eeq
with the fermionic propagator for electrons $g_{\ve{x}_i \ve{y}}$ computed at the zeroth time slice in the projected fermionic operator  (\ref{eq:M_projection}).
$A^{h.}_{\{\ve{x}\}}$ is the same with the exception of the complex conjugation of the fermionic propagator $g_{\ve{x}_i \ve{y}}$ in (\ref{eq:tilde_g}).

  \begin{figure}[]
   \centering
   \subfigure[]     {\label{fig:ThetaN}\includegraphics[width=0.35\textwidth,angle=270]{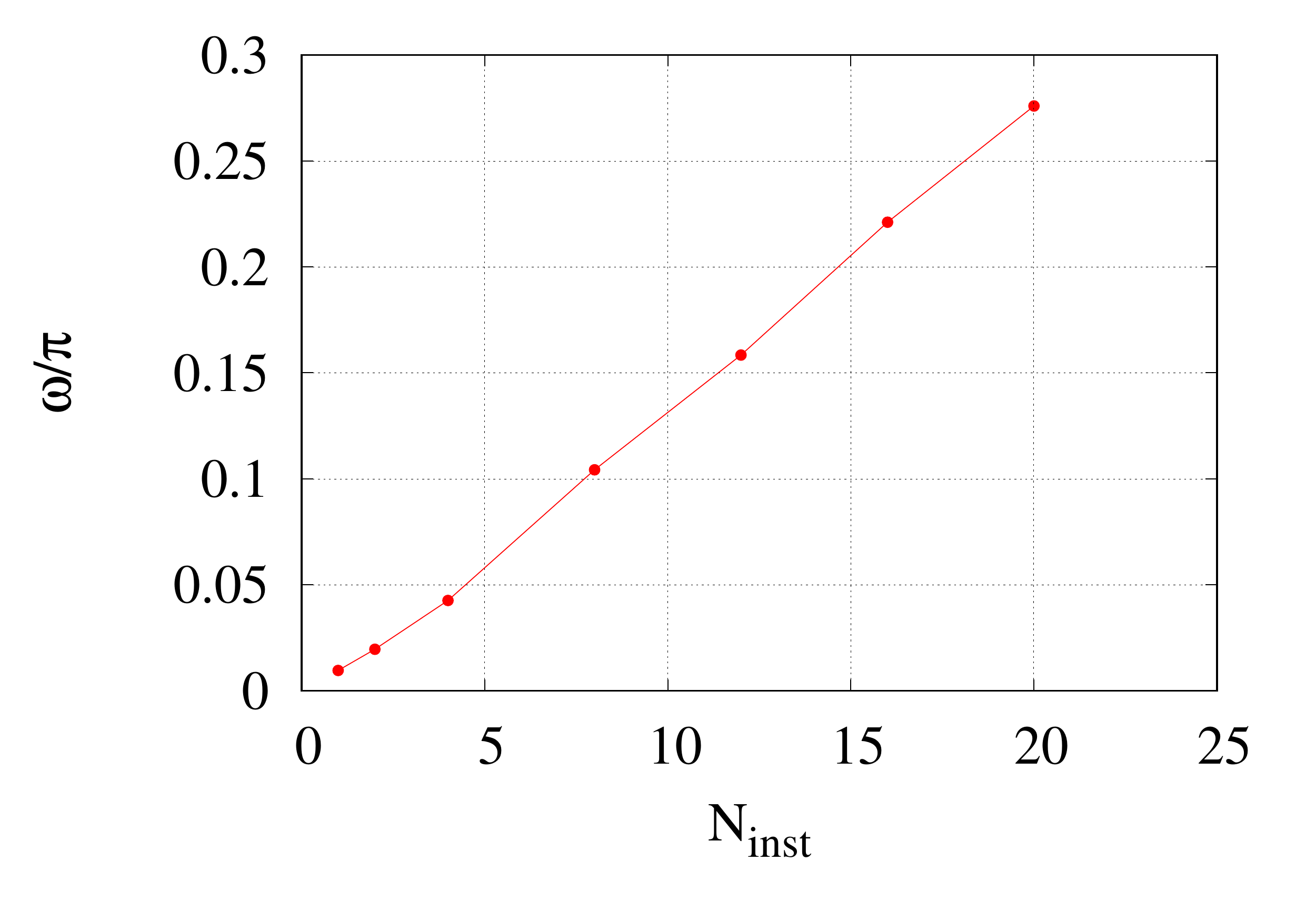}}
   \subfigure[]    {\label{fig:WavefunctionN}\includegraphics[width=0.35\textwidth,angle=270]{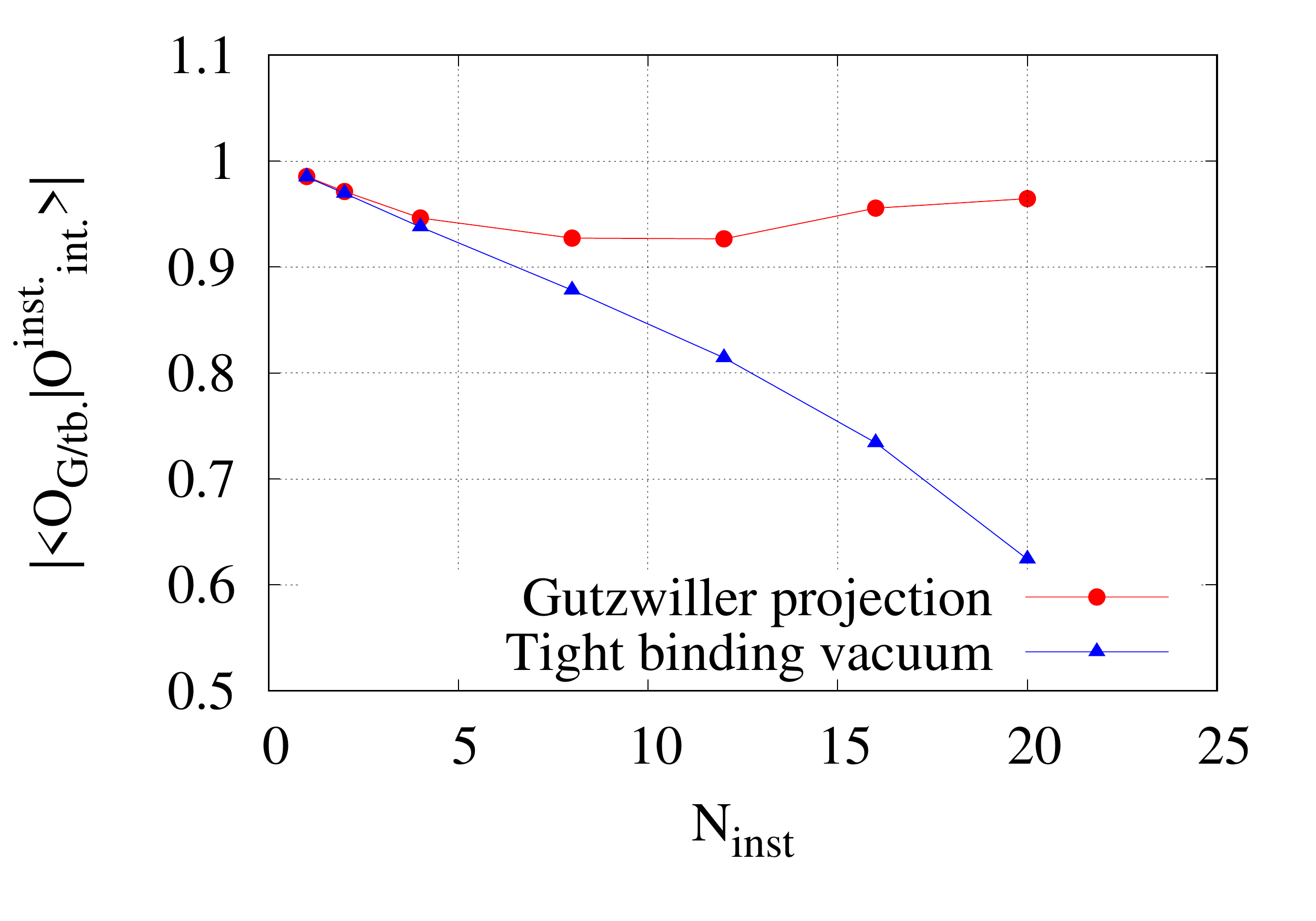}}
          \caption{(a) Dependence of the $\theta$ angle in the Gutzwiller projection for the optimal description of the $N-$instanton saddle point. All instantons are located at the same spatial site and placed equidistantly in Euclidean time.
          (b) Overlap between two variants of probe wavefunction and the wavefunction following from the $N-$ instanton saddle. These calculations were performed on a $6\times6$ lattice with $\beta \kappa=20$, $U=6 \kappa$ and $N_\tau=512$.  }
   \label{fig:GutzGoodness}
 \end{figure}

The approximation in (\ref{eq:projection_traces1}) is due to the usage of the sum over only dominant saddle point field configurations $\{\phi_{inst.} \}$ instead of the sum over all configurations of the auxiliary field. In fact we will use only the field configurations with one or more instantons at the origin in order to better understand the properties of the ground state following from the presence of these semiclassical objects. 

As the factor $\langle  O_{tb}|O_{int.} \rangle$ in (\ref{eq:projection_traces}) is some constant which is independent of the occupation numbers $\{n_{\ve{x}, e.}; n_{\ve{x}, h.} \}$, we can use the non-normalized probability distribution 
\beq
 |\mathcal{A}|^2  \sim  |\langle O^{inst.}_{int.}| \{n_{\ve{x}, e.}; n_{\ve{x}, h.} \}  \rangle|^2
  \label{eq:n_distribution}
\eeq
following from the approximate expression (\ref{eq:projection_traces1}) to generate the configurations $\{n_{\ve{x}, e.}; n_{\ve{x}, h.} \}$ employing standard MC techniques.  We finally  obtain a set of  basis vectors $|n_{\ve{x}, e.}; n_{\ve{x}, h.} \rangle$ distributed according to their weight within the interacting ground state $|O^{inst.}_{int.} \rangle $ corresponding to the instanton gas approximation. 

In order to characterize this distribution, we plot the ratio of the frequencies
\beq
\mathcal{R}_{\ve{x}}=\frac{\mathcal{F}((\uparrow)_{\ve{x}} \,\,\,\, \mbox{OR} \,\,\,\, (\downarrow)_{\ve{x}}) }{\mathcal{F}((\uparrow\downarrow)_{\ve{x}} \,\,\,\, \mbox{OR} \,\,\,\, (..)_{\ve{x}} ) }
\label{eq:RatioFrequency}
\eeq
where $\mathcal{F}((\uparrow)_{\ve{x}} \,\,\,\, \mbox{OR} \,\,\,\, (\downarrow)_{\ve{x}})$ corresponds to the frequency of the configurations $\{n_{\ve{x}, e.}; n_{\ve{x}, h.} \}$ with isolated spin up or spin down  at the site $\ve{x}$ within the whole set of such configurations generated in the Monte Carlo process. $(\uparrow \downarrow)_{\ve{x}}$ denotes the configuration where both spin orientations are present on this site.   $(..)_{\ve{x}}$  refers to the empty  site.  Both these configurations correspond  non-zero charge  at the corresponding lattice site. 
The connection of spin up and spin down indices to the occupation numbers $\{n_{\ve{x}, e.}; n_{\ve{x}, h.} \}$ can be established via (\ref{eq:el_holes_state}). 

The map of frequencies $\mathcal{R}_{\ve{x}}$ is plotted in Fig.~\ref{fig:GutzProbDistr}, where we generate configurations $\{n_{\ve{x}, e.}; n_{\ve{x}, h.} \}$ on the basis of a one-instanton saddle in  (\ref{eq:projection_traces1}). We clearly see that the configurations with single spin (up or down) at the origin (where the instanton is located) are more frequent than the configurations with non-zero charge. This result directly corroborates the results displayed in Figures \ref{fig:AnalyticalModel} and \ref{fig:SpinInstantons} in the main text where the increased spin localization was observed with increasing instanton density. 

Noting that the Gutzwiller projection can be used to describe this increasing localization, we check how well the ground state following from the saddle point approximation can be described by the local Gutzwiller  Ansatz:
\beq
| \Omega_{G}\rangle = \hat P_{\ve{x}}(\eta) | O_{tb.}\rangle,
\label{eq:Gutzwiller_vector}
\eeq
where the operator $\hat P_{\ve{x}}$ is defined as
\beq
\hat P_{\ve{x}}(\eta) = \mathcal{N} e^{-\eta \hat q_{x}^2},
\label{eq:Gutzwiller_operator}
\eeq
and the normalization constant is obtained from the condition $ \langle \Omega_{G} | \Omega_{G}\rangle =1$:
\beq
\mathcal{N} = \sqrt{\frac{2}{e^{-2 \eta}+1}}.
\label{eq:Gutzwiller_norm}
\eeq

We will consider several instantons located at a single spatial site, but separated in Euclidean time. Thus, the spatial site $\ve{x}$ in (\ref{eq:Gutzwiller_vector})  coincides with the location of the center of the instantons in the saddle point field configuration. In order to characterise the projection (\ref{eq:Gutzwiller_vector}) in a simpler way, we rewrite it in terms of the particle number operators $\hat n_{{\ve{x}}, \text{el.}}$ and $ \hat n_{{\ve{x}}, \text{h.}}$:
\beq
\hat P_{\ve{x}}(\eta) = - 2 (\hat n_{{\ve{x}}, \text{el.}} + \hat n_{{\ve{x}}, \text{h.}} - 2 \hat n_{{\ve{x}}, \text{el.}} \hat n_{{\ve{x}}, \text{h.}})  \sin \omega \\  \nonumber + \sqrt{2} \cos (\frac{\pi}{4}-\omega),
\label{eq:Gutzwiller_operator1}
\eeq
where 
\beq
\cos (\frac{\pi}{4}-\omega)=\frac{1}{\sqrt{e^{-2\eta}+1}}.
\label{eq:Gutzwiller_angle}
\eeq

The real parameter $\omega$ is tuned to maximize the overlap of the two states $ \langle O^{inst.}_{int.}| \Omega_{G}\rangle = \langle O^{inst.}_{int.}|\hat P_{\ve{x}}(\eta) | \Omega_{tb.}\rangle $. This quantity can be obtained analogously to Eqs. (\ref{eq:projection_traces}) and (\ref{eq:projection_traces1}), where the operator $\hat A$ is replaced by the operator $\hat P$ from Eq. (\ref{eq:Gutzwiller_operator1}) and the observable  $\mathcal{A}$ in (\ref{eq:projection_traces1}) is replaced by the corresponding observable for the operator $\hat P$.   The  unknown constant $\langle  O_{tb}|O^{inst.}_{int.} \rangle$ can be computed within the instanton gas approximation via the sum
\beq
| \langle  O_{tb}|O^{inst.}_{int.} \rangle |^2 = \sum_{\{\phi_{inst.} \}}  \frac{|\det M^{proj.}_{el.}(\{ \phi\})|^2 }{|\det M_{el.}(\{ \phi\})|^2}. 
 \label{eq:int_tb_sc_prod}
\eeq
As we are looking at the properties of the ground state corresponding to a set of instantons located at the same spatial site,  only this multi-instanton saddle  point field configuration is included in the sum  (\ref{eq:int_tb_sc_prod}). 

The results are shown in Figs.~\ref{fig:GutzGoodness}\textcolor{red}{(a)} and \ref{fig:GutzGoodness}\textcolor{red}{(b)}. 
First, we look at the dependence of $\omega$ on the number of instantons (Fig. ~\ref{fig:GutzGoodness}\textcolor{red}{(a)}): 
the angle grows almost linearly. Second, we plot the dependence of the scalar product $\langle  O_{G}|O^{inst.}_{int.} \rangle$ for this optimal $\omega$ on the number of instantons and compare it with  $\langle  O_{tb}|O^{inst.}_{int.} \rangle$ (Fig. ~\ref{fig:GutzGoodness}\textcolor{red}{(b)}). As one can see, the overlap with the  tight-binding vacuum quickly decays as  the instanton number increases, but the overlap with the Gutzwiller Ansatz is stable  and  takes  the  value  0.95. Thus, we can conclude that the instanton gas approximation corresponds well to the Gutzwiller projection, with the added dynamics in Euclidean time allowing us to go beyond the properties of the ground state and to look at the properties of the spectral function, as done in Fig.~\ref{fig:SpectralFunctionsU6Comparison}.

\bibliography{thimbles,fassaad,savvas}

\end{document}